\documentclass[aip,pop,reprint,showkeys,floatfix, graphics]{revtex4-2} 

\usepackage[margin=0.8in]{geometry}

\newcommand{\comment}[1]{}
\usepackage[english,french]{babel}
\usepackage{printlen}
\usepackage{color}
	\usepackage[toc,page]{appendix}

	\usepackage[T1]{fontenc}
	\usepackage[colorlinks=true,allcolors=blue]{hyperref}
	\usepackage{url}
	\usepackage[pdftex]{graphicx}
	\usepackage{amsmath}
	\usepackage{amssymb}
	\usepackage{appendix}
	\usepackage{caption}
	\usepackage{subcaption}
	\usepackage{float}
	\usepackage{amsmath}
	\usepackage{enumitem}
	\usepackage{makecell}
	\usepackage{txfonts}
	\usepackage{setspace}
	\usepackage[svgnames]{xcolor}
	\usepackage{hhline}
	\usepackage{tabu}
	\usepackage[multidot]{grffile}
	\definecolor{klein}{RGB}{0,47,167}
	\definecolor{malachite}{RGB}{31,160,85}
	\definecolor{DBlue}{rgb}{0.0, 0.0, 1.0}
	\definecolor{DGreen}{rgb}{0.5, 1.0, 0.0}

	\newcommand{\cor}{\textcolor{black}}
	\newcommand{\corNew}{\textcolor{black}}
	\renewcommand{\thefootnote}{\Roman{footnote}} 

	\newcommand{\Fref}[1]{Fig.~\ref{#1}}
	\newcommand{\Frefs}[2]{Figs.~\ref{#1}-\ref{#2}}
	\newcommand{\fref}[1]{Fig.~\ref{#1}}
	\newcommand{\frefs}[2]{Figs.~\ref{#1}-\ref{#2}}
	
	\newcommand{\Eref}[1]{Eq.~(\ref{#1})}
	
	\newcommand{\eref}[1]{Eq.~(\ref{#1})}

	\newcommand{\Cref}[1]{Chapter~(\ref{#1})}
	
	\newcommand{\cref}[1]{chapter~(\ref{#1})}

	\newcommand{\Sref}[1]{Sec.~(\ref{#1})}
	
	\newcommand{\sref}[1]{Sec.~(\ref{#1})}
	\newcommand{\srefs}[2]{Secs.~(\ref{#1})-(\ref{#2})}

	\newcommand{\curl}[1]{\nabla\times{\boldsymbol{ #1}}} 
	\newcommand{\cross}[2]{{\boldsymbol{#1}}\times{\boldsymbol{#2}}} 

	\newcommand{\GT}[1]{\nabla T_{#1}/T_{#1}}
	\newcommand{\GN}[1]{\nabla n_{#1}/n_{#1}}
	\newcommand{\RLT}[1]{-R_0\nabla T_{#1}/T_{#1}}
	\newcommand{\RLN}[1]{-R_0\nabla n_{#1}/n_{#1}}

\begin{document}
\selectlanguage{english}
\title{Linear and Nonlinear Dynamics of Self-Consistent Collisionless Tearing Modes in Toroidal Gyrokinetic Simulations}
\author{F.~Widmer}
\altaffiliation[Guest researcher at ]{Max Planck Institute for Plasma Physics, 85748 Garching, Germany}
\affiliation{Headquarters for Co-Creation Strategy, National Institutes of Natural Sciences, Minato-ku, Tokyo 105-0001, Japan}
\email[Electronic mails: \href{mailto:f.widmer@nins.jp}{f.widmer@nins.jp} and ]{fabien.widmer@ipp.mpg.de}
\author{E.~Poli}
\affiliation{Max Planck Institute for Plasma Physics, 85748 Garching, Germany}
\affiliation{Graduate School of Energy Science, Kyoto University, Uji, Kyoto 611-0011, Japan}
\author{A.~Mishchenko}
\affiliation{Max Planck Institute for Plasma Physics, 17491 Greifswald, Germany}
\author{A.~Ishizawa}
\affiliation{Graduate School of Energy Science, Kyoto University, Uji, Kyoto 611-0011, Japan}
\author{A.~Bottino}
\affiliation{Max Planck Institute for Plasma Physics, 85748 Garching, Germany}
\author{T.~Hayward-Schneider}
\affiliation{Max Planck Institute for Plasma Physics, 85748 Garching, Germany}
\date{\today}
\begin{abstract}
	We investigate tearing modes (TM) driven by current density gradient in collisionless tokamak plasmas by using the electromagnetic gyrokinetic simulation code ORB5.
We elucidate the TM width by simulations for flat profiles, as the absence of background diamagnetic flows implies a small rotation-speed, while finite-gradients are included to investigate the TM rotation.
 For flat profiles, the initial saturation width of nonlinearly driven magnetic islands is related to the TM linear growth rate; however, large islands in the initial saturation phase are prone to current density redistribution that reduces the island width in the following evolution. Island-induced $\cross{E}{B}$ and diamagnetic sheared flows develop at the separatrix, able to destabilize the Kelvin-Helmholtz instability (KHI). The KHI turbulence enhances a strong quadrupole vortex flow that reinforces the island decay, resulting in a strong reduction of the island width in an eventual steady state. This process is enhanced by trapped electrons. For finite gradients profile, the TM usually rotates in the electron diamagnetic direction, but can change direction when the ion temperature gradient dominates the other gradients. The reduced growth of the TM by diamagnetic effects results in a moderate island size, which remains almost unchanged after the initial saturation. At steady state, strong zonal flows are nonlinearly excited and dominate the island rotation, as expected from previous theoretical and numerical studies. When $\beta$ is increased, the TM mode is suppressed and a mode with the same helicity but with twisting parity, coupled with the neighboring poloidal harmonics, is destabilized, similar to the kinetic ballooning mode.
\end{abstract}
\maketitle

\section{Introduction\label{sec:Intro}}

Magnetic reconnection is an ubiquitous phenomenon in magnetized plasmas, from astrophysics to fusion plasma experiments. It changes the magnetic field topology leading to
the appearance of magnetic island shaped structures.\cite{furth_PF1973} The magnetic energy is released in the process and a lower energy plasma state is attained by the system.\cite{Yamada2010,BiskampBook00}
In fusion experiments, the nested magnetic flux surfaces
configuration is violated by the appearance of magnetic islands. As
the island grows radially in size, it provides radial shortcuts for the particle motion.
The result is a flattening of the radial pressure profile, lowering the energy confinement.\cite{SauterPoP97a,IshizawaPPCF19,ZohmBook2022}

Magnetic islands can develop in tokamaks in the nonlinear phase of a current-driven tearing mode (TM), characterized by a positive tearing parameter $\Delta'>0$.\cite{furth_PF1973} Neoclassical Tearing modes (NTM) are of particular
concern for present and future fusion devices as they can develop also for TM-stable current profiles ($\Delta'<0$) as a consequence of the bootstrap-current perturbation arising from the pressure flattening mentioned above.\cite{carrera_PoF1986}
NTMs often need an initial seed island, which is influenced by the island rotation through the polarization current, to grow,\cite{kotschenreuther_PF1985,Carreras1981a,WidmerITER2018} the classical tearing mode has the ability to further develop into an NTM by playing the role of the seed.\cite{reimerdes_prl2002}
The dynamics of non-linear islands are known to be considerably
altered by kinetic and toroidicity effects even within a drift-kinetic or neoclassical picture.\cite{Poli_PPR16}
This has prompted several studies in this sense, starting with the
analysis of the plasma response to a prescribed magnetic island.\cite{PoliNF09,PoliPPCF10a,HornsbyEPL10,HornsbyPoP2010,WaltzPoP12,ZarzosoPoP15,BanonNavarroPPCF2017,KwonPoP18,FangPoP19,MutoPoP22,LiNF23}

Meanwhile, a few self-consistent gyrokinetic investigations of the tearing instability and
nonlinear evolution are available as well.\cite{HornsbyPPCF15,HornsbyPPCF16,ZarzosoPoP19,JitsukNF24}
We mention for completeness that several gyrokinetic studies of the tearing instability in slab geometry have been performed as well.\cite{WanPoP05,RogersPoP07,PueschelPoP11,Zacharias12,YaoPLA21}
As it can be surmised, the self-consistent evolution of the toroidal tearing mode in a
turbulent environment exhibits an extremely rich dynamics.\cite{IshizawaPPCF19} At the same
time, the computational effort required by gyrokinetic simulations in this context is huge,
since the electron dynamics needs to be resolved, as well as the different time scales
associated with turbulence and tearing mode evolution. Moreover, for collisionless
tearing modes, the violation of the frozen-in condition needed for reconnection
is due to the finite electron inertia, and the collisionless skin depth of the electron must
be resolved. Thus, although such global multi-scale simulations resolving simultaneously
drift and MHD-like instabilities are becoming more and more feasible, see e.g. \cite{MishchenkoPPCF22}
and references therein, extended parameter scans are still very demanding. 

In this paper, we investigate the dynamics of current-driven tearing modes in toroidal
geometry employing gyrokinetic simulations with the gyrokinetic particles-in-cell (PIC)
code ORB5.\cite{JollietCPC2007,LantiJPCS2018} In order to counteract the computational problems mentioned above,
we follow a twofold approach. On one hand, we relax the physical constraint on the mass ratio,
increasing the electron mass by one order of magnitude in many of the simulations presented
here. On the other hand, we limit the simulations to vanishing to mild density and
temperature gradients, below the threshold for linear micro-turbulence instability.
For both linear and nonlinear dynamics of the collisionless tearing mode in a circular,
large aspect ratio tokamak,
we investigate the linear collisionless tearing mode growth rate, island saturation mechanisms and rotation
frequency through extensive parameter scans (density and temperature gradients,
mass ratio, plasma $\beta_e$, etc.). This study is not just propaedeutic to the investigation
of the interaction between turbulence and tearing modes discussed above. We will show that,
despite these simplifying assumptions, new properties of the instability are revealed;
in particular the interaction with other modes can appear both linearly at high enough
$\beta_e$ and nonlinearly due to the generation of turbulence around the island
separatrix for strongly-driven modes. This turbulence exhibits features typical of
the Kelvin-Helmholtz instability (KHI) and its evolution changes significantly
depending on the presence or absence of electron trapping in the tokamak magnetic field.

Another topic addressed in this paper is the island rotation, which
impacts the contribution both of the polarization current,\cite{Wilson1996a} which is believed to play an important role for the NTM modeling, evolution and threshold,\cite{GuenterNF03,Buttery_NF2003,PoliNF05,Waelbroeck_PPCF2008,IshizawaPoP13,IshizawaPPCF19} and of the bootstrap
current (at least in small islands).\cite{Bergmann_PoP09,SiccinoPoP11,Poli_PPR16}
Determining the TM mode drive and rotation frequency is thus necessary to properly model NTMs.
In the presence of background gradients, diamagnetic effects determine the rotation of the TM in the linear
phase,\cite{Coppi_PoF64,Biskamp_NF78} referred then as a drift-tearing mode, and
can also reduce drastically the growth rate. In the nonlinear phase, the determination of the rotation is
more subtle.\cite{Smolyakov_PPCF1993,Waelbroeck2009}
\cor{For instance, a set of reduced two-fluid equations describing drift-tearing mode demonstrated that the tearing mode
frequency was proportional to the diamagnetic drifts, the $\cross{E}{B}$ flow, the background magnetic field
geometry and the plasma resistivity.\cite{Nishimura2008} Numerical simulations of two dimensional reduced MHD
confirmed that the frequency is dictated by the diamagnetic drifts during the linear growth,\cite{MuragliaNF09}
while being controlled by the $\cross{E}{B}$ contribution at saturation.}
Additionally, the plasma $\beta$ (ratio of kinetic to magnetic pressure) was shown to contribute to the island rotation
through an amplification of the Maxwell stress.\cite{MuragliaNF09}
The rotation frequency of the tearing (linear phase) and the island
(saturation) were confirmed by two dimensional simulations of four fields models equations
in slab geometry, including parallel ion motions as well as ion and electron diamagnetic
drifts.\cite{Uzawa_PoP2010} Later work showed that at large viscosity the island poloidal rotation
is reduced.\cite{Ishizawa2012a} In the presence of micro-instabilities, the island poloidal
propagation is related to the island width $W=w/\rho_i$.\cite{IshizawaPPCF19}
Kinetic theory of TM predicts mode growth for positive tearing parameter $\Delta'$ and rotation in
the electron diamagnetic drifts direction during its linear
evolution.\cite{Drake77,Drake1983,NishimuraJPSJ07} 

Here we address the rotation of the tearing in mode in both its linear and nonlinear phase.
We show that the linear phase is usually characterized by a rotation in the electron
diamagnetic direction, as expected, unless the ion temperature gradient dominates the others.
A reversal from electron to ion direction is usually observed at the transition from the
linear to the saturated phases. The generation of an electrostatic potential that forces
the electrons to co-rotate with the island is observed, as expected from previous
theoretical and numerical results.\cite{smolyakov_PoP1995,SiccinoPoP11,Sato_PoP17}

The paper organization is as follows. In \sref{sec:ORB5} we briefly describe the ORB5 code and the tearing
mode initialization. In \sref{sec:ORB5Simul}, we compare linear simulations results for flat
density and temperature profile to kinetic and two-fluids theoretical model (\sref{sec:ValidationTM})
while the nonlinear evolution of the tearing mode and the island-induced
turbulence at its separatrix are investigated in \srefs{sec:NL_Flat}{sec:KH}. \Sref{sec:LinGradTScans} discuss the impact of
finite temperature and density gradients on the linear and nonlinear tearing mode
evolution. The linear tearing mode growth and frequency are investigated in \srefs{subsec:LinGrad}{sec:Desta_n1m3} while
\sref{sec:NLwithGradients} discusses the nonlinear island rotation frequency and its relation
to the $\cross{E}{B}$ flow for gradients that are kept below the linear threshold for
micro-instabilities.
A summary of our results is given in \sref{sec:Summary}. 
\section{ORB5 Code\label{sec:ORB5}}
\subsection{Physics Model \label{sec:Physics}}
We make use of the ORB5 code which is a global gyrokinetic particle-in-cell (PIC) code that solves the gyrokinetic Vlasov-Maxwell system of equations.\cite{Lanti2020}
 ORB5 uses the mixed-variable formulation of the gyrokinetic formulation which splits the
electromagnetic potential into a symplectic and a Hamiltonian part $A_{||}=A_{||}^{(s)} +A_{||}^{(h)}$.\cite{MishchenkoCPC19} The species distribution function $f_s$ is split into a background distribution variate $F_{0s}$, chosen as a 
shifted Maxwellian, and a time-dependent deviation from it $\delta f_s$: $f_s=F_{0s}+\delta f_s$ for species $s$ (bulk plasma ion or electrons). The distribution function $F_{0s}$ can be used as a control variate to reduce the sampling noise. The evolution of $\delta f_s$ is obtained solving the gyrokinetic Vlasov equation
\begin{equation}
	\frac{\partial \delta f_s}{\partial t} + \dot{\boldsymbol{R}}\cdot\frac{\partial\delta f_s}{\partial \boldsymbol{R}}+\dot{v_{||}}\frac{\partial \delta f_s}{\partial v_{||}}=-\dot{\boldsymbol{R}}^{(1)}\cdot\frac{\partial F_{0s}}{\partial \boldsymbol{R}}-\dot{v_{||}}^{(1)}\frac{\partial F_{0s}}{\partial v_{||}},\label{eq:Vlasov}
\end{equation}
Markers, or numerical particles, evolve accordingly to the gyrokinetic evolution equation for the gyrocenter $\boldsymbol{R}$ and the velocity parallel to the equilibrium magnetic field $v_{||}$ 
\begin{eqnarray}
&&\dot{\boldsymbol{R}}=\dot{\boldsymbol{R}}^{(0)}+\dot{\boldsymbol{R}}^{(1)}, \,\,\, \dot{v_{||}}=\dot{v_{||}}^{(0)}+\dot{v_{||}}^{(1)}\\
&&\dot{\boldsymbol{R}}^{(0)}=v_{||}\boldsymbol{b}_0^*+\frac{1}{q_sB^*_{||}}\boldsymbol{b}\times\mu\nabla B, \,\,\, \dot{v_{||}}^{(0)}=-\mu\boldsymbol{b}_0^*\cdot\nabla B\label{eq:backRandVpar}\\
&&\dot{\boldsymbol{R}}^{(1)}=\frac{\boldsymbol{b}}{B^*_{||}}\times\nabla\left\langle\phi-v_{||}\left(A_{||}^{(s)}+A_{||}^{(h)}\right)\right\rangle-\frac{q_s}{m_s}\left\langle A_{||}^{(h)}\right\rangle\boldsymbol{b}^* \\
&&\dot{v_{||}}^{(1)}=-\frac{q_s}{m_s}\left[\boldsymbol{b}^*\cdot\nabla\left\langle\phi-v_{||}A_{||}^{(h)}\right\rangle+\frac{\partial}{\partial t} \left\langle A_{||}^{(s)}\right\rangle\right]\nonumber\\
&&\textcolor{white}{\dot{v_{||}}^{(1)}=}-\mu\frac{\boldsymbol{b}\times \nabla B}{B^*_{||}}\cdot\nabla{\left\langle A_{||}^{(s)}\right\rangle} \\
&&\boldsymbol{b}^*=\boldsymbol{b}_0^*+\frac{\nabla\left\langle A_{||}^{(s)}\right\rangle\times \boldsymbol{b}}{B_{||}^*}, \,\,\, \boldsymbol{b}_0^*=\boldsymbol{b}+\frac{m_s v_{||}}{q_sB_{||}^*}\curl{b}\\
&&B^*_{||}=B+\frac{m_sv_{||}}{q_s}\boldsymbol{b}\cdot\curl{b} 
\end{eqnarray}
The superscript (0) and (1) indicate the unperturbed and perturbed variables and the gyro-average $\langle \zeta \rangle=\oint\zeta(\boldsymbol{R}+\boldsymbol{\rho})d\alpha/(2\pi)$ for $\zeta=\phi, A_{||}$ with $\boldsymbol{\rho}$ the
gyro-radius of the particle and $\alpha$ the gyro-phase. 
The symbol $\mu=m_sv_{\perp}^2/(2B)$ is the magnetic moment, $m_s$ the particle mass, $\boldsymbol{b}=\boldsymbol{B}/B$ with $B=|\boldsymbol{B}|$ and $\langle\phi\rangle$ is the perturbed electrostatic potential. The Ohm's
law is written for the symplectic part of the electromagnetic potential\cite{MishchenkoCPC19}
\begin{equation}
\frac{\partial}{\partial t} A_{||}^{(s)} +\boldsymbol{b}\cdot\nabla\phi = 0,
\end{equation}
and the field equations are
\begin{eqnarray}
&&-\nabla\cdot\left(\frac{n_0}{B\omega_{ci}}\nabla_\perp\phi\right)=\sum\limits_{s=i,e}q_s\bar{n}_{1,s}\\
&&\sum\limits_{s=i,e}\frac{\beta_s}{\rho_s^2}A_{||}^{(h)} -\nabla_\perp^2A_{||}^{(h)}=\mu_0\sum\limits_{s=i,e}\bar{J}_{||1s}+\nabla_\perp^2A_{||}^{(s)}\label{eq:Amper},
\end{eqnarray}
with $\omega_{ci}=q_iB/m_i$ the ion cyclotron frequency, $\rho_s=\sqrt{m_iT_e}/(q_iB)$ the species sound thermal gyro-radius, $q_s=eZ_s$ ($Z_s$ the atomic number, $e$ the elementary charge) the species charge, $\beta_s=\mu_0q_sn_sT_s/B^2_0$ the ratio of kinetic to magnetic pressure for species $s$, $\bar{n}_{1,s}=\int d^6Z\delta f_s\delta(\boldsymbol{R}+\boldsymbol{\rho}-\boldsymbol{x})$ and $\bar{J}_{||1s}=\int d^6Zv_{||}\delta f_s\delta(\boldsymbol{R}+\boldsymbol{\rho}-\boldsymbol{x})$ the perturbed gyro-center density and parallel current with $d^6Z=B_{||}^*d\boldsymbol{R}dv_{||}d\mu d\alpha$ the phase-space volume. \corNew{In our simulations, the electron parallel flow is much smaller than the thermal speed and the deviation of the shifted Maxwellian, with respect to equilibrium, is in the range 1.14 to 0.011 $c_s=\sqrt{T_e/m_i}\cong v_{th,i}\ll v_{th,e}$. There is no significant deviation of the marker distribution compared to the distribution used in the derivation of the weight equation.}

Electromagnetic simulations in global gyrokinetic simulations suffer from the
cancellation problem.\cite{ChenPoP01,MishchenkoPoP04} This problem is addressed in ORB5 solving the equations using a control variate mitigation technique and
a Pullback algorithm for the mixed-variables.\cite{BottinoPPCF11,MishchenkoPoP14a,MishchenkoPoP14b,MishchenkoCPC19}
In ORB5, all physical variables are normalized to the following four quantities: $m_i$ and $q_i=eZ_i$ the ion mass and charge, the magnetic field amplitude at the magnetic axis $B_0$ and the electron temperature $T_e(s_0)$ at a reference magnetic surface $s_0$.\cite{Lanti2020} In addition, the volume average electron density $\langle n_e\rangle$ is a fifth
normalised quantity used in the electromagnetic version of the code.
The normalized radial coordinate is given by $s=\sqrt{{\psi/\psi_{edge}}}$ using the initial equilibrium geometry.
\begin{figure}
\begin{center}
	\includegraphics[width=\linewidth,keepaspectratio]{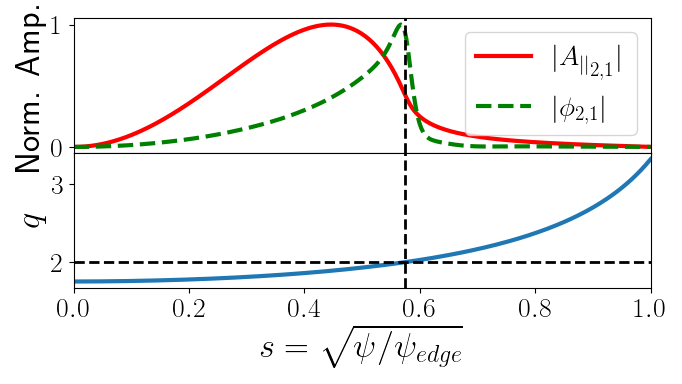}
	\caption{Radial profiles of $|{A_{||}}_{2,1}|$ and $|\phi_{2,1}|$ for an
	unstable tearing mode in toroidal geometry and safety factor profile
	$q$. The vertical dashed line indicates
        the radial position of the $q=2$ resonant surface.}
\label{fig:profiles}
\end{center}
\end{figure}
Finally, all species are treated kinetically, the ions gyrokinetically, and the electrons drift-kinetically, without fast particles, for the tearing mode simulations described below.
\subsection{Definition of Linear Simulations in ORB5}
\cor{The linearized Vlasov equation \eref{eq:Vlasov} is written as}
\begin{equation}
	\frac{\partial \delta f_s}{\partial t}+\dot{\boldsymbol{R}}^{(0)}\frac{\partial \delta f_s}{\partial \boldsymbol{R}}+\dot v_{||}^{(0)}\frac{\partial \delta f_s}{\partial v_{||}}=-\dot{\boldsymbol{R}}^{(1)}\cdot\frac{\partial F_{0s}}{\partial \boldsymbol{R}} - \dot{v}_{||}^{(1)}\frac{\partial F_{0s}}{\partial v_{||}}
\end{equation}
The left-hand-side is the Lagrangian derivative of $\delta f$ along the unperturbed trajectories
while the right-hand-side (rhs) depends on the perturbed potentials. Linear simulations can be
performed evolving the system along unperturbed trajectories by evaluating the rhs at the new
particle position at each time step.

\subsection{Tearing Mode Initialization\label{sec:TM_Init}}
In ORB5, the Maxwellian distribution function of the electrons can be shifted such that it
is not centered around $v_{||}=0$ and produces an equilibrium electron flow consistent with the desired
equilibrium current. In such a case, the electron bulk velocity is obtained as
$v_{||,0}=-J_{||}/(e n_e)$, where the equilibrium parallel current density
$J_{||}$ is calculated from the background magnetic field $\boldsymbol{B}$ as $J_{||}=\boldsymbol{b}\cdot\curl{B}$,
with $\boldsymbol{b}=\boldsymbol{B}/|\boldsymbol{B}|$.
The tearing mode is destabilized initializing a safety factor $q$ profile obtained from the Wesson analytic form of the following current density profile\cite{WessonBook}
\begin{equation}
	J_{||}=J_0\left(1-\left(\frac{r}{a}\right)^2\right)^\zeta.\label{eq:Jprof}
\end{equation}
Here $J_0$ is the current on axis, $r$ the radial position, $a$ the location of the current peaking on $r$ and $\zeta$ an integer determining the peaking of the current. In our simulations, $\zeta=1$. The safety factor $q$-profile associated to this current is
\begin{equation}
	q=q_a\frac{r^2/a^2}{1-\left(1-r^2/a^2\right)^{\zeta+1}}\label{eq:qprof},
\end{equation}
with $q_a=q(r=a)$ and the on-axis value $q_0=q(r=0)$.
In our simulations, we choose $q_0=1.75$ and $q_a/q_0=\zeta+1=2$.
The initial safety factor $q$ profile condition destabilizes a tearing mode with helicity $(m,n)=(2,1)$ at radial position $s=\sqrt{\psi/\psi_{edge}}=0.57$.
\Fref{fig:profiles} shows the profiles of the electrostatic potential $|\phi|$, the parallel component of the potential vector $|A_{||}|$ and the safety factor $q$ with respect to the normalized radial direction $s=\sqrt{\psi/\psi_{edge}}$.
The vertical dashed-line shows the resonant surface $q=2$ where the tearing mode is current unstable. The profiles are taken at the time the tearing mode has established its linear growth. The tearing parameter for this profile is $\Delta'=21.3$.
\begin{figure}
           \begin{subfigure}{0.5\linewidth}
           \begin{center}
           \includegraphics[width=\linewidth,keepaspectratio]{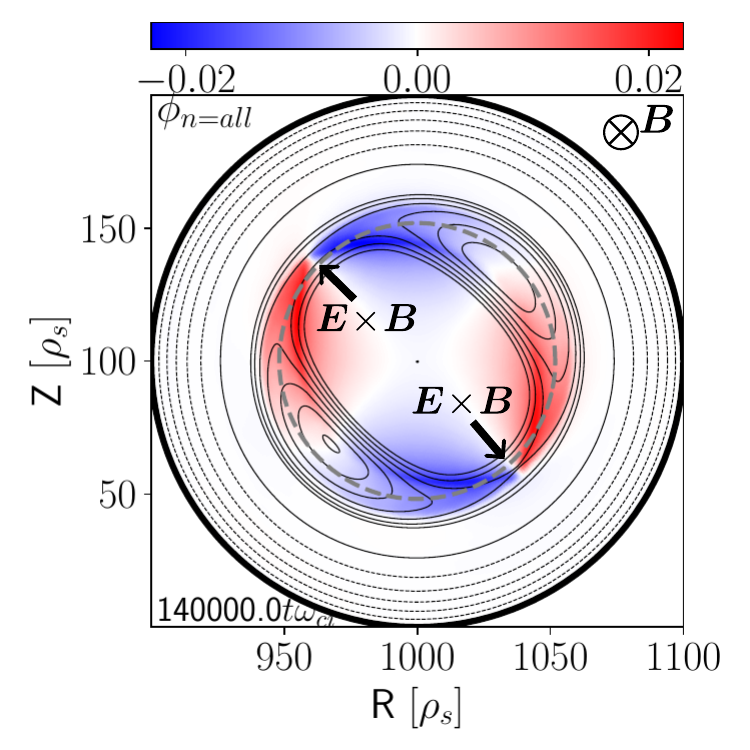}
           \caption{Electrostatic potential $\phi$.}
           \label{subfig:Phi_2D_t1_4e5_be0005}
           \end{center}
           \end{subfigure}~
           \begin{subfigure}{0.5\linewidth}
           \begin{center}
           \includegraphics[width=\linewidth,keepaspectratio]{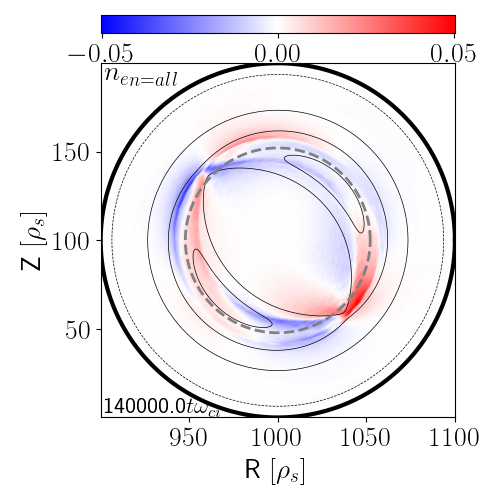}
           \caption{Electron density $n_e$.}
           \label{subfig:Density_e_2D_t1_4e5_be0005}
           \end{center}
           \end{subfigure}\\
	   \caption{Poloidal cross sections showing the tearing mode structures of normalised (a) electrostatic potential $\phi$ and (b) electron density $n_e$.
	            The potential leads to $\cross{E}{B}$ inflows towards the X-points.}
	   \label{fig:Typical2D}
\end{figure}
\begin{figure}
\begin{center}
   \includegraphics[width=\linewidth,keepaspectratio]{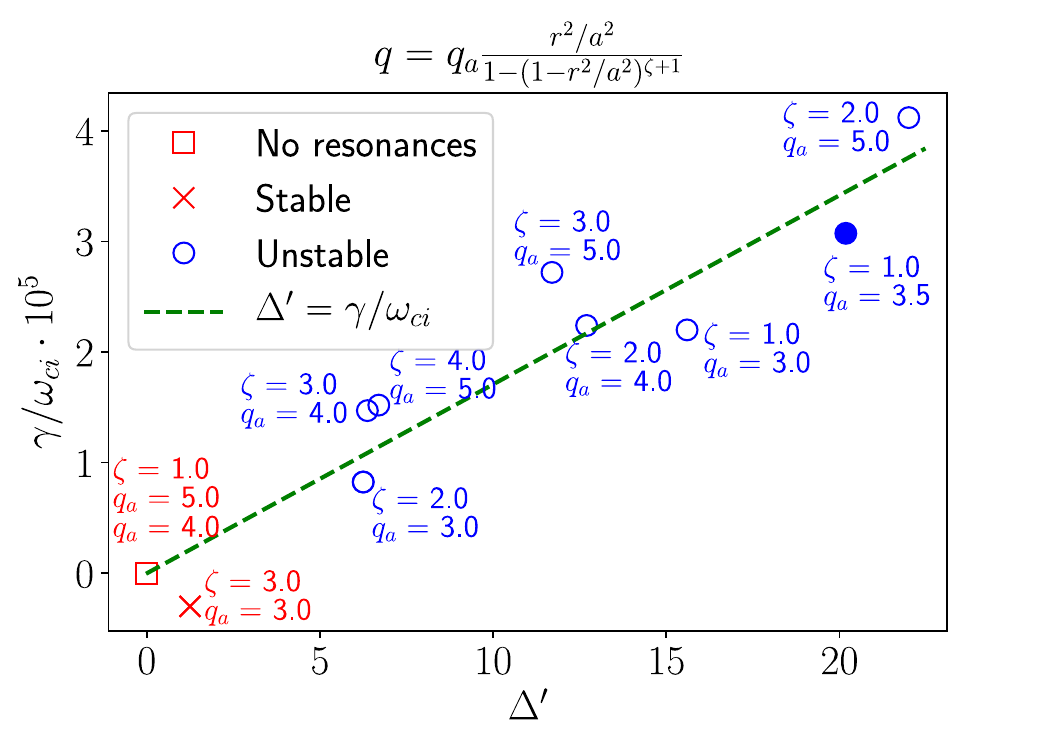}
		\caption{Growth rate $\gamma$ of the mode $m/n=2/1$ against the tearing parameter $\Delta'$
                         for several parameters $\zeta=q_a/q_0-1$ and $q_a$ of \eref{eq:qprof}. The growth rate
                         increases with the tearing parameter. The filled blue circle corresponds to the parameters selected
                         for the present investigation. $\beta=0.1\%$, $m_i/m_e=800$.}
        \label{fig:gamma_Deltap_Wesson}
\end{center}
\end{figure}

\begin{figure*}
\begin{center}
   \includegraphics[width=0.7\linewidth,keepaspectratio]{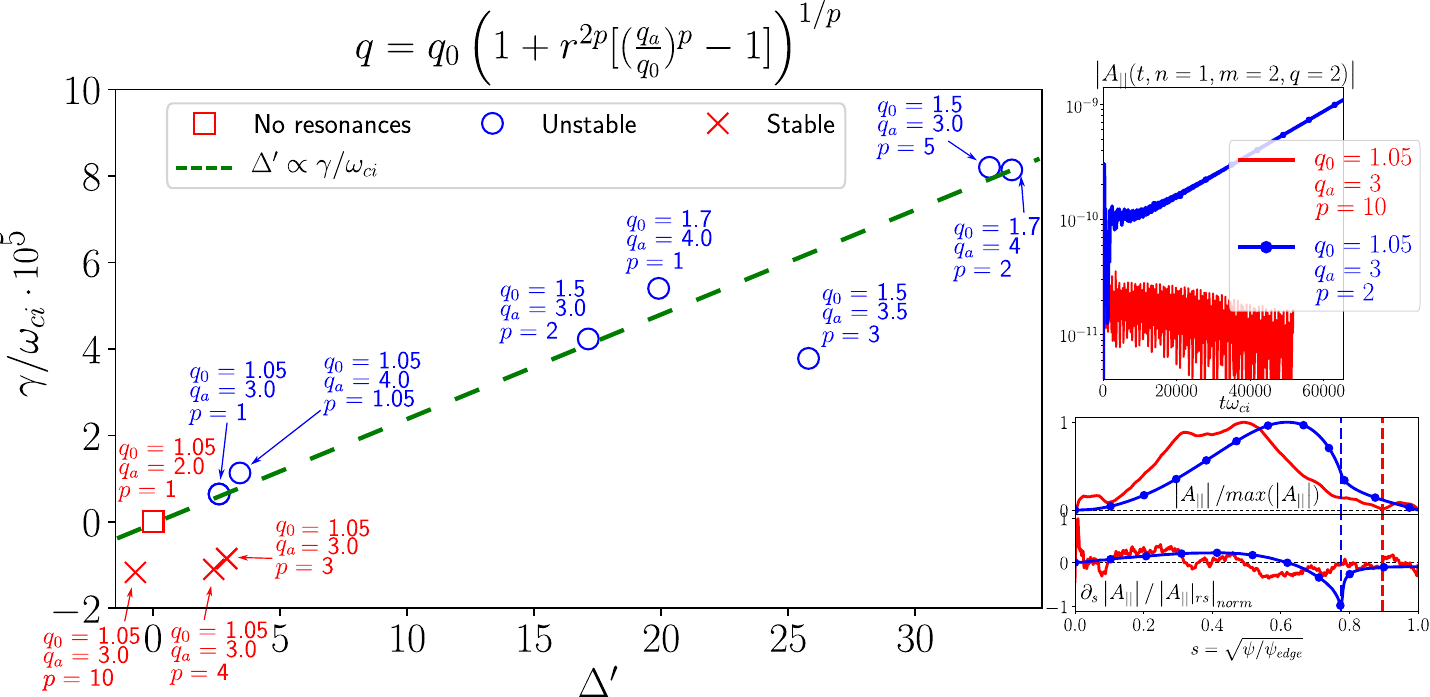}
		\caption{Growth rate $\gamma$ against the tearing parameter $\Delta'$, for the mode $m/n=2/1$,
                         for several parameters $q_0$, $q_a$ and $p$ of \eref{eq:qAk}. The time trace of $A_{||}(n=1,m=2,q=2)$, its radial profile and radial derivatives are show on the right panels for a typical stable (red) and unstable case (blue). $\beta=0.1\%$, $m_i/m_e=400$.}
        \label{fig:gamma_Deltap_Akihiro}
\end{center}
\end{figure*}

Finally, we consider a circular cross-section tokamak with a minor radius, in Larmor units $\rho_i=m_iv_{th,i}/eB$, given by the parameter $lx=2a/\rho_s=200$, a mass ratio $m_i/m_e=1/0.005=200$ and aspect ratio $\varepsilon=a/R_0=0.1$ for a minor radius $a=1$ [m] and major radius (at plasma center) $R_0=10$ [m]. The magnetic field on axis $B_0=1$ [T]. \cor{The number of
grid points are ($n_s$,$n_\phi=128$,$n_\chi=384$), where $n_s$, $n_\phi$ and $n_\chi$ are the radial, toroidal and poloidal directions. The Fourier decompositions are applied to the toroidal and poloidal directions. The radial grid number ranges from
$n_s=320$ to $900$ such that the electron skin depth, which depends on the plasma
$\beta$ and mass ratio $m_i/m_e$ values ($\eref{eq:beta_omega_e}$), is always resolved by
at least 5 grid points.\footnote{The
number is found to be sufficient by a convergence test.}
The initial load of markers
are $4\cdot10^8$ for the ions and $8\cdot10^8$ for the electrons.} 
\begin{equation}
\beta_e=\frac{\mu_0N_0T_0}{B_0^2}\label{eq:beta_electron},
\end{equation}
where $N_0=\langle n_e\rangle$ is the volume averaged electron density and $T_0=T_e(s_p)$ at radial location $s_p$ where the gradients are maximum. We obtain the linear evolution of the tearing mode that has a m/n =
2/1 helicity structure (\fref{subfig:Phi_2D_t1_4e5_be0005}), as expected by theory.\cite{BiskampBook00}
The figure further depicts the $\cross{E}{B}$ inflows
towards the X-points.
In slab geometry, the density is localized in the X-points vicinity with a quadrupole structure.\cite{KlevaPoP95,JainPoP17} In toroidal geometry, such a quadrupole structure is visible around the high-field side X-point (top-left X-point). Compared to slab geometry, the quadrupole structure gets shifted and tilted at the low-field side. A more detailed analysis of the reasons for this asymmetry is left for future work.
In our simulations, we use the reduced mass ratio $m_i/m_e=200$ in order to reduce the computational costs
and to be able to perform several parameter scans
(we also notice that the mass ratio can be used to control the growth
rate of the TM, as shown later).
In \sref{sec:NL_Flat}, we compare selected results to the more realistic mass ratio $m_i/m_e=1600$.

\section{Tearing Mode with Flat Temperature and Density Profiles\label{sec:ORB5Simul}}
Kinetic theory of collisionless tearing mode states that the tearing mode growth rate $\gamma$ depends on the plasma $\beta$ as well as the mass and temperature ratio value.\cite{RogersPoP07,Pueschel_PoP2015} We will compare the theoretical growth rate to our
simulations scanning these parameters. In ORB5, the plasma $\beta_e$ is introduced in terms of the electron pressure as:
In ORB5, varying $\beta_e$ as an input parameter corresponds to a variation of the density, which in turn implies a change in the collisionless skin depth $d_e=c/\omega_{pe}$, where $\omega_{pe}^2=N_0e^2/m_e\epsilon_0$ is the square of electron plasma frequency. Making use of the above definition, the \cor{electron} plasma $\beta$ is written as
\begin{equation}
\beta_e=\left(\frac{1}{d_e}\right)^2\frac{m_e}{m_i}\rho_s^2\label{eq:beta_omega_e},
\end{equation}
In fusion plasmas, $\rho_s$ is usually larger than the collisionless skin depth, i.e. the relevant regime is $d_e<\rho_s$.
The growth rate of the tearing mode in the collisionless limit reads:\cite{RogersPoP07}
\begin{equation}
	\gamma/\omega_{ci}=\Delta' k_\theta k_{r}\rho_s^3\frac{B_\theta}{B_\phi}\left(1+\frac{T_i}{T_e}\right)^{1/2}\left(\frac{m_e}{m_i}\right)^{1/2}\frac{1}{\beta_e},
\label{eq:gamma_cl_Rogers_betae}
\end{equation}
where $\Delta'$ is the tearing parameter,\cite{furth_PF1973} $k_\theta=m/r$ is the tearing poloidal wave number, $k_r$ is the radial wave number, $B_\theta$ and $B_\phi$ are the poloidal and toroidal magnetic field. The estimation is valid in the limit of small $\Delta'$ which is the relevant limit for Tokamak tearing modes with poloidal mode number $m>1$.\cite{RogersPoP07} \Eref{eq:gamma_cl_Rogers_betae} shows that the tearing mode growth rate in the collisionless limit is proportional to $ 1/\beta_e$ (or $d_e^2$) at fixed $T_i$, $T_e$ and mass ratio. \cor{In the remaining of the paper, we use the notation $\beta\equiv\beta_e$.}
\begin{figure}
	\begin{center}
	\begin{subfigure}{\linewidth}
	\begin{center}
		\includegraphics[width=0.75\linewidth,keepaspectratio]{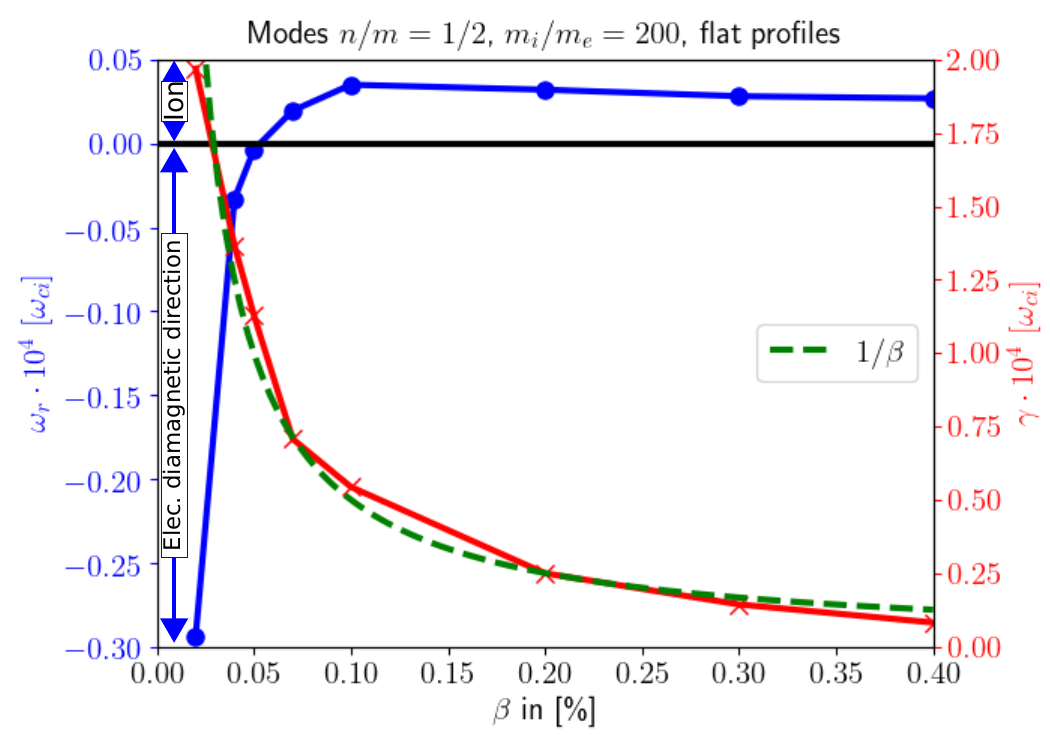}
        \caption{Growth rate $\gamma$ (red line with crosses) and frequency $\omega_r$ (blue line with circles) versus $\beta$.}
        \label{subfig:A_n1m2_gamma_LinBeta_Flat}
	\end{center}
	\end{subfigure}\\
        \begin{subfigure}{\linewidth}
	\begin{center}
		\includegraphics[width=0.75\linewidth,keepaspectratio]{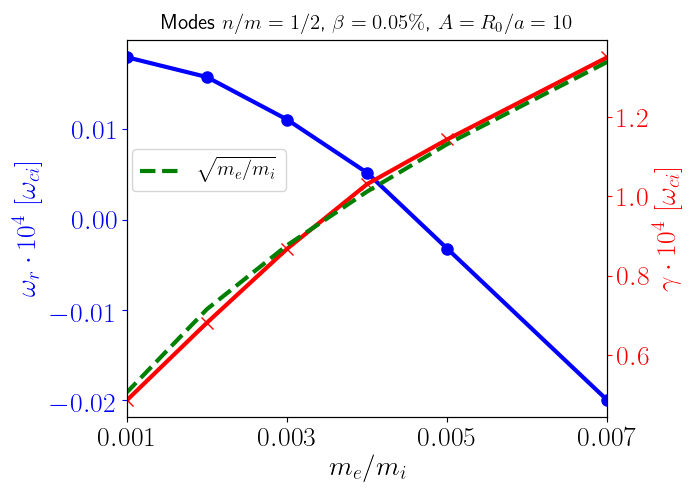}
		\caption{Growth rate $\gamma$ (red line with crosses) and frequency $\omega_r$ (blue line with circles) versus mass ratio $m_e/m_i$.}
        \label{subfig:A_n1m2_gamma_LinMass_Flat}
	\end{center}
	\end{subfigure}
		\caption{Tearing mode linear simulations ($n=1$) for flat temperature and density profiles. 
		(a): The $1/\beta$ scaling (\eref{eq:gamma_cl_Rogers_betae}) of the
		${A_{||}}_{2,1}$ growth rate $\gamma$ is obtained for $\beta<0.4\%$ (cf.~discussion in \sref{subsec:LinGrad}).
		The tearing mode has a weak rotation frequency $\omega_r$ whose sign
		depends on $\beta$. Electron diamagnetic drift direction for $\omega_r<0$.
		(b): The mode growth rate $\gamma$ decreases as the electron mass is reduced. In both plots, the green dashed-line
		represents the theoretical scaling.}
\label{fig:LinBeta_and_Mass_Scan}
\end{center}
\end{figure}

\subsection{Linear growth of the tearing mode\label{sec:ValidationTM}}
\begin{figure}
\begin{center}
	\begin{subfigure}{\linewidth}
	\begin{center}
		\includegraphics[width=0.75\linewidth,keepaspectratio]{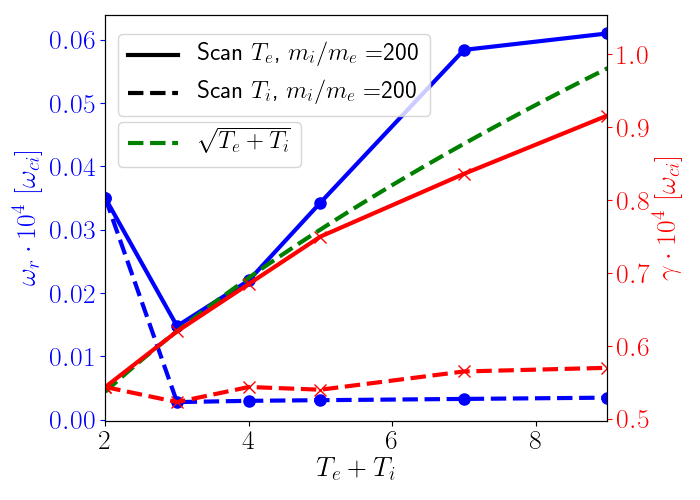}
		\caption{Temperature scan, $m_i/m_e=200$}
        \label{subfig:A_n1m2_gamma_LinTempMass200_Flat}
	\end{center}
	\end{subfigure}\\
        \begin{subfigure}{\linewidth}
	\begin{center}
		\includegraphics[width=0.75\linewidth,keepaspectratio]{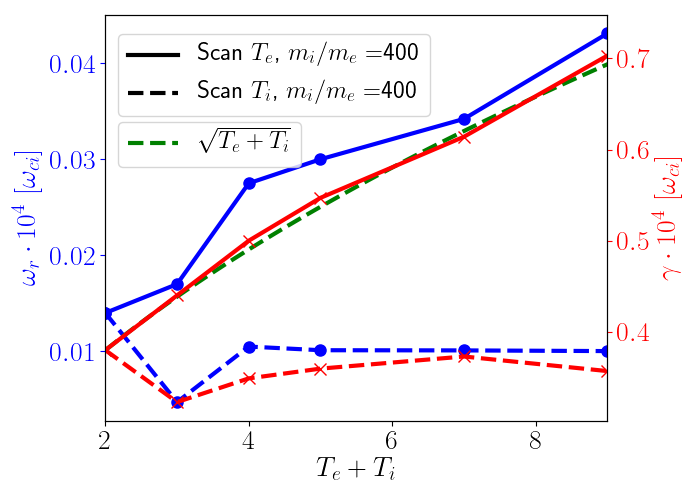}
		\caption{Temperature scan, $m_i/m_e=400$.}
        \label{subfig:A_n1m2_gamma_LinTempMass400_Flat}
	\end{center}
	\end{subfigure}

		\caption{Scans in temperatures for (a) mass ratio $m_i/m_e=200$ and
		         (b) $m_i/m_e=400$. Good agreement of the growth rate $\gamma$ (red curves)
			 with theoretical scaling $\sqrt{T_e+T_i}$ is obtained for $T_e$
			 scans. The disagreement with theoretical scaling
			 changing $T_i$ is related to the limitations
			 of the theoretical model, confirming the results in
			 \citet{PueschelPoP11}}
\label{fig:LinTempScan}
\end{center}
\end{figure}

We start with a code-theory comparison initializing a tearing mode at the rational surface $q=m/n=2/1$
and performing linear simulations varying \corNew{the tearing parameter $\Delta'$,} the plasma $\beta$ and the mass ratio $m_i/m_e$
for flat temperature and density profiles.
The parameters have the following values, except when varied:
$\beta=0.2\%$, $m_e/m_i=0.005$, $T_e/T_i=1$ and $\nabla T/T=\nabla n/n=0$.
The linear simulations are performed for the toroidal mode number $n=1$ with 
poloidal mode numbers in the intersection of two filters: a) $nq+m\leq\delta m$,
with $\delta m=5$ and b) $[-m_{max}; m_{max}]$ where
$m_{max}\approx n_{max}q_{max}\approx 4$.
\corNew{The tearing parameter $\Delta'$ is determined by the shape of the eigenfunction of $A_{||}$ which is varied through the parameters $q_a$ and $q_0$ of \eref{eq:Jprof}. Since we are interested in the
tearing mode with $m/n=2/1$, we restrict the parameter space to $1\leq q_0<2$ and $2<q_a<5$. We estimate the tearing
parameter $\Delta'$ by means of a shooting algorithm and plot the $m/n=2/1$ mode growth rate versus $\Delta'$ (\fref{fig:gamma_Deltap_Wesson}). The $m/n=2/1$ tearing mode grows proportionally with $\Delta'$. However, it is difficult to obtain a stable $m/n=2/1$ tearing mode with a rational surface not too close to the boundaries with the profile of \eref{eq:qprof}. For this reason, we use additionally the following safety factor profile (for our $\Delta'$ scan only) }
\begin{equation}
q = q_0\left(1+\left(r/a\right)^{2p}\left[\left(\frac{q_a}{q_0}\right)^p-1\right]\right)^{1/p},\label{eq:qAk}
\end{equation}
\corNew{which exhibits less restrictions on the parameter space as the power $p$ is not related to the ratio $q_a/q_0$.\cite{Ishizawa2009} Using this profile, we obtain a stable $m/n=2/1$ tearing mode due to a negative $\Delta'$, using $q_0=1.0$, $q_a=3$ and $p=10$, (\fref{fig:gamma_Deltap_Akihiro}) which has the expected $A_{||}$ eigenfunction for a stable tearing mode in toroidal geometry.\cite{ZohmBook2022} We further obtain that the mode is unstable for $q_0=1.7$, $q_a=4$, $p=4$ and marginally stable for $q_0=1.0$, $q_a=3$ and $p=1$ as already shown by a reduced set
of two-fluid equations.\cite{Ishizawa2009} The fact that stable cases are observed for small but positive values of $\Delta'$ is a consequence of the fact
that the tearing parameter does not accurately account for the tearing stability in toroidal systems with asymmetric current profiles.\cite{ArcisPLA05,ArcisPoP06}} 
\corNew{We further compare our simulations} with the theoretical
trends derived from \eref{eq:gamma_cl_Rogers_betae}.
\cor{We obtain that the tearing mode growth rate $\gamma$ decreases with an increased
value of the plasma $\beta$ (\fref{subfig:A_n1m2_gamma_LinBeta_Flat}) or
mass ratio $m_i/m_e$ (\fref{subfig:A_n1m2_gamma_LinMass_Flat}) as theoretically
estimated by \eref{eq:gamma_cl_Rogers_betae}.}
\cor{The mass ratio allows us to change the tearing drive instead of changing the tearing parameter
$\Delta'$ which would require a modification of the safety factor profile $q$.\cite{HornsbyPoP15}
But, changing the $q$ profile might lead to several rational surface where the tearing mode is unstable, a situation we want to avoid in order to investigate the nonlinear evolution of the tearing mode at a single rational surface.} Finally, we scan the temperature ratio. We obtain the expected scaling, $\gamma\propto\sqrt{T_i+T_e}$, when changing the electron
temperature $T_e$, while it is much weaker when $T_i$ is varied (\fref{fig:LinTempScan}).
This gyrokinetic behavior was already observed previously and associated to the limitation of the model for $m_e/m_i<\beta$.\cite{RogersPoP07}
Good agreement was reported at much higher values of the mass ratio.\cite{PueschelPoP11}
\begin{figure}
\begin{center}
   \includegraphics[width=0.75\linewidth,keepaspectratio]{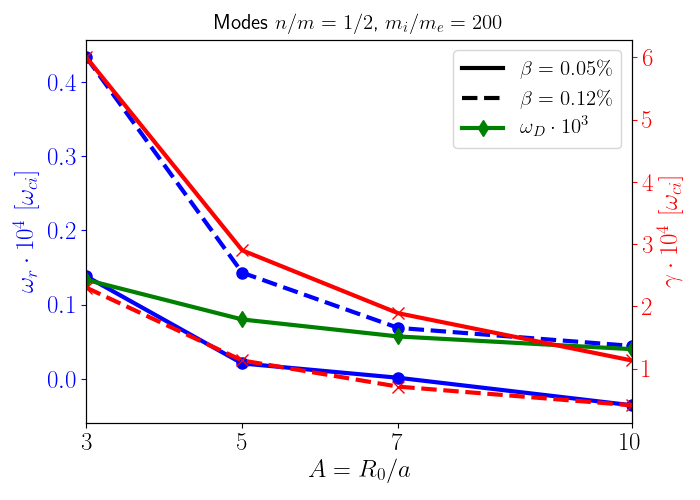}
		\caption{Growth rate $\gamma$ and real frequency $\omega_r$ ($<0$ electron diamagnetic drift direction) versus the aspect ratio $A=R_0/a$, flat density and temperature profile. Increasing $A$ reduces $\gamma$ and $\omega_r$. The frequency is negative at $\beta=0.05\%$ and $A=10$. The magnetic drift $\omega_D^*=-k_\theta T_0/(eB_0)\nabla B \times B/B^2$ is represented in green.}
        \label{fig:gamma_freq_aspect_Flat}
\end{center}
\end{figure}
\Fref{fig:profiles}, \fref{fig:LinBeta_and_Mass_Scan} and \fref{fig:LinTempScan} show that our initialization provides a physically consistent $m/n=2/1$ tearing mode. 
Furthermore, a successful benchmark with the GENE code,\cite{JenkoPoP00,PueschelPoP11} in which a shifted Maxwellian for the electrons and for the ions were recently implemented, showed also a good agreement.\cite{JitsukNF24}
A small but finite rotation frequency of the tearing mode of the order of $10^{-6}\omega_{ci}$ is observed as the plasma $\beta$ is varied (\fref{subfig:A_n1m2_gamma_LinBeta_Flat}).
In our normalization, a negative frequency corresponds to a rotation in the electron diamagnetic direction. The mode rotates
in the electron diamagnetic direction\footnote{Although the equilibrium pressure is flat in these
simulations, we refer to the diamagnetic direction as the direction in which a given particle species
would rotate in the presence of a finite gradient.}
for $\beta=<0.07\%$ and in the ion diamagnetic drift direction otherwise. The mode rotation decreases as the mass ratio is increased at fixed $\beta$.
This weak rotation of the mode for flat temperature and density profiles
has been ascribed to toroidicity effects,\cite{HornsbyPoP15} but its dependence on $\beta$ was not investigated.
Finite magnetic field gradient lead to a grad-B drift $\omega_D^*=-k_\theta T_0/(eB_0)\nabla B \times B/B^2$ due to circular cross-section.
A two-fluid model gives the relation $\omega_{TM}\cong\omega^*_n +\alpha\omega^*_T+\omega_{\cross{E}{B}}+\omega^*_D+\omega^*_\eta$, where $\omega_{TM}$ is the tearing frequency, $\omega_n^*$ and $\omega_T^*$ the density and temperature diamagnetic drift frequencies, $\alpha$ a ions species dependent parameter (hydrogen, deuterium, $\ldots$), $\omega_{\cross{E}{B}}=E\times B/(B^2r)$ the zonal flow frequency, $\omega^*_D$ the grad-B drift and $\omega^*_\eta$ a resistivity related drift.\cite{Nishimura2008}
For our parameters, \cor{the linear rotation frequency} can be estimated as $\omega_{TM}\cong\omega^*_D$.
\Fref{fig:gamma_freq_aspect_Flat} shows that the growth rate and the frequency decrease for larger aspect ratios $A=R/a$ \cor{as $\gamma\propto (1+2/A)$}.\cite{ZarzosoPoP19} \cor{Removing the mirror force from the electron dynamics, shows that the contribution of the trapped electron population to the growth rate, for our $\Delta'$, is weak a low $A$ and negligible at large $A$. The frequency of
the tearing mode is reduced in presence of trapped electrons. Still, the frequency decreases with increasing $A$.} We conclude that the weak rotation frequency of the tearing mode is related to curvature effects and the plasma $\beta$.

\subsection{Nonlinear Island Decay\label{sec:NL_Flat}}
\begin{figure}
\begin{center}
           \includegraphics[width=0.75\linewidth,keepaspectratio]{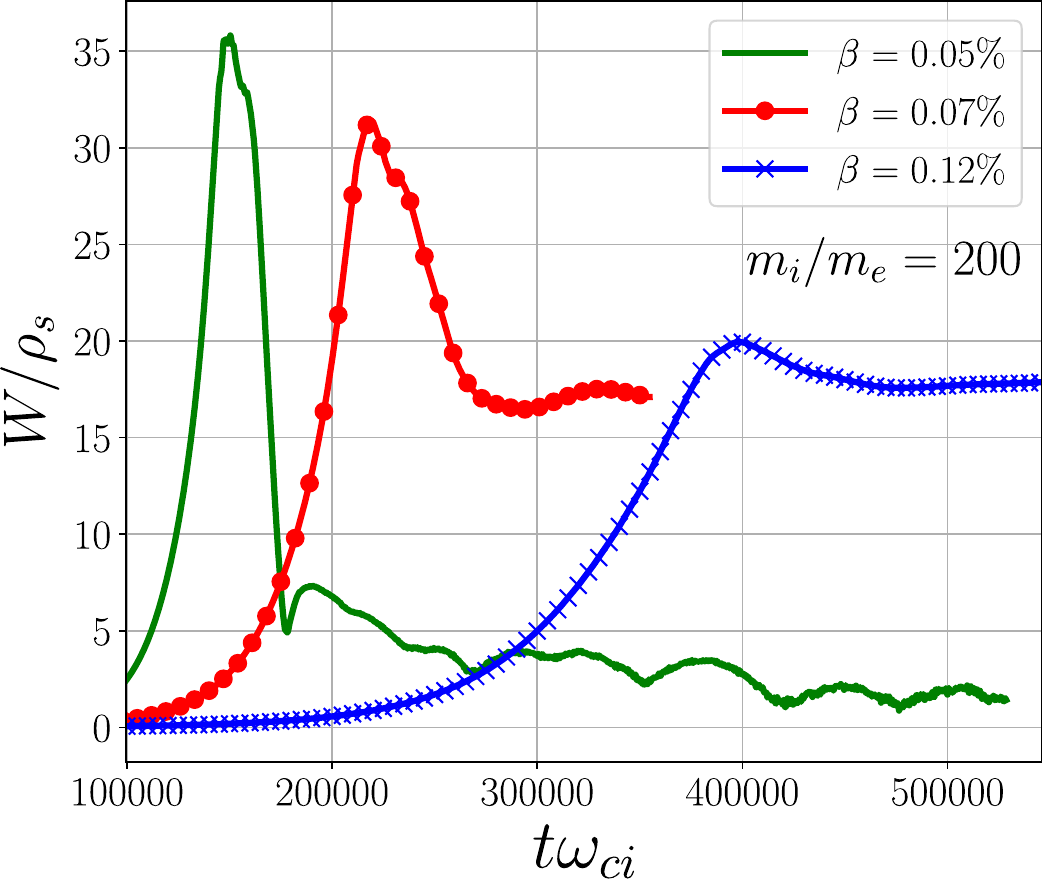}
           \caption{Time evolution of the normalised magnetic island width $W$, for different values of $\beta_e$ with mass ratio $m_i/m_e=200$.}
           \label{fig:Apar_n1m2_vs_Time_mime200_beta_all}
\end{center}
\end{figure}

We perform nonlinear simulations with flat density and temperature profiles, allowing for the back reaction of the perturbed fields on the particles trajectories (excluded in the linear simulations presented in the previous section), using $n\in[0,30]$, with $n\in\mathbb{N}$, toroidal modes. The poloidal modes are $m\in[nq-5,nq+5]$, with $m\in \mathbb{Z}$, as explained in \sref{sec:TM_Init}.
The time evolution \cor{of the island width } 
for three values of the plasma $\beta$ are compared in \fref{fig:Apar_n1m2_vs_Time_mime200_beta_all}. These
simulations have an unstable tearing mode whose linear growth reduces with
increasing plasma $\beta$ as $1/\beta$ (\eref{eq:gamma_cl_Rogers_betae}).
\cor{The magnetic island width is $W=w/\rho_s=\sqrt{2q|{A_{||}}_{2,1}|/s}$, with $s$ the normalised magnetic shear length and $\rho_s=\rho_i/\sqrt{2}$.\cite{HornsbyNF16} Since the time evolution of ${A_{||}}_{2,1}$ is in arbitrary units, the
island width can be computed as $W=C\sqrt{|{A_{||}}_{2,1}|}$ with a conversion factor $C=626.68$.} The maximum island size is observed to diminish as the
linear growth reduces. After the peak in the island size, the island evolution
differs depending on $\beta$. At the lowest value $\beta=0.05\%$, a strong decay of
the island size $W$ is observed after its peak which is then followed by a weaker decrease.
This island size reduction process weakens as the tearing drive is
reduced by increasing $\beta$.
\begin{figure}
           \begin{subfigure}{\linewidth}
           \begin{center}
           \includegraphics[width=\linewidth,keepaspectratio]{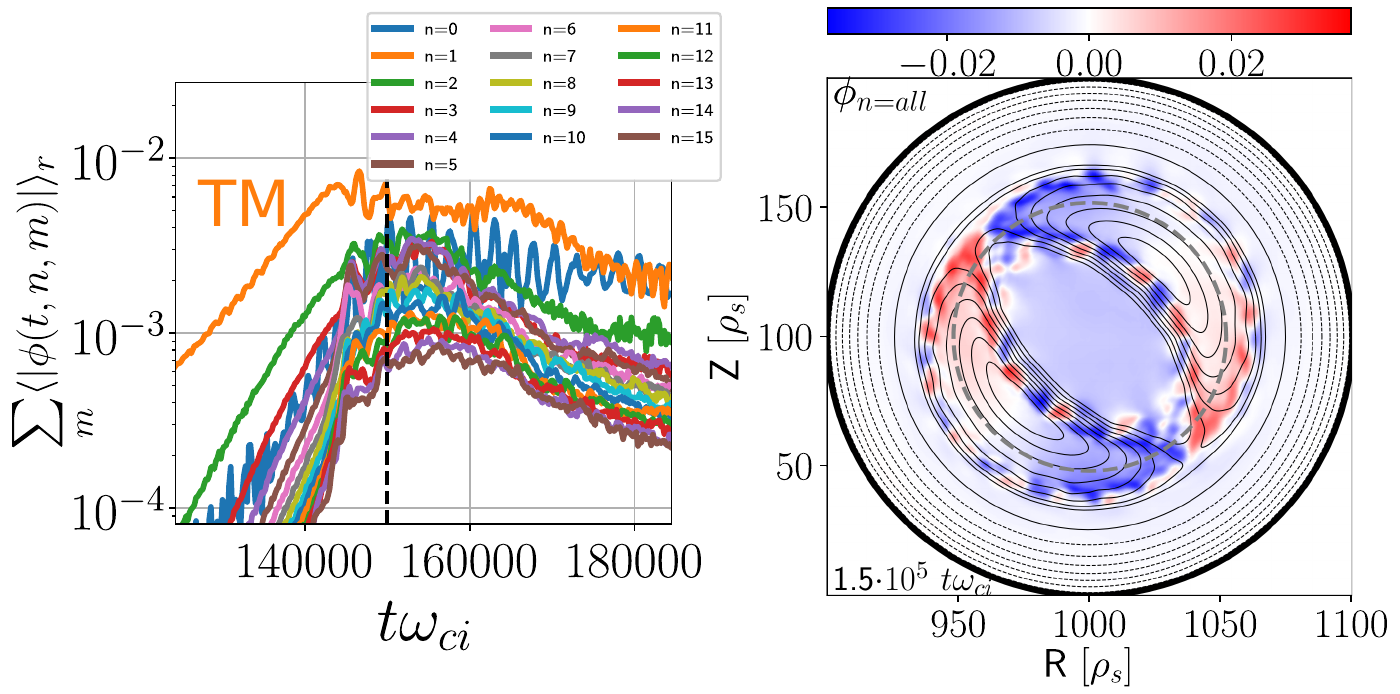}
           \caption{Electrostatic potential $\phi$.}
           \label{subfig:Phi_n_vs_time_and_Contour_t1e5}
           \end{center}
           \end{subfigure}\\           
           \begin{subfigure}{\linewidth}
           \begin{center}
           \includegraphics[width=\linewidth,keepaspectratio]{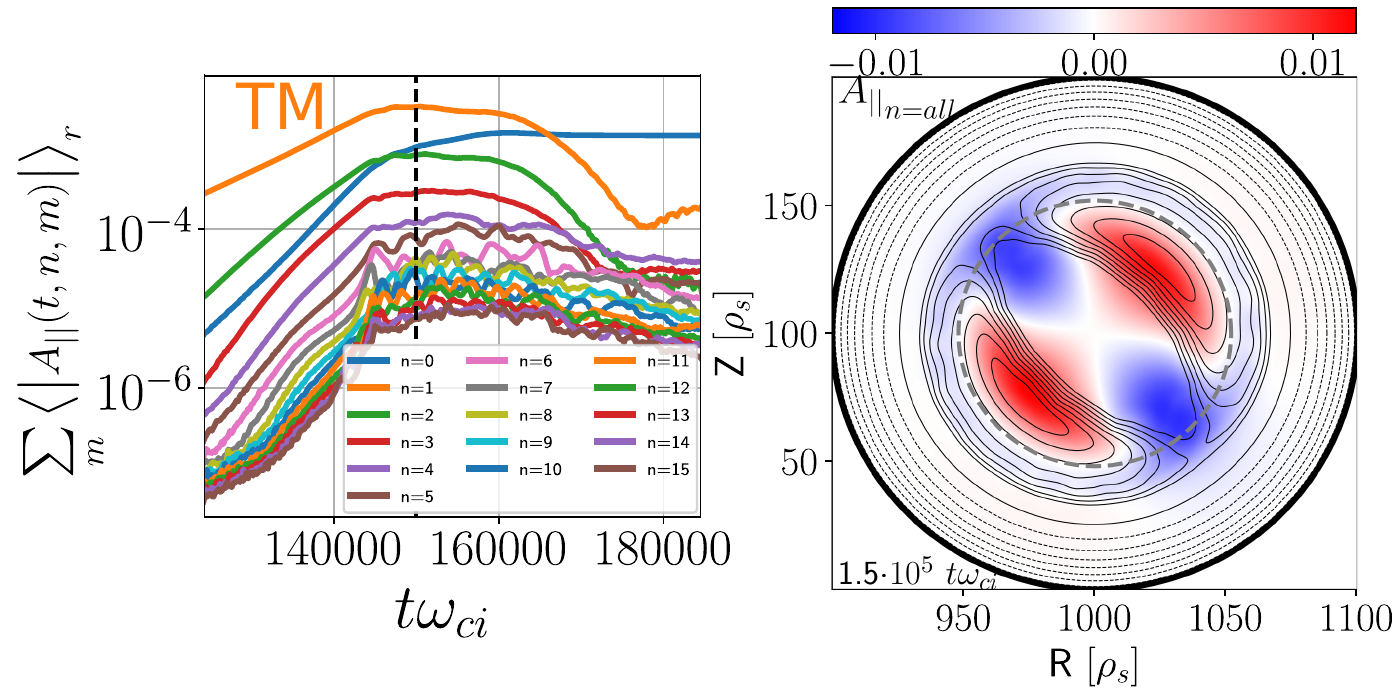}
           \caption{Vector potential $A_{||}$.}
           \label{subfig:Apar_n_vs_time_and_Contour_t1e5}
           \end{center}
           \end{subfigure}           
\caption{Time evolution of the toroidal mode amplitudes, summed over
         the poloidal modes and averaged
	 over the radius, and poloidal cuts for normalised (a) $\phi$
         and (b) $A_{||}$ including all toroidal modes. The vertical dashed
         line corresponds to the time of the contour plots.
         Plasma $\beta=0.05\%$ and $m_i/m_e=200$.}
\label{fig:Apar_Phi_n_vs_time_and_Contour_Be0005}
\end{figure}
In particular, for $\beta=0.12\%$ there is almost no island decay after the maximum amplitude is reached, and the island saturation size
is similar to that obtained after a weak island decay for $\beta=0.07\%$ (\fref{fig:Apar_n1m2_vs_Time_mime200_beta_all}).
Due to this nonlinear decay process, there are no direct correlations between the tearing linear growth and the island
size at the quasi-steady state. However, a strongly driven tearing mode impacts the maximum size that the
island can reach before the beginning of the decay process.
\begin{figure}
           \begin{subfigure}{\linewidth}
           \begin{center}
           \includegraphics[width=\linewidth,keepaspectratio]{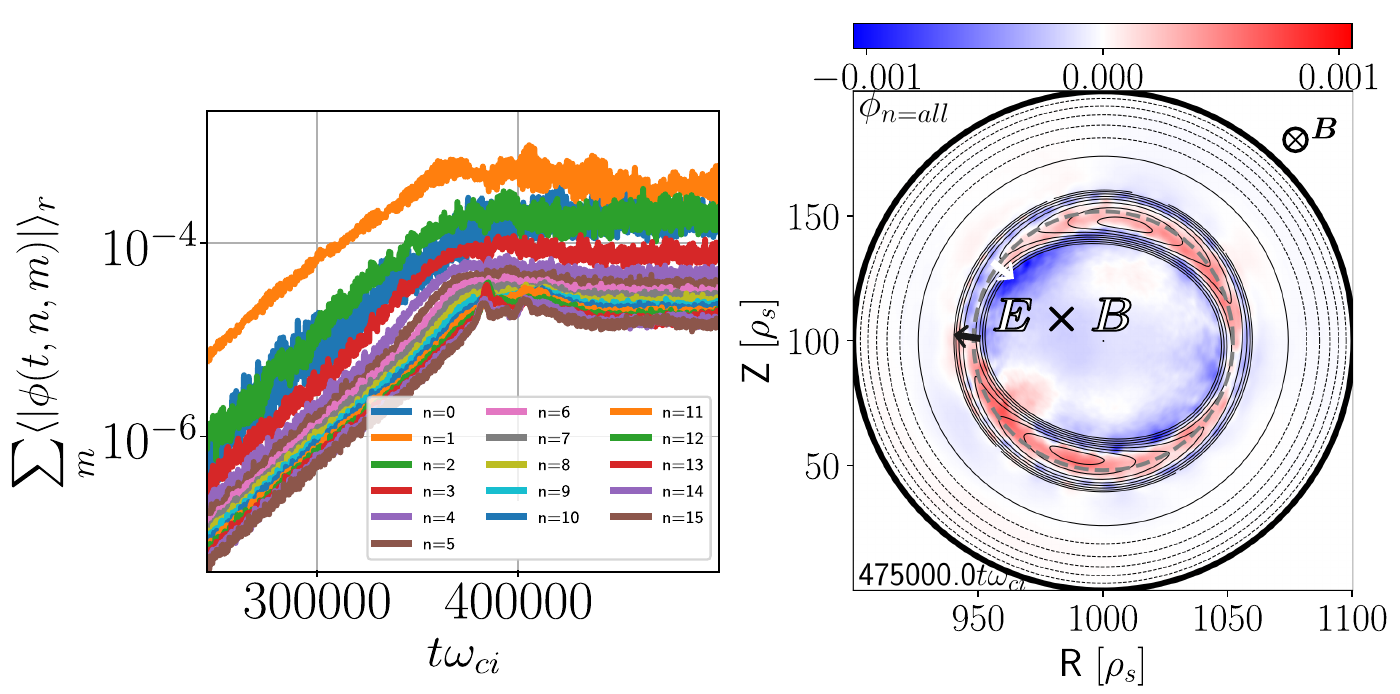}
           \caption{Electrostatic potential $\phi$.}
           \label{subfig:Phi_n_vs_time_Contour_Be0_0012}
           \end{center}
           \end{subfigure}\\           
           \begin{subfigure}{\linewidth}
           \begin{center}
           \includegraphics[width=\linewidth,keepaspectratio]{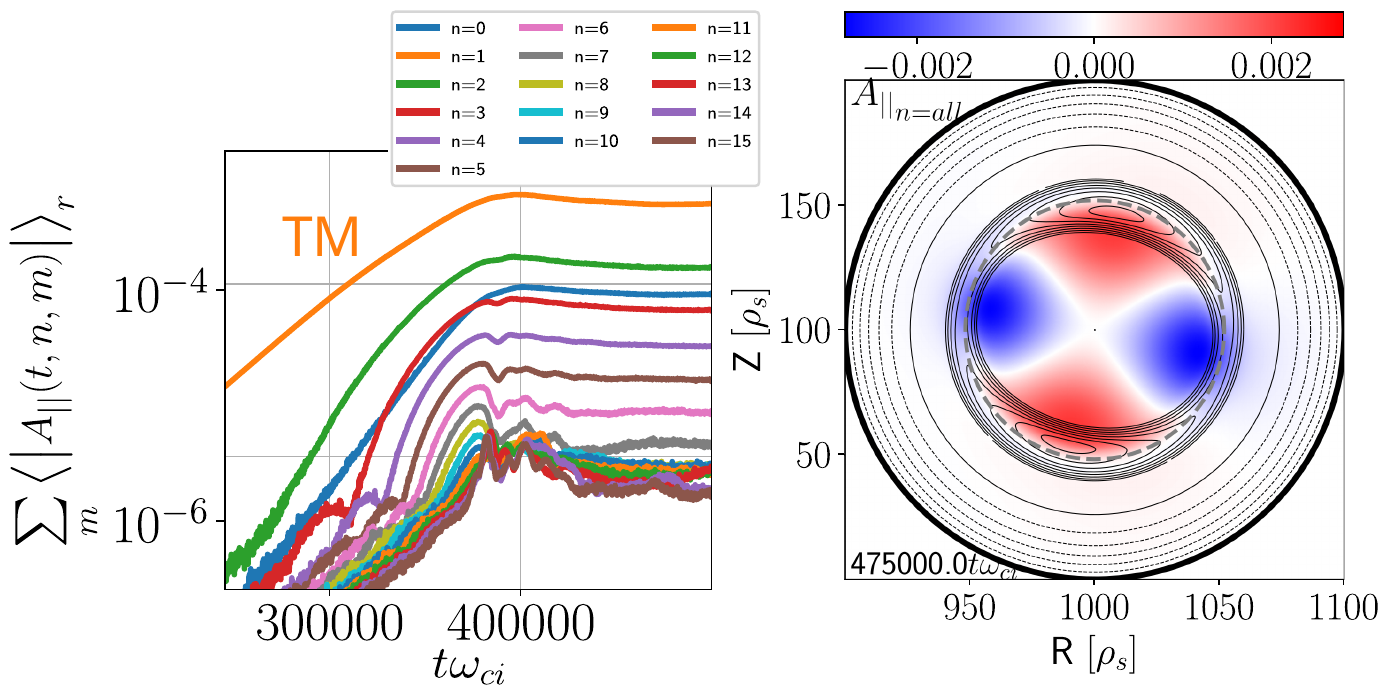}
           \caption{Vector potential $A_{||}$.}
           \label{subfig:Apar_n_vs_time_Contour_Be0_0012}
           \end{center}
           \end{subfigure}           
\caption{Time evolution of the toroidal mode amplitudes, summed over
         the poloidal modes and averaged
	 over the radius, and poloidal cuts for normalised (a) $\phi$
	 and (b) $A_{||}$ 
         including all toroidal modes. The contour plots are taken at
	 the simulation end. Plasma $\beta=0.12\%$ and $m_i/m_e=200$.}
\label{fig:Apar_Phi_n_vs_time_and_Contour_Be0012}
\end{figure}
The temporal evolution of the toroidal Fourier amplitudes of $\phi$ and $A_\parallel$, averaged over the radius,
and poloidal cuts of the potentials for $\beta=0.05\%$ and $0.12\%$ are depicted in \frefs{fig:Apar_Phi_n_vs_time_and_Contour_Be0005}{fig:Apar_Phi_n_vs_time_and_Contour_Be0012}.
As the island reaches its maximum size, strong electrostatic turbulence is produced at the
island separatrix and strong zonal currents (${A_{||}}_{n=0}$) develop when the island
decay process begins for $\beta=0.05\%$ (\fref{subfig:Apar_n_vs_time_and_Contour_t1e5} left). Oppositely, much weaker
zonal currents are produced
and almost no turbulence develops for $\beta=0.12\%$ (\fref{subfig:Apar_n_vs_time_Contour_Be0_0012}).
Two mechanisms seem to be at play for the nonlinear evolution of the island: I) current density redistribution and II)
nonlinear electrostatic instabilities at the island separatrix.
We investigate the physics of the island decay process for $\beta=0.05\%$
comparing the island evolution by a) decreasing the electron mass $m_e$ by a factor of $8$, b) filtering out toroidal mode number larger than $n\in[0,2]$, and c) removing the
mirror force setting the time derivative of the
unperturbed parallel velocity (\eref{eq:backRandVpar}, right) for the
electrons to zero (\fref{fig:Apar_n1m2_vs_Time_beta0005_all}). We observe no decay when
the electron mass is reduced ($m_i/m_e=1600$), a small island decay process
using $n\in[0,2]$, and a moderate decay removing the mirror force (MF $=0$).
This behavior is discussed in \sref{sec:KH}. We address the current redistribution first.
\subsection{Current Redistribution and Island Size Reduction.\label{sec:CurrentRedis}}
\begin{figure}
\begin{center}
   \includegraphics[width=\linewidth,keepaspectratio]{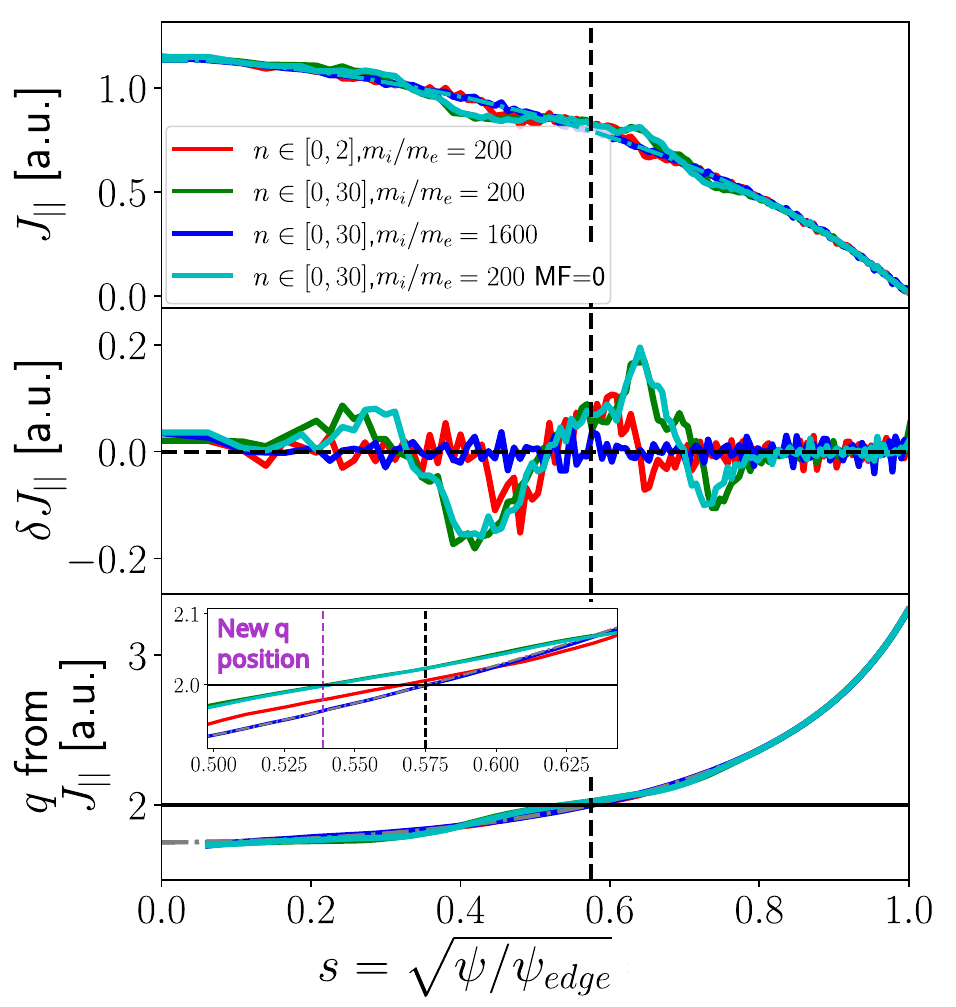}
   \caption{Poloidally and toroidally averaged parallel current profile $J_{||}$,
            its perturbation $\delta J_{||}= J_{||}-J_{||,t=0}$ and safety factor
            $q$ computed from the current profiles shortly after the islands
            reach their maximum size.\newline
            Red curve: $n\in[0,2]$, $m_i/m_e=200$ at $t\omega_{ci}=1.55\cdot10^5$.
            Green curve: $n\in[0,30]$, $m_i/m_e=200$ at $t\omega_{ci}=1.55\cdot10^5$.
            Blue curve: $n\in[0,30]$, $m_i/m_e=1600$ at $t\omega_{ci}=4.1\cdot10^5$.
            Cyan curve: $n\in[0,30]$, $m_i/m_e=200$ at $t\omega_{ci}=1.55\cdot10^5$
            with electron mirror force (MF) set to $0$.
	    The black dashed-line represents the equilibrium $q=2$ resonant surface.}
           \label{fig:q_from_J_Flat_Prof_be0005_allcases}
\end{center}
\end{figure}

In the presence of large magnetic islands, strong zonal fields $A_{||}(m=0,n=0)$
develop when toroidal modes up to 30 are retained in the simulation.
Their drive occurs at twice the growth rate of the dominant tearing instability, as expected,
so these fields are weaker for reduced linear growth rate of the tearing mode, for instance increasing the
plasma $\beta$ or reducing the electron mass $m_e$.
\begin{figure}
\begin{center}
           \includegraphics[width=0.75\linewidth,keepaspectratio]{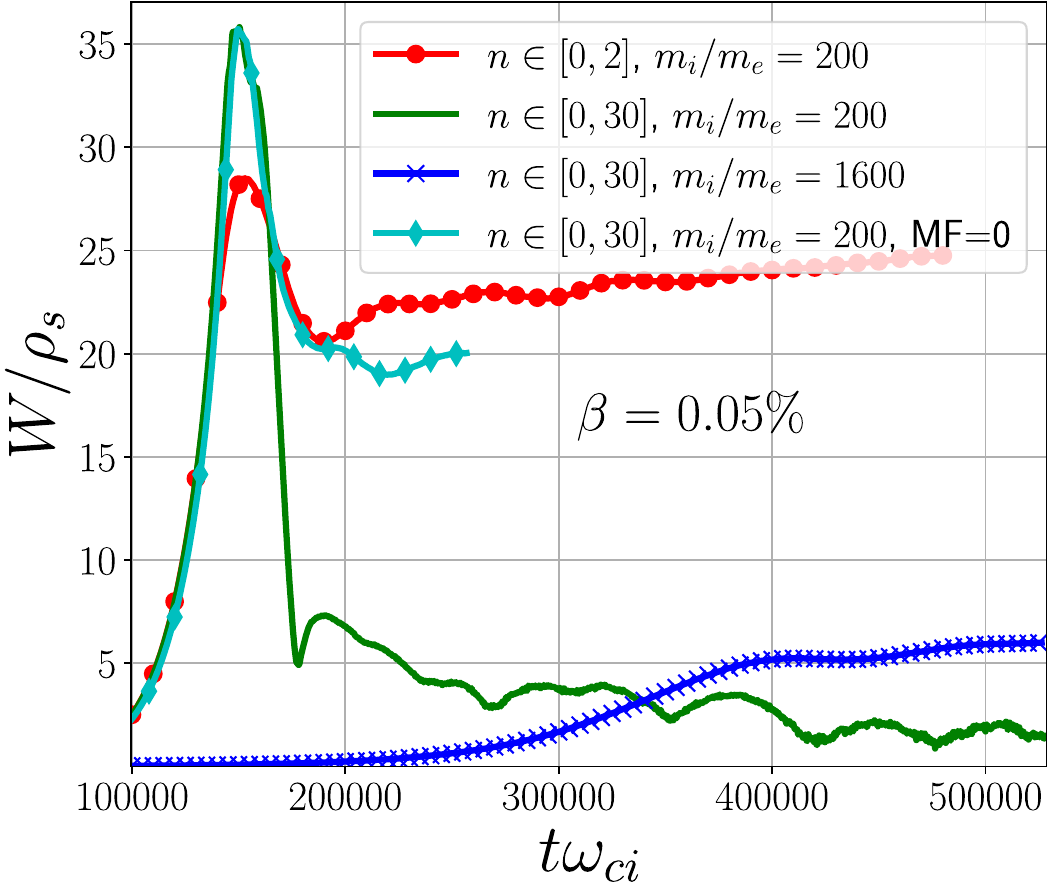}
	   \caption{Time evolution of the normalised island width $W$ for mass ratio
            $m_i/m_e=200$ and $1600$ with $n\in[0,30]$, $m_i/m_e=200$ with $n\in[0,2]$
            and artificially removing the mirror force (MF) for $m_i/m_e=200$
            with $n\in[0,30]$.}
           \label{fig:Apar_n1m2_vs_Time_beta0005_all}
\end{center}
\end{figure}
The zonal magnetic fields are found to
impact the island evolution through
the modification of the background current density $J_{||}$
and the safety factor $q$ profile.
\Fref{fig:q_from_J_Flat_Prof_be0005_allcases} compares the modification of these
profiles for simulations with $\beta=0.05\%$, $n\in[0,2]$ (red curve) and $[0,30]$
(green curve) with mass ratio $m_i/m_e=200$, $n\in[0,30]$ with $m_i/m_e=1600$
(blue curve) and $n\in[0,30]$ with $m_i/m_e=200$ setting the mirror force (MF)
for the electrons to zero (cyan curve).
While in the case $m_i/m_e=1600$ the island saturates at a quite low level,
see \fref{fig:Apar_n1m2_vs_Time_beta0005_all}, with no visible
modification of the current profile, a clear flattening of the radial current
profile is obtained for $m_i/m_e=200$.
The radial range of the flattening is quite broad if toroidal modes up to 30
are retained and does not seem to be influenced by the presence (or absence)
of trapped electrons. If only modes from $0$ to $2$ are retained, the island size, zonal
currents and the flattening of the current profiles are reduced.
The safety factor profile is modified as the island reaches its maximum width
and the rational surface $q=2$ moves from $s=0.57$ to $0.53$, closer to
the plasma center.
The fact that the modification of the current profile is almost identical
with and without electron trapping, together with the fact that both cases
exhibit the same dynamics until the maximum island amplitude is reached,
suggests that the current perturbation is the main effect stopping the island
growth and initiating its decrease in these strongly-driven cases. Its further
evolution is associated to the modification of the flows, which is different
depending on whether the mirror force is included or not, as explained in
\sref{sec:KH}. As stated before, the growth of
the zonal currents (related to the axisymmetric part of $A_\parallel$) show
that they are nonlinearly driven by the tearing instability and
saturate with it. In other words, the current profile flattening is not a result of the excitation of the electrostatic turbulence, described in
the next subsection, as it is also present for the simulation including toroidal modes 0 to 2. 

\subsection{Kelvin-Helmholtz Instability (KHI) and Island Saturation Mechanism.\label{sec:KH}}
We investigate the instability at the island separatrix scrutinizing the
electron density $n_e$ and $\phi$ fluctuations for the case with $\beta=0.05\%$ and $n\in[0,30]$,
including the mirror force for the electrons, shown in \fref{fig:Apar_Phi_n_vs_time_and_Contour_Be0005}.
\begin{figure}

           \begin{subfigure}{0.5\linewidth}
           \begin{center}
           \includegraphics[width=\linewidth,keepaspectratio]{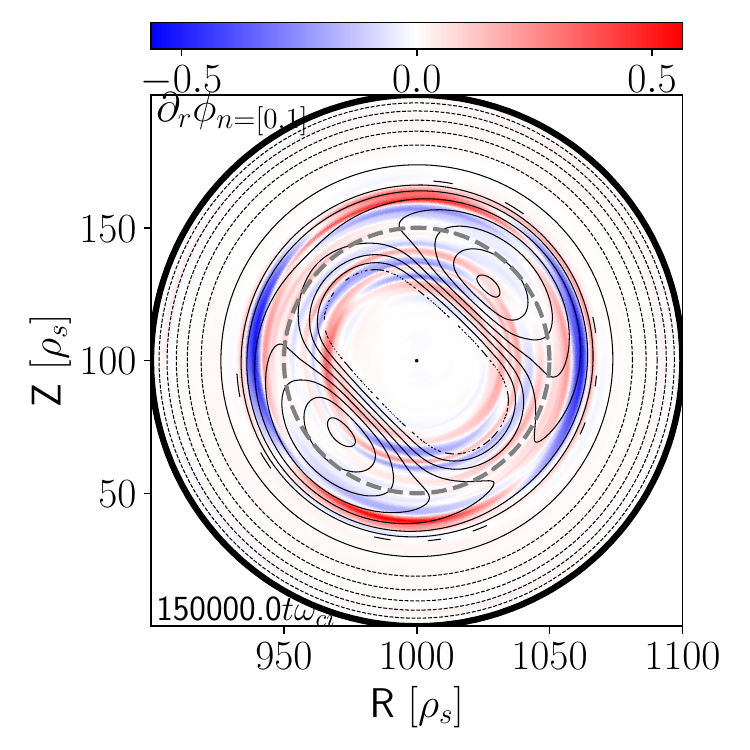}
           \caption{Zonal ($n=0$) plus island ($n=1$) flows.}  
           \label{subfig:ZF01_0_flat_profs}
           \end{center} 
           \end{subfigure}~
           \begin{subfigure}{0.5\linewidth}
           \begin{center}
           \includegraphics[width=\linewidth,keepaspectratio]{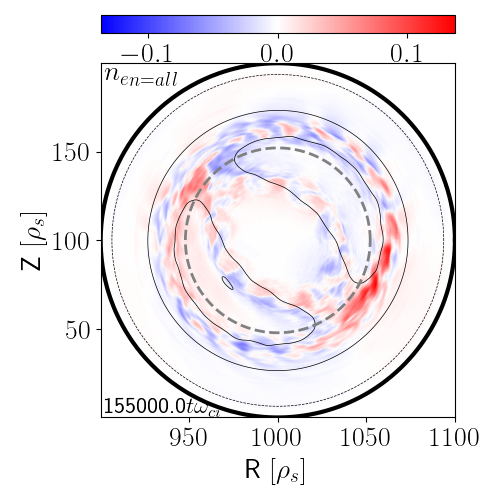}
           \caption{Electron density fluctuations $n_e$.}
           \label{subfig:Density_e_2D_t1_55e5_be0005}
           \end{center}
           \end{subfigure}           
	\caption{Poloidal cross-sections representing: (a)
	         the radial derivative of the sum of the zonal ($n=0$) and
	         island-related ($n=1$) components of the normalised electrostatic potential, 
		 exhibiting a dominant $m/n=2/1$ helicity at the separatrix
		 and (b) the normalised electron density fluctuations $n_e$
		 \cor{showing vortex flow similar to the Kelvin-Helmholtz instability},
		 particularly visible in the inner region.
		 Plasma $\beta=0.05\%$ and $m_i/m_e=200$.}
           \label{fig:Density_e_2D_be0005}
\end{figure}
We observe that the localized $\phi$ and $n_e$ structures along the island separatrix
(\fref{fig:Typical2D}) imply strong $\cross{E}{B}$ and diamagnetic flows, which
can become unstable to the Kelvin-Helmholtz instability (KHI). Excitation of KHI has been observed before in magnetic
reconnection, and particularly in collisionless reconnection with asymmetric current profiles, see e.g. \cite{GrassoPoP20,BorgognoAPJ22} and references therein, and in particular very recently at the separatrix of
strongly driven magnetic islands.\cite{GranierPoP24}
\Frefs{subfig:ZF01_0_flat_profs}{subfig:Density_e_2D_t1_55e5_be0005} show that 
the island develops strong $n=1$ potential structure at its separatrix, before
the growth of fluctuations.
\begin{figure}
\begin{center}
        \begin{subfigure}{0.5\linewidth}
        \begin{center}
           \includegraphics[width=\linewidth,keepaspectratio]{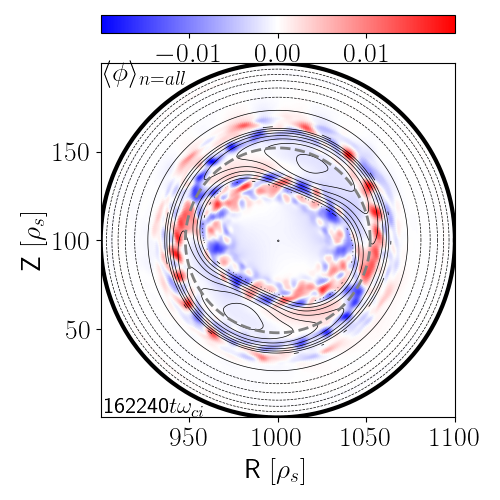}
           \caption{$t\omega_{ci}\in[1.6,1.65]\cdot10^5$\newline MF $=0$}  
           \label{fig:Phi_2D_tave_1_6e5_1_65e5MF0}
	\end{center}
	\end{subfigure}~
        \begin{subfigure}{0.5\linewidth}
        \begin{center}
        \includegraphics[width=\linewidth,keepaspectratio]{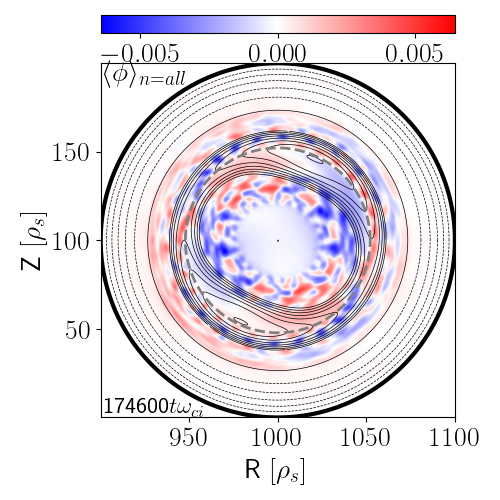}
        \caption{$t\omega_{ci}\in[1.65,1.85]\cdot10^5$\newline MF $=0$}  
	\label{subfig:Phi_2D_tave_1_65e5_1_85e5MF0}
	\end{center}
	\end{subfigure}\\
        \begin{subfigure}{0.5\linewidth}
        \begin{center}
           \includegraphics[width=\linewidth,keepaspectratio]{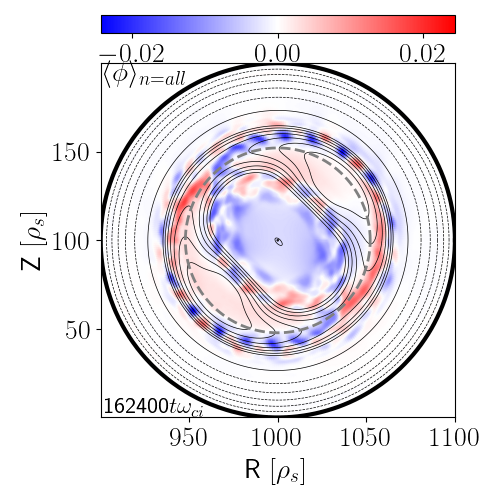}
        \caption{$t\omega_{ci}\in[1.6,1.65]\cdot10^5$\newline MF $\neq0$}  
	\label{subfig:Phi_2D_tave_1_6e5_1_65e5}
	\end{center}
	\end{subfigure}~
        \begin{subfigure}{0.5\linewidth}
        \begin{center}
        \includegraphics[width=\linewidth,keepaspectratio]{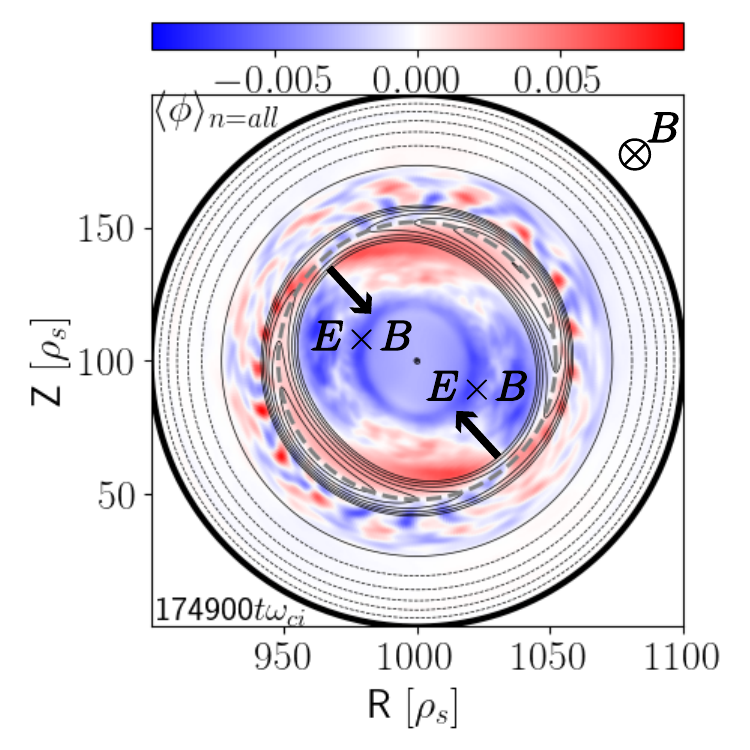}
		\caption{$t\omega_{ci}\in[1.65,1.85]\cdot10^5$\newline MF $\neq0$}  
	\label{subfig:Phi_2D_tave_1_65e5_1_85e5}
	\end{center}
	\end{subfigure}

        \caption{Time averaged of the normalised electrostatic potential $\langle \phi \rangle$ for $\beta=0.05\%$,
	         $n\in[0,30]$ and $m_i/m_e=200$.
                 Top line: without mirror force MF $=0$.
                 Bottom line: with mirror force MF $\neq 0$.
		 (d): Turbulence structures are merged into larger ones with
		 helicity $m/n=2/1$. The local $\cross{E}{B}$ outflows flows are
		 depicted by the black arrows.}
        \label{fig:ContourPhi_MF_NoMF}
\end{center}
\end{figure}
The related sheared flows destabilize the KHI at the island separatrix, more clearly in the
inner region (between the island and the magnetic axis), as shown by the vortices in the density
(\fref{subfig:Density_e_2D_t1_55e5_be0005}). \cor{Additionally, we obtained that the velocity jump across the separatrix,
computed using the $\cross{E}{B}$ flow as a proxy for the velocity, is larger than the Alfvén speed, as expected for an unstable
Kelvin-Helmholtz instability.\cite{FermoPRL12}}

Since the island evolution differs starting 
from $t\omega_{ci}\cong1.7\cdot10^5$ retaining the mirror force or not, we investigate
in what manners the electrostatic
turbulence changes in the presence (or absence) of trapped particles.
\Fref{fig:ContourPhi_MF_NoMF} depicts the time average of the electrostatic potential over
$t\omega_{ci}\in[1.6,1.65]\cdot 10^5$ and $[1.65,1.85]\cdot 10^5$.
A much stronger $n=1$ component of $\phi$ can be observed when electron trapping is retained. The associated $\cross{E}{B}$ flows
correspond to a shrinking of the magnetic island, as they point away from the island X-point, as can be checked by a comparison with the flows during the growing phase of the island shown in \fref{subfig:Phi_2D_t1_4e5_be0005}, resulting in the
reversed reconnection.
\begin{figure*}
\begin{center}
       \begin{subfigure}{0.4\linewidth}
       \begin{center}
           \includegraphics[width=\linewidth,keepaspectratio]{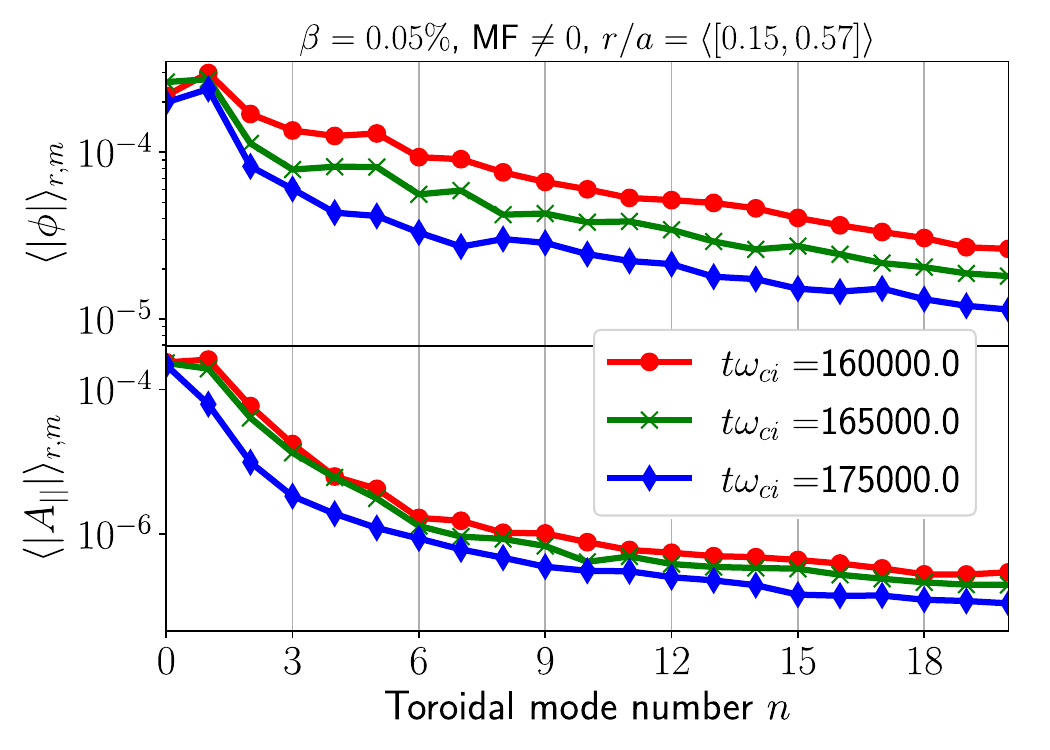}
        \caption{MF$\neq0$, $r\in[0.15,0.57]$.}  
	\label{subfig:Spect_rintMF}
	\end{center}
	\end{subfigure}~
        \begin{subfigure}{0.4\linewidth}
        \begin{center}
           \includegraphics[width=\linewidth,keepaspectratio]{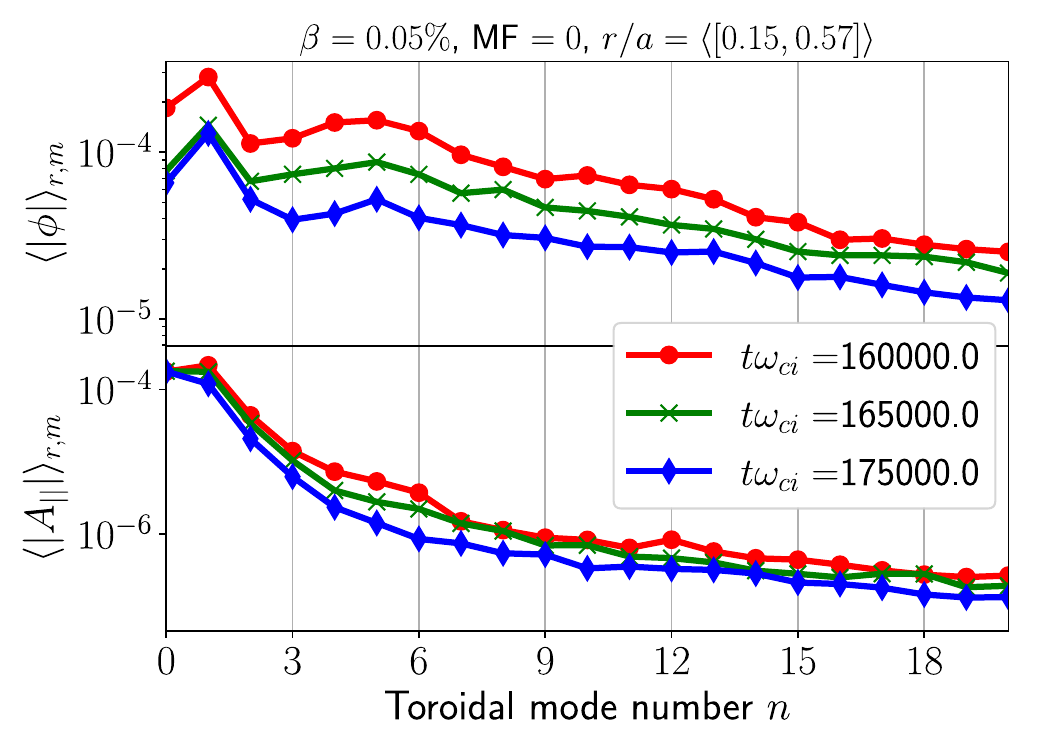}
        \caption{MF$=0$, $r\in[0.15,0.57]$.}  
	\label{subfig:Spect_rint}
	\end{center}
	\end{subfigure}\\
        \begin{subfigure}{0.4\linewidth}
        \begin{center}
           \includegraphics[width=\linewidth,keepaspectratio]{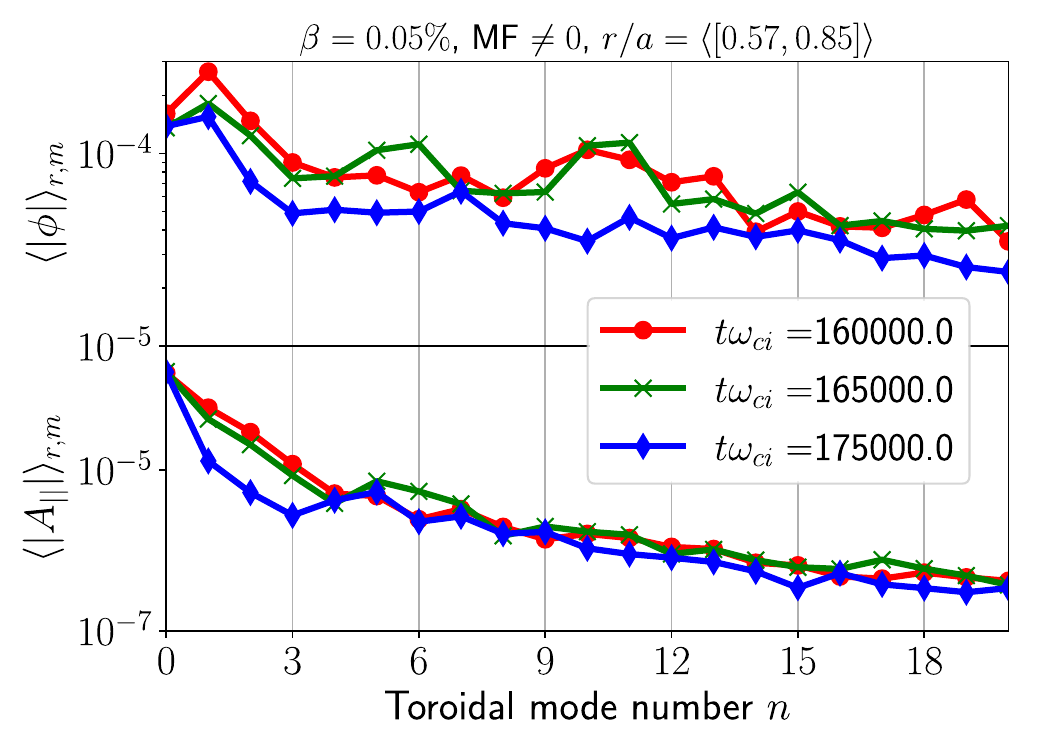}
        \caption{MF$\neq0$, $r\in[0.57,0.85]$.}  
	\label{subfig:Spect_rextMF}
	\end{center}
	\end{subfigure}~
        \begin{subfigure}{0.4\linewidth}
        \begin{center}
        \includegraphics[width=\linewidth,keepaspectratio]{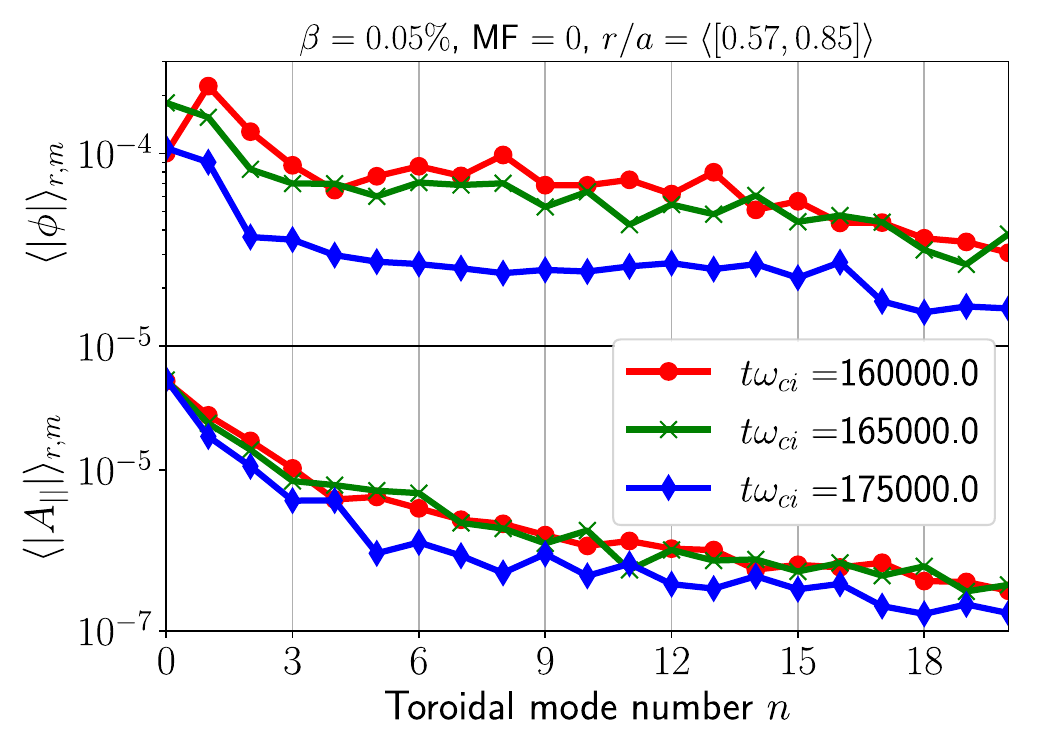}
        \caption{MF$=0$, $r\in[0.57,0.85]$.} 
	\label{subfig:Spect_rext}
	\end{center}
	\end{subfigure}
        \caption{Poloidal and partial radial average amplitudes for
		 $\phi$ and $A_{||}$ showing the spectra for the inner region $s=[0.15,0.57]$
		 when the electron mirror force MF is (a) included in electron dynamics (MF $\neq 0$)
		 and (b) excluded (MF $= 0$). The spectra for the outer region $s=[0.57,0.85]$
		 is shown for (c) MF $\neq 0$ and (d) MF$= 0$.}
       \label{fig:Spec_in_out_MF}
\end{center}
\end{figure*}
Without the mirror force, the $n=1$ vortices are weaker, the higher-$n$
turbulent structures persist and dominate the flow dynamics.
Because the nature of turbulence is
different with and without trapped electrons,
we investigate the amplitude of the toroidal modes, summed over poloidal modes, and 
averaged separately over the inner and outer regions, i.e., for $s\in[0.15,0.57]$ and
$s\in[0.57,0.85]$ respectively (\Fref{fig:Spec_in_out_MF}), at $t\omega_{ci}=1.6\cdot10^5$,
$1.65\cdot10^5$ and $1.75\cdot10^5$.
In the inner region, the amplitude of $\phi_{n=0,1}$, corresponding to the
zonal and island generated flows, remain about the same during the rapid island
decay phase when the mirror force is included, while they decrease after the island maximum size
if electron trapping is excluded (\frefs{subfig:Spect_rintMF}{subfig:Spect_rint}).
The reduced flows without the mirror force seem to be related to the peak
around $n=5$ for $\phi$, likely due to KHI driven turbulence,
which almost disappear in time when the mirror force is included.
This is clearly visible in \frefs{subfig:Phi_2D_tave_1_6e5_1_65e5}{subfig:Phi_2D_tave_1_65e5_1_85e5},
where turbulent structures persist without the mirror force.

\cor{In the outer region, additional modes peaking at $n=10$, $13$ and $19$, whose intensity reduces in time,
are visible in presence of the trapped electrons populations (MF $\neq0$) (\fref{subfig:Spect_rextMF}).}
Without the mirror force, the amplitude of $\phi_{n=1}$ drops significantly
while the amplitude of $\phi_{n=0}$ increases, indicating that the island-induced
flows are not enhanced.
In comparison, the $\cross{E}{B}$ flows for $\beta=0.12\%$ ($n\in[0,30]$) in the saturated phase
seem to compensate around the low-field-side X-point while no in- or out-flows
are observed at the high-field-side one (\fref{subfig:Phi_n_vs_time_Contour_Be0_0012} right panel).
The trapped electrons driven instability and KHI are less destabilized
for this weaker driven tearing mode due to larger plasma $\beta$, due to the reduced size
of the potential and density perturbations in this case.
The results reported in this section might be relevant to the
evolution of strongly-driven magnetic islands, as they develop for instance during sawtooth reconnection.
This point is further addressed in the Conclusions. Here, the nonlinear flow dynamics is found to be responsible for a
violent shrinking of the island, which is particularly strong in the presence of trapped electrons.
A discussion of the differences in the nonlinear energy exchange processes with and without electron trapping,
which would require dedicated code diagnostics that are presently not available, is left for future work.

We conclude that combination of current density redistribution by
strongly driven tearing mode, the mutual interaction of the grown island with
island-generated Kelvin-Helmholtz turbulence and
the trapped electrons driven instability 
is responsible for the island decay. These effects are
weakened as the tearing mode growth rate is decreased either
reducing the electron mass $m_e$ or increasing the plasma $\beta$.
Another reduction of the tearing mode can be obtained for non-flat
profiles of the temperature or the density.

\section{Tearing Mode with Temperature and Density Gradients\label{sec:LinGradTScans}}
\cor{In this section, we investigate the evolution of the drift-tearing-mode including finite density and temperature gradients. In such a case, }kinetic
theory predicts that the real frequency of the tearing mode should
be in electron diamagnetic drift direction for finite density
gradient.\cite{Drake77} The rotation direction can change depending
on the ratio $T_i/T_e$.\cite{Porcelli_PRL91}
For hot ions, the frequency was numerically obtained to be in the
ion diamagnetic drift direction.\cite{Tassi_JPhCS10} In this section,
the linear tearing mode growth rate and frequency, as well as the island frequency
during the nonlinear phase, are investigated including finite temperature and density gradients. \cor{Here, the gradients are the drive of the tearing mode rotation and the contribution of $\omega_D$ (see \fref{fig:gamma_freq_aspect_Flat}) is negligible.}
\subsection{Linear Growth of the Tearing Mode \label{subsec:LinGrad}}
\begin{figure}
	\begin{center}
        \begin{subfigure}{0.75\linewidth}
	\begin{center}
		\includegraphics[width=\linewidth,keepaspectratio]{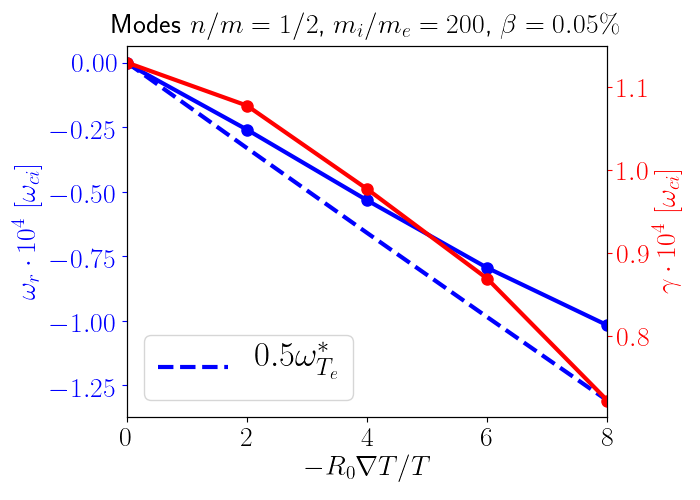}
		\caption{$\RLT{}$ scan, $\beta=0.05\%$.}
        \label{subfig:gamma_Freq_vs_GT_GN0_Beta005}
	\end{center}
	\end{subfigure}\\
        \begin{subfigure}{0.75\linewidth}
	\begin{center}
		\includegraphics[width=\linewidth,keepaspectratio]{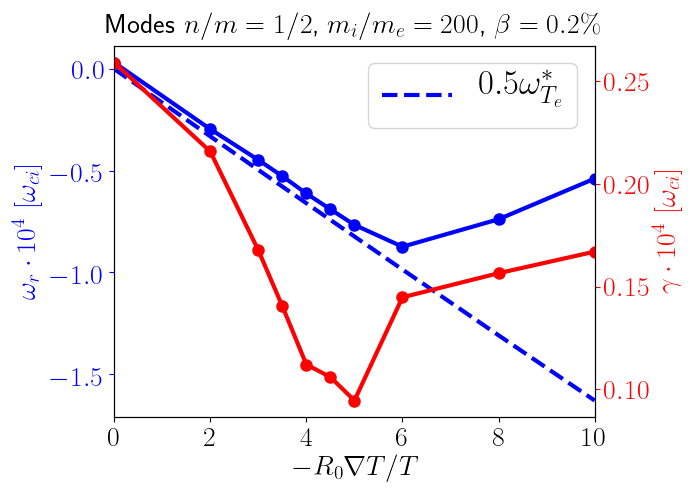}
		\caption{$\RLT{}$ scan, $\beta=0.2\%$.}
        \label{subfig:gamma_Freq_vs_GT_GN0_Beta02}
	\end{center}
	\end{subfigure}
		\caption{$\left|{A_{||}}_{2,1}\right|$ growth rate $\gamma$ and frequency $\omega_r$ versus $-R_0\GT{}$ with $\GN{}=0$. 
		(a): $\beta=0.05\%$, $\gamma$ decreases with $\RLT{}$. The frequency
		$\omega_r$ is in the electron diamagnetic direction ($\omega_r<0$),
		its absolute value increasing with $\RLT{}$. 
		(b): $\beta=0.2\%$, for $\RLT{}>5$ an increase of the growth rate and a decrease in absolute value of the mode frequency are observed.}
	\label{fig:GN0_GT_scan_beta0_0005_and_0_002}
	\end{center}
\end{figure}
We perform first simulations scanning the temperature
gradients of both species simultaneously
($\GT{e}=\GT{i}\equiv\GT{}$), for flat equilibrium density and for
two values of the plasma $\beta$ \cor{and an inverse aspect ratio $A=R_0/a=10$}. 
At $\beta=0.05\%$, the mode growth $\gamma$ is reduced increasing
the temperature gradient while the real frequency increases in
absolute value. The tearing frequency follows the kinetic
estimation $\omega_{TM}=\omega^*_n+0.5\omega^*_{T_e}$ and the
pressure gradient has a stabilizing effect on the mode growth.
A different situation is however observed for $\beta=0.2\%$.
While the growth rate $\gamma$ decreases with an increased pressure
gradient for $-R_0\GT{}<5$ with a frequency $\omega_{TM}\cong\omega^*_n+0.5\omega^*_{T_e}$,
the growth rate increases for larger values of the normalized logarithmic gradient and
the frequency deviates from $\omega^*_n+0.5\omega^*_{T_e}$ (\fref{fig:GN0_GT_scan_beta0_0005_and_0_002}).
\begin{figure}
   \begin{center}
      \begin{subfigure}{\linewidth}
      \begin{center}
        \includegraphics[width=0.75\linewidth,keepaspectratio]{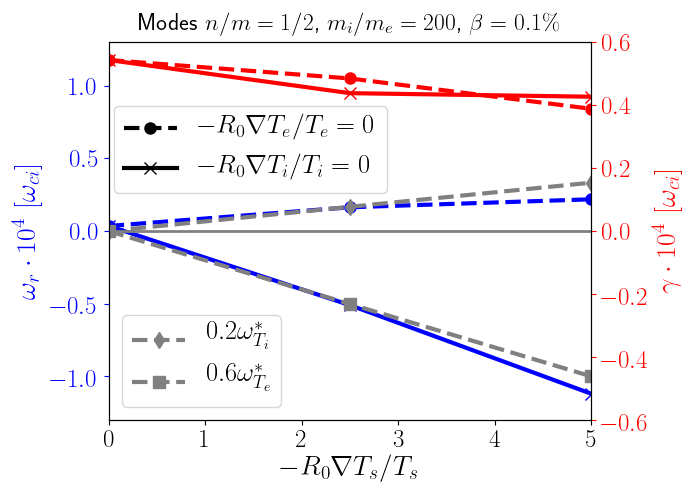}
        \caption{Scan in $\RLT{s}$, $s=i$ or $e$ and $\beta=0.1\%$.}
        \label{subfig:Lin_growth_Aparn1m2_Be0.001}
      \end{center}
      \end{subfigure}\\
      \begin{subfigure}{\linewidth}
      \begin{center}
        \includegraphics[width=0.75\linewidth,keepaspectratio]{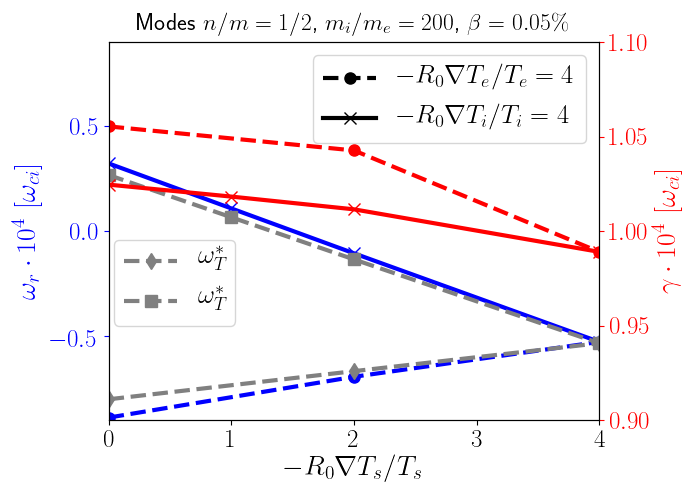}
	      \caption{Scan in $\RLT{s}$, $s=i$ or $e$, $\beta=0.05\%$ and
	      $\omega_T^*=0.2\omega^*_{T_i}+0.6\omega^*_{T_e}$ }
        \label{subfig:Lin_growth_Aparn1m2_Be0.0005}
      \end{center}
      \end{subfigure}
	   \caption{Tearing growth rate $\gamma$ and frequency $\omega_r$ as a function of the
		    logarithmic temperature gradient of one species ($e$ or $i$),
		    keeping the logarithmic gradient of the other species fixed either to
		    (a) $\RLT{}=0$ or (b) $\RLT{}=4$.}
\label{fig:Lin_Apar_GammaFreq_GTs_Scan_be0.001}
\end{center}
\end{figure}
The tearing frequency $\omega_{TM}$ remains, however, in the electron diamagnetic
drift frequency $(\omega_{TM}<0)$ for the selected gradients and $\beta$ values.
The (very small) positive frequency at zero gradient is related to the value of the
plasma $\beta$ and aspect ratio ($>0.07\%$, see \sref{sec:ValidationTM}).
We will show in \sref{sec:Desta_n1m3} that increasing the pressure gradient not only stabilizes the
tearing mode, but equally leads to the destabilization of a pressure gradient driven mode at high
plasma $\beta$.
We next isolate the temperature gradient contributions of each species on the tearing mode growth rate
and real frequency using $\GT{i}\neq\GT{e}$.
\Fref{subfig:Lin_growth_Aparn1m2_Be0.001} shows that the
tearing mode rotates in the ion diamagnetic drift direction ($\omega_{TM}>0$) when
$\GT{i}\neq0$ and $\GT{e}=0$, while it drifts in the electron diamagnetic drift direction
($\omega_{TM}<0$) when $\GT{i}=0$ and $\GT{e}\neq0$.
It can be clearly seen that the electron temperature gradient has a stronger impact on the rotation
as compared to the ion temperature gradient, so that the rotation is in the electron diamagnetic
direction if both gradients are equal.
The tearing growth rate is decreased by both $\GT{e}$ and $\GT{i}$.

\begin{figure}
\begin{center}
	\includegraphics[width=0.75\linewidth,keepaspectratio]{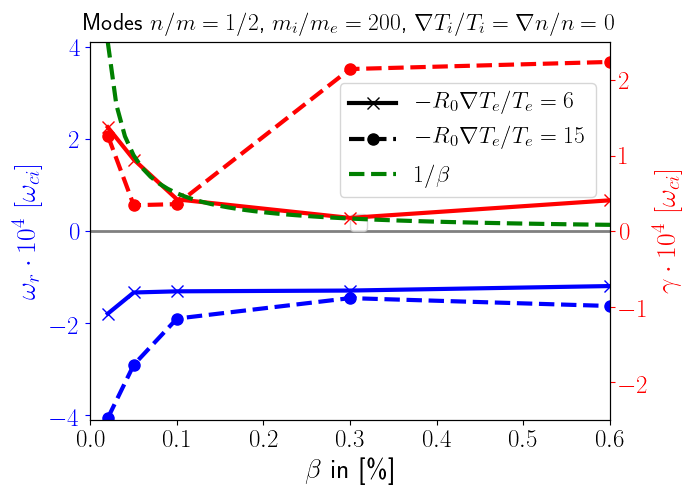}
        \caption{Plasma $\beta$ scan for $\RLT{}=6$ and 15.
	Mode $m/n=2/1$ growth rate $\gamma$ and frequency $\omega_r$ are displayed.
        The $1/\beta$ scaling is obtained for
        $\RLT{e}=6$ for $\beta\leq0.3\%$ while the growth rate rapidly increases with $\beta$ for $\RLT{e}=15$.
        The mode frequency is in the electron diamagnetic drift
        direction $(\omega_r<0)$, decreasing in absolute value with $\beta$.}
\label{fig:GTe0.6_1.5_Beta_Scans}
\end{center}
\end{figure}

The role of $\beta$ is illustrated
on \fref{fig:GTe0.6_1.5_Beta_Scans}. The $m/n=2/1$ mode growth follows the
$1/\beta$ scaling until it grows again beyond a given threshold in $\beta$. This threshold decreases
as the temperature gradient is increased.
As observed for flat profiles, the mode frequency reduces in absolute value with increasing
plasma $\beta$. For $\RLT{e}=15$, it rapidly slows down, deviating from the electron diamagnetic
frequency $\omega^*_e\cong-4.5\cdot10^{-4}\omega_{ci}$, but remains in the electron diamagnetic drift direction (here $\GT{i}=\GN{}=0$).
A similar analysis has been performed scanning the density gradient for flat electron and ion
temperatures and different values of $\beta$ (not shown for the sake of brevity).
Similarly to \fref{fig:GN0_GT_scan_beta0_0005_and_0_002}, a transition in growth rate and
frequency of the $m/n=2/1$ mode can be observed if the density gradient exceeds a ($\beta$-dependent)
threshold. The rotation of the mode turns to the ion diamagnetic direction for large density
gradients (with $\GT{}=0$). This behavior is discussed in the next subsection.
\subsection{Destabilization of the $m/n=3/1$ Mode\label{sec:Desta_n1m3}}
\begin{figure}
\begin{center}
        \begin{subfigure}{\linewidth}
	\begin{center}
	\includegraphics[width=0.75\linewidth,keepaspectratio]{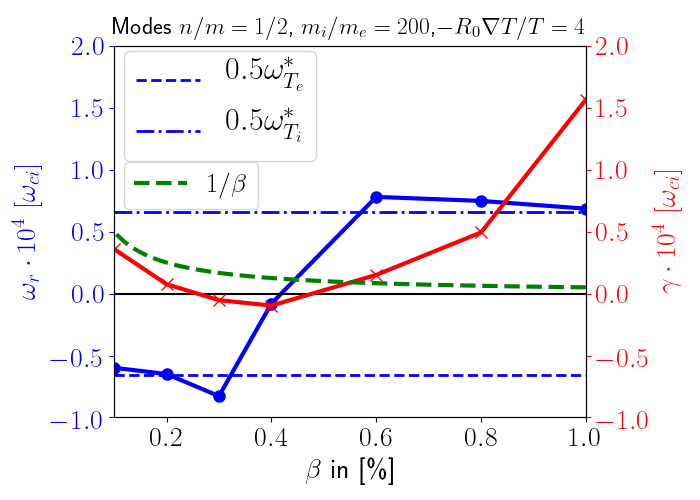}
        \caption{Growth rate $\gamma$ and frequency $\omega_r$ versus $\beta$}
        \label{subfig:gamma_Freq_vs_beta_LinGT0.4}
	\end{center}
	\end{subfigure}\\
	\begin{subfigure}{\linewidth}
	\begin{center}
		\includegraphics[width=0.75\linewidth,keepaspectratio]{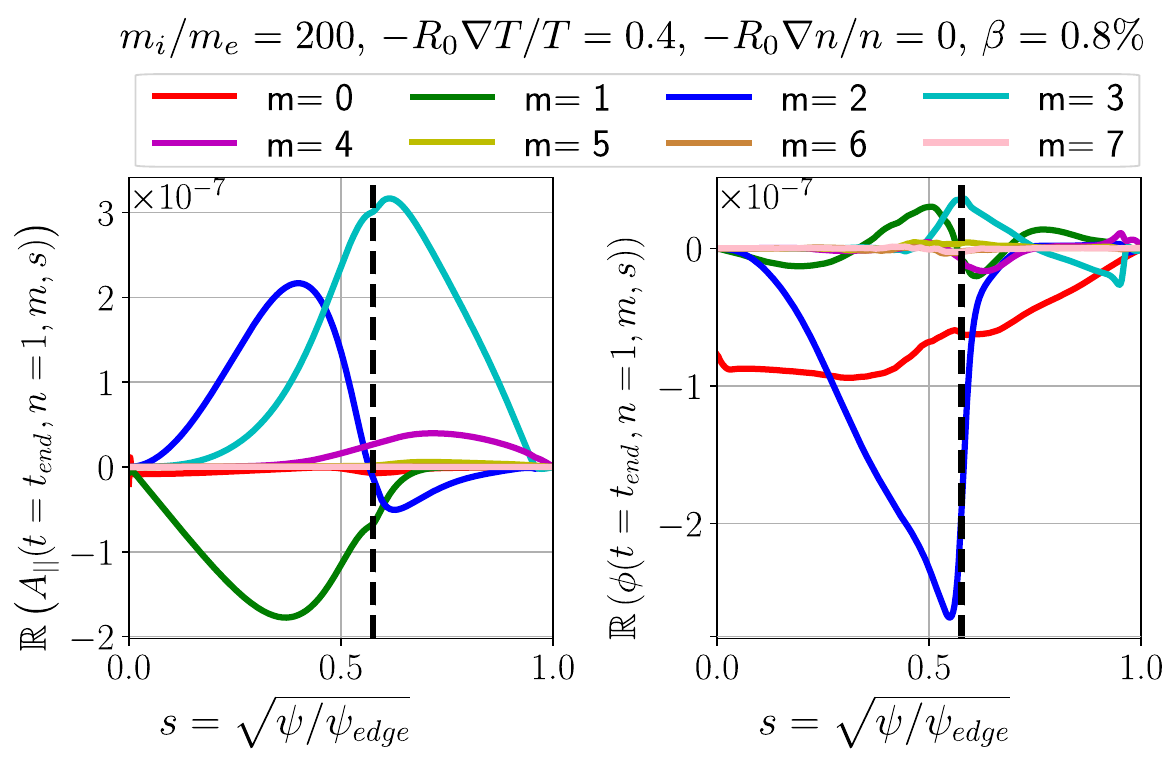}
        \caption{$\mathbb{R}\left(\phi\right)$ and $\mathbb{R}\left(A_{||}\right)$ poloidal modes radial structures of the eigenfunctions.}
        \label{subfig:Phi_Apar_n1_mall_Time_and_Struct_GT0.4GN0_beta0_008}
	\end{center}
	\end{subfigure}
	\caption{Parameters: $\GN{}=0$, $\GT{}=0.4$, $m_i/m_e=200$. The tearing is stabilized by an increase of $\beta$ until $\beta= 0.4\%$, where the
          frequency changes from the electron ($\omega_r<0$) to the ion ($\omega_r>0$) diamagnetic drift direction (a). On the $q=2$ surface, the
          dominant poloidal mode of the $n=1$ amplitude of $A_\parallel$ is $m=3$ for $\omega_r>0$, as shown in the left panel of (b).}
\label{fig:Lin_Apar_rotation_GT04_GN0_beAll}
\end{center}
\end{figure}
In order to understand the deviation of the growth rate from the $1/\beta$ scaling and of the
frequency from $\omega_{TM}=\omega^*_n+0.5\omega^*_{T_e}$ as the pressure gradient increases,
we perform a $\beta$ scan for $-R_0\GT{}=4$ (same logarithmic gradient for both species) and flat equilibrium density.
\Fref{fig:Lin_Apar_rotation_GT04_GN0_beAll} shows that the growth rate $\gamma$ decreases
with $\beta$ until $\beta=0.4\%$, where the mode is stable, and increases again for higher values of $\beta$.
The $m/n=2/1$ mode frequency changes from the electron $(\omega_{2,1}<0)$ to ion
$(\omega_{2,1}>0)$ diamagnetic direction. The tearing is the branch where the mode frequency is
close to $0.5\omega_{T_e}^*$ while the second branch has a frequency close to $0.5\omega_{T_i}^*$.
The dominant instability at the $q=2$ rational surface is, in this case, a pressure gradient driven
electromagnetic mode, with rotation frequency $\omega\cong 0.5\omega_{T_i}^*$ in the ion diamagnetic direction.
Inspection of the radial structures of the poloidal modes for the toroidal mode $n=1$,
\fref{subfig:Phi_Apar_n1_mall_Time_and_Struct_GT0.4GN0_beta0_008}, reveals that the $m=2$ has a
twisting (not tearing) parity in this high-$\beta$ branch with $\omega_{2,1}>0$. The amplitude of the
electromagnetic potential $A_\parallel$ exhibits a strong coupling of the $(2,1)$ component to its
neighboring poloidal harmonics. In particular, the amplitude $(3,1)$ component is stronger than the
$(2,1)$ component around the $q=2$ surface. The electrostatic potential is still dominated by its $(2,1)$ component.
Gyrokinetic simulations of collisionless tearing mode in slab geometry showed that a
mode driven by a high pressure gradient, at large plasma $\beta$, can enhance the tearing mode growth rate when $\tilde{\boldsymbol{B}}_{||}$ is non-zero.\cite{PueschelPoP11,Pueschel_PoP2015} In such a case, the $\nabla B$ drifts are balancing the $\cross{E}{B}$ drifts and the tearing mode is enhanced and has a rotation frequency in the ion diamagnetic drift direction. Here, the high pressure driven mode is similar but the tearing mode is stable when the pressure gradient mode is growing. In our case, the destabilization of the mode with the pressure gradient, predominantly at large
plasma $\beta$, is similar to the kinetic ballooning mode (KBM)
instability, so that the process resembles the well-known ITG-KBM transition with increasing $\beta$.\cite{Maeyama_PoP2014,IshizawaJPP15} Another possibility for the instability are Alfvén ion temperature gradient (AITG) modes,
destabilized by pressure gradients at rational surfaces and with large frequencies in the ion diamagnetic direction.\cite{ZoncaPoP99,FalchettoPoP03} While additional work is required to fully understand the nature of this mode, we performed
additional scans in mass ratio $m_i/m_e$ and aspect ratio $A=R_0/a$ for $\beta=0.8\%$ and $\RLT{}=4$.
The $m/n=3/1$ mode growth rate slightly diminishes with more realistic mass ratio $m_i/m_e$ while a larger
aspect ratio $A$ augments it. The frequency moderately increases and reduces with $m_i/m_e$ and
$A$. The mode $m/n=3/1$ is thus behaving in an opposite manner to the $m/n=2/1$ tearing mode which becomes more
and more stable with increasing pressure gradient, $\beta$, mass ratio $m_i/m_e$ and
aspect ratio $A$.


\subsection{Propagation of Nonlinear Islands \label{sec:NLwithGradients}}

As mentioned in \sref{sec:Intro}, the island rotation frequency is an important quantity for the determination of the stability of magnetic islands at the beginning of their nonlinear phase, as it influences both the polarization and the bootstrap current. The polarization current is proportional to $(\omega-\omega_e^*)(\omega-\omega_i^*)$ if the island size is below the typical ion-orbit size and the ion response can be assumed to be adiabatic (unmagnetized), while it is proportional to
$\omega(\omega-\omega_i^*)$ in the large-island limit (flattened profiles).\cite{Smolyakov_PPCF1993} For the bootstrap current, when the island size is comparable with the ion orbit size, the adiabatic ion response implies that the bootstrap current perturbation is strong when the island rotates in the ion diamagnetic direction and weak when it rotates in the electron direction.\cite{Bergmann_PoP09}
For a collisionless drift tearing mode, the rotation frequency $\omega$ of a magnetic island is given by
\begin{equation}
	\omega\cong \langle \omega^*\rangle +\langle\tilde\omega_{E\times B}\rangle\label{eq:omegaFreq}
\end{equation}
where $\langle\cdots\rangle$ is a radial average over the island width and $\omega^*=\omega^*_{lin}+\tilde\omega^*$ the diamagnetic frequency,\cite{Nishimura2008} $\omega_{lin}$ is the contribution from the gradients before
the nonlinear interaction of the island with the surrounding plasma. The zonal flows and the pressure profile flattening play important
roles in the island evolution. In particular, there is a relation
between the flattening of the pressure profile within the island and
the nonlinear generation of zonal flows, proportional to the Reynolds and Maxwell stresses, but also
to the nonlinear ion diamagnetic stress related to the pressure profile
gradient.\cite{waelbroeck_PRL2005,lahaye_PoP2003,Uzawa_PoP2010}
The latter becomes important in the zonal flow generation mechanism for large islands as pressure profiles are strongly flattened.\cite{Uzawa_PoP2010} 
Furthermore, the electrostatic potential has the role of providing an $\cross{E}{B}$ flow that ensures
$E_{||}=0$ within the island.\cite{smolyakov_PoP1995} In the case of a magnetic island with imposed
rotation velocity, the total electron frequency within the island is
$\omega_{tot,e}=\omega_{p_e}^* +\omega_{ExB}$ which simplifies to $\omega_{tot,e}\cong\omega_{ExB}$
when the pressure profile is flattened.\cite{SiccinoPoP11}

\begin{figure}
\begin{center}
           \includegraphics[width=\linewidth,keepaspectratio]{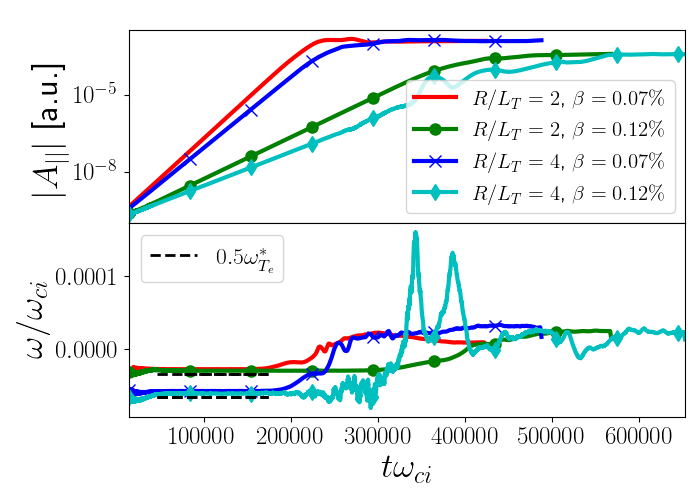}
		   \caption{$|A_{||,21}|$ and its real frequency $\omega_r$ time evolution for $\beta=0.07\%$, $0.12\%$, $\RLT{}=2$ and $4$. The plasma $\beta$ and $\RLT{}$ decrease the growth rate. The mode frequency transitions from the electron to the ion drift diamagnetic direction between the linear and nonlinear evolution. The black dashed lines represents the linear electron diamagnetic drift $0.5\omega_{T_e}^*$.}
           \label{subfig:Apar_Freq_vs_Time_RLT2_4_Be07_12}
\end{center}
\end{figure}

The nonlinear evolution of the tearing mode is investigated for finite temperature gradients using
$-R_0\nabla T_i/T_i=-R_0\nabla T_e/T_e\equiv -R_0/L_T=2$ and $4$
(for the density we take $\RLN{}=0$),
for $\beta=0.07\%$ and $0.012\%$. The gradients are selected such that ITG, TEM or KBM modes are kept
under their linear stability threshold.\cite{Horton_RMP99,RyterPRL05,DannertPoP2005,Maeyama_PoP2014}
\Fref{subfig:Apar_Freq_vs_Time_RLT2_4_Be07_12} shows the time evolution of the tearing mode amplitude
and its real frequency $\omega_{TM}$. The growth rate and saturation size decrease with
increasing $\beta$ and/or temperature gradients, as expected from the results presented above. 
The increased stability of the mode implies that the strong island decay observed for
$\beta<0.07\%$ with flat density and temperature profile is slightly seen for $-R_0\GT{}=2$ and
$\beta=0.07\%$ only. The rotation frequency reduces (in absolute value) when the island approaches saturation,
then it changes to the ion diamagnetic direction. This behavior has been observed before in two-fluid
simulations.\cite{MuragliaNF09,IshizawaPPCF19}

In order to investigate the change of rotation
frequency, we scrutinize the pressure profile and zonal flows as the island appears and grows in size.
We perform nonlinear simulations with $n\in[0,30]$ in order to compare the island size $W\propto\sqrt{|A_{||,21}|}$, its rotation frequency, the time evolution
of the zonal flows and $\omega_*$ averaged over the island width or close to the rational surface.
\begin{figure}
\begin{center}
           \includegraphics[width=\linewidth,keepaspectratio]{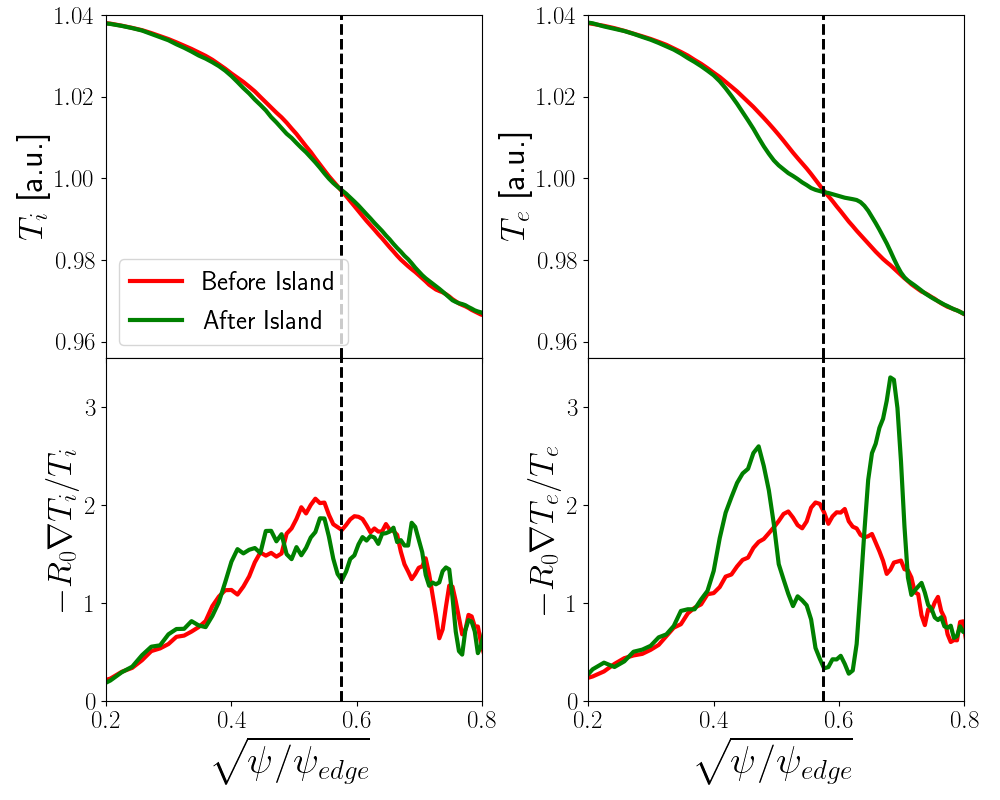}
           \caption{Radial profiles of the ion and electron temperature and gradients before and after the island appearance. The electrons strongly flattens while the ions weakly respond to islands. A flux surface average is used to compute the temperature profiles.}
           \label{fig:RLT_vs_rad_tend_kT0.2_beta0.0007}
\end{center}
\end{figure}

\begin{figure}
\begin{center}
           \includegraphics[width=\linewidth,keepaspectratio]{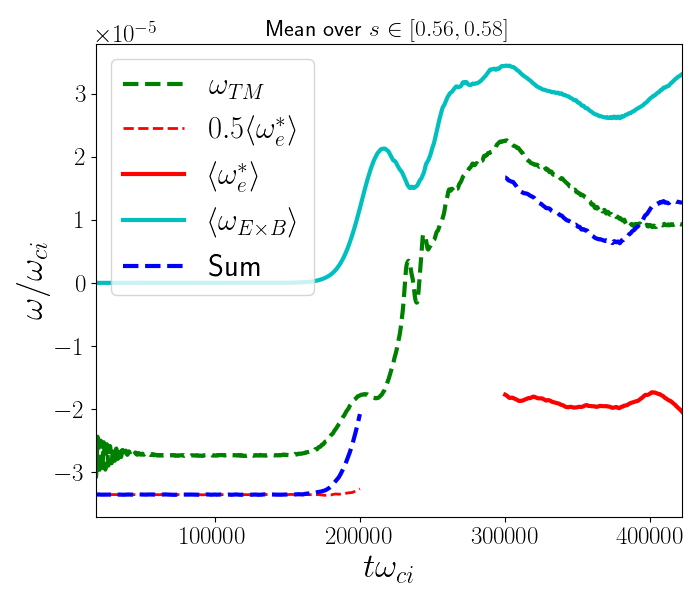}
	   \caption{Time trace of the tearing mode ($\omega_{TM}$) in green, electron diamagnetic ($\omega_e^*$) frequencies in red, zonal flows ($\omega_{\cross{E}{B}}$) in cyan and
	   the sum $\omega_e^*+\omega_{\cross{E}{B}}$ in blue for $\beta=0.07\%$ and $\RLT{}=2$. A flux surface average is applied on the zonal flow and the temperatures used to compute diamagnetic frequencies. Linearly $\omega_{TM}$ is close to $0.5\langle\omega_{T_e}^*\rangle$ while nonlinearly $\omega_{TM}$ approaches $\langle\omega_{\cross{E}{B}}\rangle+\langle\omega_{T_e}^*\rangle$.}
           \label{fig:Freqs_vs_time_RmeanZF_kT0.2_beta007}
\end{center}
\end{figure}

\begin{figure}
\begin{center}
           \includegraphics[width=\linewidth,keepaspectratio]{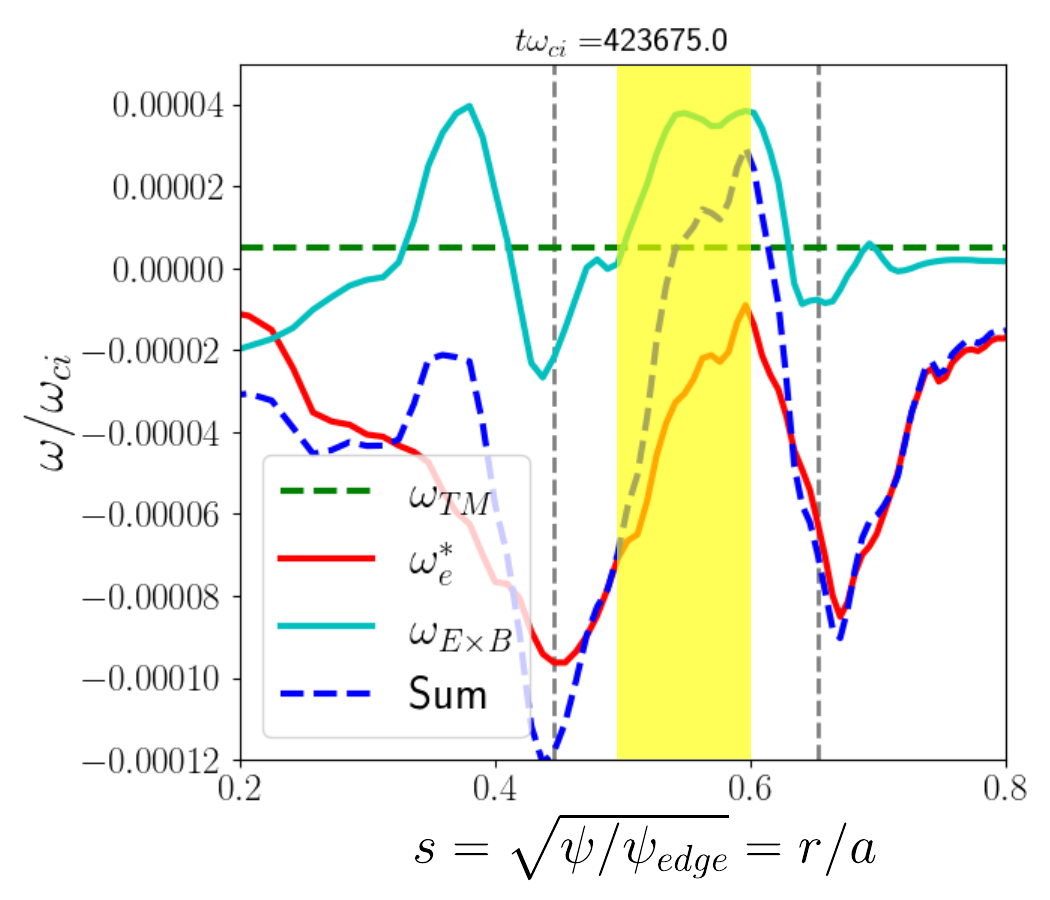}
	   \caption{Tearing mode ($\omega_{TM}$ in green), electron diamagnetic ($\omega_e^*$ in red)
	   and zonal flows ($\omega_{\cross{E}{B}}$ in cyan) frequencies radial profiles at the end of the simulations $\beta=0.07\%$ and $\RLT{}=2$. The resonant surface $q=2$ is at $s=\sqrt{\psi/\psi_{edge}}=0.57$ and vertical gray dashed-lines represent the island separatrix. The blue-dashed line is the sum of the zonal flow and the electron diamagnetic drift $\omega_{\cross{E}{B}}+\omega_e^*$.}
           \label{fig:Freqs_vs_Radius_time1e9_kT0.2_beta007}
\end{center}
\end{figure}

First, we investigate profile flattening which is linked to the diamagnetic frequencies (first term in \eref{eq:omegaFreq}) for the case $\beta=0.07\%$ with $\RLT{}=2$ (red curves in \fref{subfig:Apar_Freq_vs_Time_RLT2_4_Be07_12}).
The temperature profiles and their gradients for both species are depicted in
\fref{fig:RLT_vs_rad_tend_kT0.2_beta0.0007}. At the end of the simulation, the electron
temperature profile clearly flattens over the island while the ions show a weak response.
It should be stressed that the pressure profile flattening within the island is underestimated
in the plots, because the flux surface average considered in the temperature computation includes
contributions from the X-points. 
The weak response of the ion temperature profile is furthermore related to the
ion banana width at the resonant surface $q=2$
\begin{equation}
w_b=q_s\rho_i\sqrt{2A_s}, \label{eq:Banana}
\end{equation}
with $A_s=R_0/r_s$ and $q_s$ the aspect ratio and safety factor at the resonant
surface. Using $q_s=2$ and $r_s=0.57$, \eref{eq:Banana} gives $w_b\cong 10\rho_i$ while
the full island width $W\cong20\rho_i$. The overlap between the trapped ions and the magnetic
island is significant, the ion response to the island is weak as the temperature profile
flattening.\cite{SiccinoPoP11} \cor{In particular, only few ions can be trapped within the island and their contribution to the pressure profile flattening is
weak while the trapped ions were shown to be responsible for the density profile
flattening in gyrokinetic simulations, using slab geometry together with a model to include torodial effects, of imposed
magnetic islands for which $w_{b_i}\ll W$.\cite{ZarzosoNF15}}
\Fref{fig:Freqs_vs_time_RmeanZF_kT0.2_beta007} shows the time evolution of the zonal flow (cyan curve), tearing mode (green dashed curve),
electron diamagnetic frequency (red curve) and the sum of zonal flow with the electron drifts (dark blue dashed-curve). A flux surface average
is considered to compute the zonal flows as well as the temperatures of the species that are used to compute the diamagnetic frequencies.
These quantities are further radially averaged around the rational surface while the island frequency is computed at the rational surface. 
In the exponential growth (``linear'') phase, the contribution of the
$E\times B$ velocity to the island rotation is negligible, and the
rotation frequency is close to $0.5\omega_{T_e}^*$, as expected
from theory and found also in previous gyrokinetic simulations.\cite{Drake77,HornsbyPoP15,HornsbyNF16} As the
tearing mode leaves the linear phase and approaches saturation, its frequency turns to the ion
diamagnetic direction, following the evolution of zonal flows time variation. The final island velocity is comparable to the
sum of the zonal flow and the electron diamagnetic frequency (dark blue dashed-curve). \cor{The island frequency obtained here is different than previous gyrokinetic simulations, in slab geometry, that found an almost zero frequency of the island for gradients below the linear threshold of micro-instability.\cite{HornsbyNF16} The absence of collisions in our simulations explains this difference as there is no damping of the poloidal rotation without
neoclassical viscous force.\cite{SmolyakovPoP95} In such a case, the poloidal
velocity of the island is mostly related to $\phi$.}
\Fref{fig:Freqs_vs_Radius_time1e9_kT0.2_beta007} shows the flux surface average
radial profiles of the $\cross{E}{B}$ shear flow, tearing and electron diamagnetic drift frequencies at the simulation end.
The potential that develops starting from $t\omega_{ci}\cong2\cdot10^5$ is consistent with the
frozen-in condition for the electrons, i.e., the sum (dark blue curve) of electron diamagnetic velocity (red curve) and
$\cross{E}{B}$ velocity (cyan curve) matches closely the island speed (green dashed curve) \cor{when averaged in the vicinity of the rational surface $s\in[0.5, 0.6]$ (yellow shaded box)}.
This had been verified in previous gyrokinetic results,\cite{SiccinoPoP11} where, however, the
rotation of the island was imposed in the simulations and not self-consistently determined.
\section{Summary \label{sec:Summary}}
In this paper, we presented a numerical gyrokinetic study of the tearing mode in a
large-aspect-ratio tokamak. The mode develops because of a tearing-unstable current profile,
initialized through a shifted Maxwellian as equilibrium distribution function for the electrons.
The equilibrium is prescribed in such a way that the safety-factor profile is consistent with the
corresponding equilibrium parallel electron flow. The simulations were performed in the collisionless
limit, so that the electron inertia is responsible for the breaking of the frozen-in condition
$E_{||}=0$ around the rational surface. To reduce the computational burden and allow wide
parameter scans, the mass ratio was fixed to $m_i/m_e=200$ in most of the simulations.
The initial temperature and density gradients were kept below the threshold for turbulent instabilities to
develop, so that we could focus on the linear and nonlinear dynamics of the tearing instability.
There are two aspects of the dynamics of magnetic island caused by tearing modes. One is its width,
the other is the rotation. The former is investigated
performing mainly simulations for flat profiles, because the rotation frequency is very small compared to the
island growth in the absence of background diamagnetic flow effects, and the later is addressed
carrying out simulations for finite gradient profiles as diamagnetic effects are responsible
for the large rotation speed of the mode, at least in its linear phase.

We showed that, in the linear phase, the tearing mode follows the theoretical scaling for the growth rate
and found that it exhibits a weak rotation frequency even without pressure gradients. For our parameters,
the sign of this residual frequency, related to curvature effects, depends on $\beta$, being negative
(what would be the electron diamagnetic direction in the presence of gradients)
for $\beta\leq0.07\%$ and positive otherwise.
Nonlinear simulations demonstrate that, for strongly driven modes (which in our setup are realized
through $\beta\leq0.07\%$ and $m_i/m_e=200$), the magnetic island width is dramatically reduced shortly
after the initial saturation of its growth. The mechanism is composed of two subsequent processes:
\begin{enumerate}
\item Current redistribution: a strongly driven tearing mode produces large magnetic islands that
  generate nonlinearly significant zonal fields $A_{||}(m=0,n=0)$. The related redistribution of the
  current profile lead to a reduced island drive and an initial shrinking of the island.
\item Island-induced flows: strongly driven tearing modes generate intense
  $\cross{E}{B}$ and diamagnetic flows, with helicity $m/n=2/1$, at the island separatrix
  because of localized electrostatic potential
  and quadrupolar density distribution around the X-points. These fields
  eventually become unstable to Kelvin-Helmholtz instability which then modifies
  the flow patterns. Strong outflows from the X-points, particularly intense in the presence
  of trapped electrons, lead to a violent shrinking of the island size.
\end{enumerate}
While the current redistribution process, which initiates the island decrease, independent of the generation of the electrostatic turbulence, the latter is crucial for
the further island decay, as it enhances the island generated
flows responsible for its saturation. It is remarked that retaining
a sufficient number of toroidal modes in the simulation turns out to be essential for
capturing the current redistribution as well as the turbulence generation in their full extent.
These strongly-driven scenarios are of interest for strongly driven magnetic islands like in sawtooth
reconnection.\cite{BiskampPoP97} Obviously, the parameters considered in our simulations cannot be
considered realistic and the electron mass, in particular, is artificially increased for numerical
convenience. This, however, can be regarded as a way to achieve a strong drive of the mode, which in realistic
situations may be enforced through other physical processes.

Including density and temperature gradients, whose values are selected to avoid the
destabilization of micro-instabilities, linear simulations reveal that the tearing mode
growth rate is reduced by diamagnetic
effects increasing $-R_0\GT{i}$ and $-R_0\GN{}$ at low
plasma $\beta$ value.
For $\GT{i}\gg\GT{e}$ the tearing rotates in the ion diamagnetic
direction, while it rotates in the electron direction if the ion
and electron pressure gradients are comparable, or larger for the
electrons. At large plasma $\beta$ and pressure gradients, the tearing mode is
suppressed and is replaced by a mode with twisting parity coupled
to a mode with helicity $m/n=3/1$, whose growth rate and frequency
increase with $\beta$ and the pressure gradient. Its frequency is close to
$0.5\omega_{T_i}^*$ when $\RLN{}=0$. Additional work is required to determined
the nature of this mode which bears similarities with KBM or AITG modes.
The reduction of the linear growth rate in the presence of finite pressure gradients
results also in a smaller island width at saturation. An island decay, as observed for
flat profiles, is found in weaker form only for
$-R_0\GT{}=2$, $\beta=0.05\%$ and mass ratio $m_i/m_e=200$, i.e., for parameters close to those
used in flat-profile simulations. Concerning the mode rotation, nonlinear simulations with equal
electron and ion temperature gradients were performed. The tearing
mode is found to rotate in the electron diamagnetic drift direction, regardless of the plasma $\beta$ value,
When the island grows, the rotation frequency (in the electron direction in the linear phase) is
reduced in absolute value as the electron pressure profile flattens within the island.
The ion temperature profile weakly changes in response to the island because there is a
significant overlap of the ion banana and island width $W/w_b\cong 2$. At saturation,
the island rotates in the ion diamagnetic drift direction and this change of rotation direction is
associated to the nonlinear production of zonal flows. We verified that the $\cross{E}{B}$ flows generated
by the island are consistent with the fulfillment of the frozen-in condition.
This reversal of the rotation direction during its evolution plays a role in its neoclassical dynamics and
will be further investigated in the near future in the presence of micro-turbulence.
On a longer run, inclusion of collisions in the numerical setup will lead to
an exploration of the neoclassical processes that affect the dynamics of the tearing mode
and allow a comparison with theory, see e.g. \cite{DudkovskaiaPPCF21} and references therein.

\section{Acknowledgments}
This work was carried out (partially) using supercomputer resources, JFRS-1 provided under the EU-JA Broader Approach collaboration in the
Computational Simulation Centre of International Fusion Energy Research Centre (IFERC-CSC). Furthermore, part of this work has been carried
out within the framework of the EUROfusion Consortium, funded by the European Union via the Euratom Research and Training Programme
(Grant Agreement No. 101052200—EUROfusion). Views and opinions expressed are however those of the authors only and do not necessarily
reflect those of the European Union or the European Commission. Neither the European Union nor the European Commission can be held responsible
for them. This work was supported in part by the Swiss National Science Foundation. Simulations presented in this work were performed partially 
on the MARCONI FUSION HPC system at CINECA. We acknowledge PRACE for awarding us access to Marconi100 at CINECA, Italy. We
acknowledge PRACE for awarding us access to Joliot-Curie at GENCI@CEA, France. One of the author (FW) would like to thanks M.~Muraglia,
O.~Agullo for fruitful discussions during the first European Conference on Magnetic Reconnection in Plasma in May 2023, Marseille, France. FW would also like to thank A. Poyé from the Aix-Marseille University for providing is shooting algorithm and his help using it. Additionally, we are grateful to M.J. Pueschel and T. Jitsuk for the
code comparison with the GENE code and enlightening discussions. This research was carried out during the time the author (FW) was employed by the National Institutes of Natural Sciences (NINS). EP would like to thank
the Graduate School of Energy Science of the Kyoto University for its hospitality and
support during the finalization of this study.
EP is partially supported by the Eurofusion Enabling Research Project
(CfP-FSD-AWP24-ENR T-RecS). Finally, AM, AB and THS were partially
supported by Eurofusion Theory and Simulation Verification and
Validation Task 10.

\bibliography{libraryFW}

\begin{thebibliography}{94}%
\makeatletter
\providecommand \@ifxundefined [1]{%
 \@ifx{#1\undefined}
}%
\providecommand \@ifnum [1]{%
 \ifnum #1\expandafter \@firstoftwo
 \else \expandafter \@secondoftwo
 \fi
}%
\providecommand \@ifx [1]{%
 \ifx #1\expandafter \@firstoftwo
 \else \expandafter \@secondoftwo
 \fi
}%
\providecommand \natexlab [1]{#1}%
\providecommand \enquote  [1]{``#1''}%
\providecommand \bibnamefont  [1]{#1}%
\providecommand \bibfnamefont [1]{#1}%
\providecommand \citenamefont [1]{#1}%
\providecommand \href@noop [0]{\@secondoftwo}%
\providecommand \href [0]{\begingroup \@sanitize@url \@href}%
\providecommand \@href[1]{\@@startlink{#1}\@@href}%
\providecommand \@@href[1]{\endgroup#1\@@endlink}%
\providecommand \@sanitize@url [0]{\catcode `\\12\catcode `\$12\catcode
  `\&12\catcode `\#12\catcode `\^12\catcode `\_12\catcode `\%12\relax}%
\providecommand \@@startlink[1]{}%
\providecommand \@@endlink[0]{}%
\providecommand \url  [0]{\begingroup\@sanitize@url \@url }%
\providecommand \@url [1]{\endgroup\@href {#1}{\urlprefix }}%
\providecommand \urlprefix  [0]{URL }%
\providecommand \Eprint [0]{\href }%
\providecommand \doibase [0]{https://doi.org/}%
\providecommand \selectlanguage [0]{\@gobble}%
\providecommand \bibinfo  [0]{\@secondoftwo}%
\providecommand \bibfield  [0]{\@secondoftwo}%
\providecommand \translation [1]{[#1]}%
\providecommand \BibitemOpen [0]{}%
\providecommand \bibitemStop [0]{}%
\providecommand \bibitemNoStop [0]{.\EOS\space}%
\providecommand \EOS [0]{\spacefactor3000\relax}%
\providecommand \BibitemShut  [1]{\csname bibitem#1\endcsname}%
\let\auto@bib@innerbib\@empty
\bibitem [{\citenamefont {Furth}, \citenamefont {Rutherford},\ and\
  \citenamefont {Selberg}(1973)}]{furth_PF1973}%
  \BibitemOpen
  \bibfield  {author} {\bibinfo {author} {\bibfnamefont {H.~P.}\ \bibnamefont
  {Furth}}, \bibinfo {author} {\bibfnamefont {P.~H.}\ \bibnamefont
  {Rutherford}},\ and\ \bibinfo {author} {\bibfnamefont {H.}~\bibnamefont
  {Selberg}},\ }\bibfield  {title} {\enquote {\bibinfo {title} {{Tearing mode
  in the cylindrical tokamak}},}\ }\href {https://doi.org/10.1063/1.1694467}
  {\bibfield  {journal} {\bibinfo  {journal} {Physics of Fluids}\ }\textbf
  {\bibinfo {volume} {16}},\ \bibinfo {pages} {1054--1063} (\bibinfo {year}
  {1973})}\BibitemShut {NoStop}%
\bibitem [{\citenamefont {Yamada}, \citenamefont {Kulsrud},\ and\ \citenamefont
  {Ji}(2010)}]{Yamada2010}%
  \BibitemOpen
  \bibfield  {author} {\bibinfo {author} {\bibfnamefont {M.}~\bibnamefont
  {Yamada}}, \bibinfo {author} {\bibfnamefont {R.}~\bibnamefont {Kulsrud}},\
  and\ \bibinfo {author} {\bibfnamefont {H.}~\bibnamefont {Ji}},\ }\bibfield
  {title} {\enquote {\bibinfo {title} {Magnetic reconnection},}\ }\href
  {https://doi.org/10.1103/RevModPhys.82.603} {\bibfield  {journal} {\bibinfo
  {journal} {Reviews of Modern Physics}\ }\textbf {\bibinfo {volume} {82}},\
  \bibinfo {pages} {603--664} (\bibinfo {year} {2010})}\BibitemShut {NoStop}%
\bibitem [{\citenamefont {{Biskamp}}(2000)}]{BiskampBook00}%
  \BibitemOpen
  \bibfield  {author} {\bibinfo {author} {\bibfnamefont {D.}~\bibnamefont
  {{Biskamp}}},\ }\href@noop {} {\emph {\bibinfo {title} {{Magnetic
  Reconnection in Plasmas}}}},\ Vol.~\bibinfo {volume} {3}\ (\bibinfo {year}
  {2000})\BibitemShut {NoStop}%
\bibitem [{\citenamefont {Sauter}\ \emph {et~al.}(1997)\citenamefont {Sauter},
  \citenamefont {{La Haye}}, \citenamefont {Chang}, \citenamefont {Gates},
  \citenamefont {Kamada}, \citenamefont {Zohm}, \citenamefont {Bondeson},
  \citenamefont {Boucher}, \citenamefont {Callen}, \citenamefont {Chu},
  \citenamefont {Gianakon}, \citenamefont {Gruber}, \citenamefont {Harvey},
  \citenamefont {Hegna}, \citenamefont {Lao}, \citenamefont {Monticello},
  \citenamefont {Perkins}, \citenamefont {Pletzer}, \citenamefont {Reiman},
  \citenamefont {Rosenbluth}, \citenamefont {Strait}, \citenamefont {Taylor},
  \citenamefont {Turnbull}, \citenamefont {Waelbroeck}, \citenamefont {Wesley},
  \citenamefont {Wilson},\ and\ \citenamefont {Yoshino}}]{SauterPoP97a}%
  \BibitemOpen
  \bibfield  {author} {\bibinfo {author} {\bibfnamefont {O.}~\bibnamefont
  {Sauter}}, \bibinfo {author} {\bibfnamefont {R.~J.}\ \bibnamefont {{La
  Haye}}}, \bibinfo {author} {\bibfnamefont {Z.}~\bibnamefont {Chang}},
  \bibinfo {author} {\bibfnamefont {D.~A.}\ \bibnamefont {Gates}}, \bibinfo
  {author} {\bibfnamefont {Y.}~\bibnamefont {Kamada}}, \bibinfo {author}
  {\bibfnamefont {H.}~\bibnamefont {Zohm}}, \bibinfo {author} {\bibfnamefont
  {A.}~\bibnamefont {Bondeson}}, \bibinfo {author} {\bibfnamefont
  {D.}~\bibnamefont {Boucher}}, \bibinfo {author} {\bibfnamefont {J.~D.}\
  \bibnamefont {Callen}}, \bibinfo {author} {\bibfnamefont {M.~S.}\
  \bibnamefont {Chu}}, \bibinfo {author} {\bibfnamefont {T.~A.}\ \bibnamefont
  {Gianakon}}, \bibinfo {author} {\bibfnamefont {O.}~\bibnamefont {Gruber}},
  \bibinfo {author} {\bibfnamefont {R.~W.}\ \bibnamefont {Harvey}}, \bibinfo
  {author} {\bibfnamefont {C.~C.}\ \bibnamefont {Hegna}}, \bibinfo {author}
  {\bibfnamefont {L.~L.}\ \bibnamefont {Lao}}, \bibinfo {author} {\bibfnamefont
  {D.~A.}\ \bibnamefont {Monticello}}, \bibinfo {author} {\bibfnamefont
  {F.}~\bibnamefont {Perkins}}, \bibinfo {author} {\bibfnamefont
  {A.}~\bibnamefont {Pletzer}}, \bibinfo {author} {\bibfnamefont {A.~H.}\
  \bibnamefont {Reiman}}, \bibinfo {author} {\bibfnamefont {M.}~\bibnamefont
  {Rosenbluth}}, \bibinfo {author} {\bibfnamefont {E.~J.}\ \bibnamefont
  {Strait}}, \bibinfo {author} {\bibfnamefont {T.~S.}\ \bibnamefont {Taylor}},
  \bibinfo {author} {\bibfnamefont {A.~D.}\ \bibnamefont {Turnbull}}, \bibinfo
  {author} {\bibfnamefont {F.}~\bibnamefont {Waelbroeck}}, \bibinfo {author}
  {\bibfnamefont {J.~C.}\ \bibnamefont {Wesley}}, \bibinfo {author}
  {\bibfnamefont {H.~R.}\ \bibnamefont {Wilson}},\ and\ \bibinfo {author}
  {\bibfnamefont {R.}~\bibnamefont {Yoshino}},\ }\bibfield  {title} {\enquote
  {\bibinfo {title} {{Beta limits in long-pulse tokamak discharges}},}\ }\href
  {https://doi.org/10.1063/1.872270} {\bibfield  {journal} {\bibinfo  {journal}
  {Physics of Plasmas}\ }\textbf {\bibinfo {volume} {4}},\ \bibinfo {pages}
  {1654--1664} (\bibinfo {year} {1997})}\BibitemShut {NoStop}%
\bibitem [{\citenamefont {{Ishizawa}}, \citenamefont {{Kishimoto}},\ and\
  \citenamefont {{Nakamura}}(2019)}]{IshizawaPPCF19}%
  \BibitemOpen
  \bibfield  {author} {\bibinfo {author} {\bibfnamefont {A.}~\bibnamefont
  {{Ishizawa}}}, \bibinfo {author} {\bibfnamefont {Y.}~\bibnamefont
  {{Kishimoto}}},\ and\ \bibinfo {author} {\bibfnamefont {Y.}~\bibnamefont
  {{Nakamura}}},\ }\bibfield  {title} {\enquote {\bibinfo {title} {{Multi-scale
  interactions between turbulence and magnetic islands and parity
  mixture{\textemdash}a review}},}\ }\href
  {https://doi.org/10.1088/1361-6587/ab06a8} {\bibfield  {journal} {\bibinfo
  {journal} {Plasma Physics and Controlled Fusion}\ }\textbf {\bibinfo {volume}
  {61}},\ \bibinfo {eid} {054006} (\bibinfo {year} {2019})}\BibitemShut
  {NoStop}%
\bibitem [{Zoh(2014)}]{ZohmBook2022}%
  \BibitemOpen
  \enquote {\bibinfo {title} {Resistive mhd stability},}\ in\ \href
  {https://doi.org/https://doi.org/10.1002/9783527677375.ch8} {\emph {\bibinfo
  {booktitle} {Magnetohydrodynamic Stability of Tokamaks}}}\ (\bibinfo
  {publisher} {John Wiley \& Sons, Ltd},\ \bibinfo {year} {2014})\
  Chap.~\bibinfo {chapter} {8}, pp.\ \bibinfo {pages} {123--140},\ \Eprint
  {https://arxiv.org/abs/https://onlinelibrary.wiley.com/doi/pdf/10.1002/9783527677375.ch8}
  {https://onlinelibrary.wiley.com/doi/pdf/10.1002/9783527677375.ch8}
  \BibitemShut {NoStop}%
\bibitem [{\citenamefont {Carrera}, \citenamefont {Hazeltine},\ and\
  \citenamefont {Kotschenreuther}(1986)}]{carrera_PoF1986}%
  \BibitemOpen
  \bibfield  {author} {\bibinfo {author} {\bibfnamefont {R.}~\bibnamefont
  {Carrera}}, \bibinfo {author} {\bibfnamefont {R.~D.}\ \bibnamefont
  {Hazeltine}},\ and\ \bibinfo {author} {\bibfnamefont {M.}~\bibnamefont
  {Kotschenreuther}},\ }\bibfield  {title} {\enquote {\bibinfo {title} {{Island
  bootstrap current modification of the nonlinear dynamics of the tearing
  mode}},}\ }\href {https://doi.org/10.1063/1.865682} {\bibfield  {journal}
  {\bibinfo  {journal} {Physics of Fluids}\ }\textbf {\bibinfo {volume} {29}},\
  \bibinfo {pages} {899--902} (\bibinfo {year} {1986})}\BibitemShut {NoStop}%
\bibitem [{\citenamefont {Kotschenreuther}, \citenamefont {Hazeltine},\ and\
  \citenamefont {Morrison}(1985)}]{kotschenreuther_PF1985}%
  \BibitemOpen
  \bibfield  {author} {\bibinfo {author} {\bibfnamefont {M.}~\bibnamefont
  {Kotschenreuther}}, \bibinfo {author} {\bibfnamefont {R.~D.}\ \bibnamefont
  {Hazeltine}},\ and\ \bibinfo {author} {\bibfnamefont {P.~J.}\ \bibnamefont
  {Morrison}},\ }\bibfield  {title} {\enquote {\bibinfo {title} {{Nonlinear
  dynamics of magnetic islands with curvature and pressure}},}\ }\href
  {https://doi.org/10.1063/1.865200} {\bibfield  {journal} {\bibinfo  {journal}
  {Physics of Fluids}\ }\textbf {\bibinfo {volume} {28}},\ \bibinfo {pages}
  {294--302} (\bibinfo {year} {1985})}\BibitemShut {NoStop}%
\bibitem [{\citenamefont {Carreras}, \citenamefont {Hicks},\ and\ \citenamefont
  {Lee}(1981)}]{Carreras1981a}%
  \BibitemOpen
  \bibfield  {author} {\bibinfo {author} {\bibfnamefont {B.}~\bibnamefont
  {Carreras}}, \bibinfo {author} {\bibfnamefont {H.~R.}\ \bibnamefont
  {Hicks}},\ and\ \bibinfo {author} {\bibfnamefont {D.~K.}\ \bibnamefont
  {Lee}},\ }\bibfield  {title} {\enquote {\bibinfo {title} {{Effects of
  toroidal coupling on the stability of tearing modes}},}\ }\href
  {https://doi.org/10.1063/1.863247} {\bibfield  {journal} {\bibinfo  {journal}
  {Physics of Fluids}\ }\textbf {\bibinfo {volume} {24}},\ \bibinfo {pages}
  {66--77} (\bibinfo {year} {1981})}\BibitemShut {NoStop}%
\bibitem [{\citenamefont {{Widmer}}\ \emph {et~al.}(2019)\citenamefont
  {{Widmer}}, \citenamefont {{Maget}}, \citenamefont {{F{\'e}vrier}},
  \citenamefont {{L{\"u}tjens}},\ and\ \citenamefont
  {{Garbet}}}]{WidmerITER2018}%
  \BibitemOpen
  \bibfield  {author} {\bibinfo {author} {\bibfnamefont {F.}~\bibnamefont
  {{Widmer}}}, \bibinfo {author} {\bibfnamefont {P.}~\bibnamefont {{Maget}}},
  \bibinfo {author} {\bibfnamefont {O.}~\bibnamefont {{F{\'e}vrier}}}, \bibinfo
  {author} {\bibfnamefont {H.}~\bibnamefont {{L{\"u}tjens}}},\ and\ \bibinfo
  {author} {\bibfnamefont {X.}~\bibnamefont {{Garbet}}},\ }\bibfield  {title}
  {\enquote {\bibinfo {title} {{Non-linear simulations of neoclassical tearing
  mode control by externally driven RF current and heating, with application to
  ITER}},}\ }\href {https://doi.org/10.1088/1741-4326/ab300f} {\bibfield
  {journal} {\bibinfo  {journal} {Nuclear Fusion}\ }\textbf {\bibinfo {volume}
  {59}},\ \bibinfo {eid} {106012} (\bibinfo {year} {2019})}\BibitemShut
  {NoStop}%
\bibitem [{\citenamefont {Reimerdes}\ \emph {et~al.}(2002)\citenamefont
  {Reimerdes}, \citenamefont {Sauter}, \citenamefont {Goodman},\ and\
  \citenamefont {Pochelon}}]{reimerdes_prl2002}%
  \BibitemOpen
  \bibfield  {author} {\bibinfo {author} {\bibfnamefont {H.}~\bibnamefont
  {Reimerdes}}, \bibinfo {author} {\bibfnamefont {O.}~\bibnamefont {Sauter}},
  \bibinfo {author} {\bibfnamefont {T.}~\bibnamefont {Goodman}},\ and\ \bibinfo
  {author} {\bibfnamefont {A.}~\bibnamefont {Pochelon}},\ }\bibfield  {title}
  {\enquote {\bibinfo {title} {{From Current-Driven to Neoclassically Driven
  Tearing Modes}},}\ }\href {https://doi.org/10.1103/PhysRevLett.88.105005}
  {\bibfield  {journal} {\bibinfo  {journal} {Phys. Rev. Lett.}\ }\textbf
  {\bibinfo {volume} {88}},\ \bibinfo {pages} {105005} (\bibinfo {year}
  {2002})}\BibitemShut {NoStop}%
\bibitem [{\citenamefont {{Poli}}\ \emph {et~al.}(2016)\citenamefont {{Poli}},
  \citenamefont {{Bergmann}}, \citenamefont {{Casson}}, \citenamefont
  {{Hornsby}}, \citenamefont {{Peeters}}, \citenamefont {{Siccinio}},\ and\
  \citenamefont {{Zarzoso}}}]{Poli_PPR16}%
  \BibitemOpen
  \bibfield  {author} {\bibinfo {author} {\bibfnamefont {E.}~\bibnamefont
  {{Poli}}}, \bibinfo {author} {\bibfnamefont {A.}~\bibnamefont {{Bergmann}}},
  \bibinfo {author} {\bibfnamefont {F.~J.}\ \bibnamefont {{Casson}}}, \bibinfo
  {author} {\bibfnamefont {W.~A.}\ \bibnamefont {{Hornsby}}}, \bibinfo {author}
  {\bibfnamefont {A.~G.}\ \bibnamefont {{Peeters}}}, \bibinfo {author}
  {\bibfnamefont {M.}~\bibnamefont {{Siccinio}}},\ and\ \bibinfo {author}
  {\bibfnamefont {D.}~\bibnamefont {{Zarzoso}}},\ }\bibfield  {title} {\enquote
  {\bibinfo {title} {{Kinetic effects on the currents determining the stability
  of a magnetic island in tokamaks}},}\ }\href
  {https://doi.org/10.1134/S1063780X16050135} {\bibfield  {journal} {\bibinfo
  {journal} {Plasma Physics Reports}\ }\textbf {\bibinfo {volume} {42}},\
  \bibinfo {pages} {450--464} (\bibinfo {year} {2016})}\BibitemShut {NoStop}%
\bibitem [{\citenamefont {{Poli}}, \citenamefont {{Bottino}},\ and\
  \citenamefont {{Peeters}}(2009)}]{PoliNF09}%
  \BibitemOpen
  \bibfield  {author} {\bibinfo {author} {\bibfnamefont {E.}~\bibnamefont
  {{Poli}}}, \bibinfo {author} {\bibfnamefont {A.}~\bibnamefont {{Bottino}}},\
  and\ \bibinfo {author} {\bibfnamefont {A.~G.}\ \bibnamefont {{Peeters}}},\
  }\bibfield  {title} {\enquote {\bibinfo {title} {{Behaviour of turbulent
  transport in the vicinity of a magnetic island}},}\ }\href
  {https://doi.org/10.1088/0029-5515/49/7/075010} {\bibfield  {journal}
  {\bibinfo  {journal} {Nuclear Fusion}\ }\textbf {\bibinfo {volume} {49}},\
  \bibinfo {eid} {075010} (\bibinfo {year} {2009})}\BibitemShut {NoStop}%
\bibitem [{\citenamefont {Poli}\ \emph {et~al.}(2010)\citenamefont {Poli},
  \citenamefont {Bottino}, \citenamefont {Hornsby}, \citenamefont {Peeters},
  \citenamefont {Ribeiro}, \citenamefont {Scott},\ and\ \citenamefont
  {Siccinio}}]{PoliPPCF10a}%
  \BibitemOpen
  \bibfield  {author} {\bibinfo {author} {\bibfnamefont {E.}~\bibnamefont
  {Poli}}, \bibinfo {author} {\bibfnamefont {A.}~\bibnamefont {Bottino}},
  \bibinfo {author} {\bibfnamefont {W.~A.}\ \bibnamefont {Hornsby}}, \bibinfo
  {author} {\bibfnamefont {A.~G.}\ \bibnamefont {Peeters}}, \bibinfo {author}
  {\bibfnamefont {T.}~\bibnamefont {Ribeiro}}, \bibinfo {author} {\bibfnamefont
  {B.~D.}\ \bibnamefont {Scott}},\ and\ \bibinfo {author} {\bibfnamefont
  {M.}~\bibnamefont {Siccinio}},\ }\bibfield  {title} {\enquote {\bibinfo
  {title} {{Gyrokinetic and gyrofluid investigation of magnetic islands in
  tokamaks}},}\ }\href {https://doi.org/10.1088/0741-3335/52/12/124021}
  {\bibfield  {journal} {\bibinfo  {journal} {Plasma Physics and Controlled
  Fusion}\ }\textbf {\bibinfo {volume} {52}},\ \bibinfo {pages} {124021}
  (\bibinfo {year} {2010})}\BibitemShut {NoStop}%
\bibitem [{\citenamefont {{Hornsby}}\ \emph
  {et~al.}(2010{\natexlab{a}})\citenamefont {{Hornsby}}, \citenamefont
  {{Peeters}}, \citenamefont {{Poli}}, \citenamefont {{Siccinio}},
  \citenamefont {{Snodin}}, \citenamefont {{Casson}}, \citenamefont
  {{Camenen}},\ and\ \citenamefont {{Szepesi}}}]{HornsbyEPL10}%
  \BibitemOpen
  \bibfield  {author} {\bibinfo {author} {\bibfnamefont {W.~A.}\ \bibnamefont
  {{Hornsby}}}, \bibinfo {author} {\bibfnamefont {A.~G.}\ \bibnamefont
  {{Peeters}}}, \bibinfo {author} {\bibfnamefont {E.}~\bibnamefont {{Poli}}},
  \bibinfo {author} {\bibfnamefont {M.}~\bibnamefont {{Siccinio}}}, \bibinfo
  {author} {\bibfnamefont {A.~P.}\ \bibnamefont {{Snodin}}}, \bibinfo {author}
  {\bibfnamefont {F.~J.}\ \bibnamefont {{Casson}}}, \bibinfo {author}
  {\bibfnamefont {Y.}~\bibnamefont {{Camenen}}},\ and\ \bibinfo {author}
  {\bibfnamefont {G.}~\bibnamefont {{Szepesi}}},\ }\bibfield  {title} {\enquote
  {\bibinfo {title} {{On the nonlinear coupling between micro turbulence and
  mesoscale magnetic islands in a plasma}},}\ }\href
  {https://doi.org/10.1209/0295-5075/91/45001} {\bibfield  {journal} {\bibinfo
  {journal} {EPL (Europhysics Letters)}\ }\textbf {\bibinfo {volume} {91}},\
  \bibinfo {pages} {45001} (\bibinfo {year} {2010}{\natexlab{a}})}\BibitemShut
  {NoStop}%
\bibitem [{\citenamefont {{Hornsby}}\ \emph
  {et~al.}(2010{\natexlab{b}})\citenamefont {{Hornsby}}, \citenamefont
  {{Peeters}}, \citenamefont {{Snodin}}, \citenamefont {{Casson}},
  \citenamefont {{Camenen}}, \citenamefont {{Szepesi}}, \citenamefont
  {{Siccinio}},\ and\ \citenamefont {{Poli}}}]{HornsbyPoP2010}%
  \BibitemOpen
  \bibfield  {author} {\bibinfo {author} {\bibfnamefont {W.~A.}\ \bibnamefont
  {{Hornsby}}}, \bibinfo {author} {\bibfnamefont {A.~G.}\ \bibnamefont
  {{Peeters}}}, \bibinfo {author} {\bibfnamefont {A.~P.}\ \bibnamefont
  {{Snodin}}}, \bibinfo {author} {\bibfnamefont {F.~J.}\ \bibnamefont
  {{Casson}}}, \bibinfo {author} {\bibfnamefont {Y.}~\bibnamefont {{Camenen}}},
  \bibinfo {author} {\bibfnamefont {G.}~\bibnamefont {{Szepesi}}}, \bibinfo
  {author} {\bibfnamefont {M.}~\bibnamefont {{Siccinio}}},\ and\ \bibinfo
  {author} {\bibfnamefont {E.}~\bibnamefont {{Poli}}},\ }\bibfield  {title}
  {\enquote {\bibinfo {title} {{The nonlinear coupling between gyroradius scale
  turbulence and mesoscale magnetic islands in fusion plasmas}},}\ }\href
  {https://doi.org/10.1063/1.3467502} {\bibfield  {journal} {\bibinfo
  {journal} {Physics of Plasmas}\ }\textbf {\bibinfo {volume} {17}},\ \bibinfo
  {eid} {092301} (\bibinfo {year} {2010}{\natexlab{b}})}\BibitemShut {NoStop}%
\bibitem [{\citenamefont {{Waltz}}\ and\ \citenamefont
  {{Waelbroeck}}(2012)}]{WaltzPoP12}%
  \BibitemOpen
  \bibfield  {author} {\bibinfo {author} {\bibfnamefont {R.~E.}\ \bibnamefont
  {{Waltz}}}\ and\ \bibinfo {author} {\bibfnamefont {F.~L.}\ \bibnamefont
  {{Waelbroeck}}},\ }\bibfield  {title} {\enquote {\bibinfo {title}
  {{Gyrokinetic simulations with external resonant magnetic perturbations:
  Island torque and nonambipolar transport with plasma rotation}},}\ }\href
  {https://doi.org/10.1063/1.3692222} {\bibfield  {journal} {\bibinfo
  {journal} {Physics of Plasmas}\ }\textbf {\bibinfo {volume} {19}},\ \bibinfo
  {eid} {032508} (\bibinfo {year} {2012})}\BibitemShut {NoStop}%
\bibitem [{\citenamefont {{Zarzoso}}\ \emph
  {et~al.}(2015{\natexlab{a}})\citenamefont {{Zarzoso}}, \citenamefont
  {{Casson}}, \citenamefont {{Hornsby}}, \citenamefont {{Poli}},\ and\
  \citenamefont {{Peeters}}}]{ZarzosoPoP15}%
  \BibitemOpen
  \bibfield  {author} {\bibinfo {author} {\bibfnamefont {D.}~\bibnamefont
  {{Zarzoso}}}, \bibinfo {author} {\bibfnamefont {F.~J.}\ \bibnamefont
  {{Casson}}}, \bibinfo {author} {\bibfnamefont {W.~A.}\ \bibnamefont
  {{Hornsby}}}, \bibinfo {author} {\bibfnamefont {E.}~\bibnamefont {{Poli}}},\
  and\ \bibinfo {author} {\bibfnamefont {A.~G.}\ \bibnamefont {{Peeters}}},\
  }\bibfield  {title} {\enquote {\bibinfo {title} {{Verification of a magnetic
  island in gyro-kinetics by comparison with analytic theory}},}\ }\href
  {https://doi.org/10.1063/1.4908549} {\bibfield  {journal} {\bibinfo
  {journal} {Physics of Plasmas}\ }\textbf {\bibinfo {volume} {22}},\ \bibinfo
  {eid} {022127} (\bibinfo {year} {2015}{\natexlab{a}})}\BibitemShut {NoStop}%
\bibitem [{\citenamefont {{Ba{\~n}{\'o}n Navarro}}\ \emph
  {et~al.}(2017)\citenamefont {{Ba{\~n}{\'o}n Navarro}}, \citenamefont
  {{Bard{\'o}czi}}, \citenamefont {{Carter}}, \citenamefont {{Jenko}},\ and\
  \citenamefont {{Rhodes}}}]{BanonNavarroPPCF2017}%
  \BibitemOpen
  \bibfield  {author} {\bibinfo {author} {\bibfnamefont {A.}~\bibnamefont
  {{Ba{\~n}{\'o}n Navarro}}}, \bibinfo {author} {\bibfnamefont
  {L.}~\bibnamefont {{Bard{\'o}czi}}}, \bibinfo {author} {\bibfnamefont
  {T.~A.}\ \bibnamefont {{Carter}}}, \bibinfo {author} {\bibfnamefont
  {F.}~\bibnamefont {{Jenko}}},\ and\ \bibinfo {author} {\bibfnamefont {T.~L.}\
  \bibnamefont {{Rhodes}}},\ }\bibfield  {title} {\enquote {\bibinfo {title}
  {{Effect of magnetic islands on profiles, flows, turbulence and transport in
  nonlinear gyrokinetic simulations}},}\ }\href
  {https://doi.org/10.1088/1361-6587/aa557e} {\bibfield  {journal} {\bibinfo
  {journal} {Plasma Physics and Controlled Fusion}\ }\textbf {\bibinfo {volume}
  {59}},\ \bibinfo {eid} {034004} (\bibinfo {year} {2017})}\BibitemShut
  {NoStop}%
\bibitem [{\citenamefont {{Kwon}}\ \emph {et~al.}(2018)\citenamefont {{Kwon}},
  \citenamefont {{Ku}}, \citenamefont {{Choi}}, \citenamefont {{Chang}},
  \citenamefont {{Hager}}, \citenamefont {{Yoon}}, \citenamefont {{Lee}},\ and\
  \citenamefont {{Kim}}}]{KwonPoP18}%
  \BibitemOpen
  \bibfield  {author} {\bibinfo {author} {\bibfnamefont {J.-M.}\ \bibnamefont
  {{Kwon}}}, \bibinfo {author} {\bibfnamefont {S.}~\bibnamefont {{Ku}}},
  \bibinfo {author} {\bibfnamefont {M.~J.}\ \bibnamefont {{Choi}}}, \bibinfo
  {author} {\bibfnamefont {C.~S.}\ \bibnamefont {{Chang}}}, \bibinfo {author}
  {\bibfnamefont {R.}~\bibnamefont {{Hager}}}, \bibinfo {author} {\bibfnamefont
  {E.~S.}\ \bibnamefont {{Yoon}}}, \bibinfo {author} {\bibfnamefont {H.~H.}\
  \bibnamefont {{Lee}}},\ and\ \bibinfo {author} {\bibfnamefont {H.~S.}\
  \bibnamefont {{Kim}}},\ }\bibfield  {title} {\enquote {\bibinfo {title}
  {{Gyrokinetic simulation study of magnetic island effects on neoclassical
  physics and micro-instabilities in a realistic KSTAR plasma}},}\ }\href
  {https://doi.org/10.1063/1.5027622} {\bibfield  {journal} {\bibinfo
  {journal} {Physics of Plasmas}\ }\textbf {\bibinfo {volume} {25}},\ \bibinfo
  {eid} {052506} (\bibinfo {year} {2018})}\BibitemShut {NoStop}%
\bibitem [{\citenamefont {{Fang}}\ and\ \citenamefont
  {{Lin}}(2019)}]{FangPoP19}%
  \BibitemOpen
  \bibfield  {author} {\bibinfo {author} {\bibfnamefont {K.~S.}\ \bibnamefont
  {{Fang}}}\ and\ \bibinfo {author} {\bibfnamefont {Z.}~\bibnamefont {{Lin}}},\
  }\bibfield  {title} {\enquote {\bibinfo {title} {{Global gyrokinetic
  simulation of microturbulence with kinetic electrons in the presence of
  magnetic island in tokamak}},}\ }\href {https://doi.org/10.1063/1.5096962}
  {\bibfield  {journal} {\bibinfo  {journal} {Physics of Plasmas}\ }\textbf
  {\bibinfo {volume} {26}},\ \bibinfo {eid} {052510} (\bibinfo {year}
  {2019})}\BibitemShut {NoStop}%
\bibitem [{\citenamefont {{Muto}}, \citenamefont {{Imadera}},\ and\
  \citenamefont {{Kishimoto}}(2022)}]{MutoPoP22}%
  \BibitemOpen
  \bibfield  {author} {\bibinfo {author} {\bibfnamefont {M.}~\bibnamefont
  {{Muto}}}, \bibinfo {author} {\bibfnamefont {K.}~\bibnamefont {{Imadera}}},\
  and\ \bibinfo {author} {\bibfnamefont {Y.}~\bibnamefont {{Kishimoto}}},\
  }\bibfield  {title} {\enquote {\bibinfo {title} {{Effects of magnetic island
  on profile formation in flux-driven ITG turbulence}},}\ }\href
  {https://doi.org/10.1063/5.0081125} {\bibfield  {journal} {\bibinfo
  {journal} {Physics of Plasmas}\ }\textbf {\bibinfo {volume} {29}},\ \bibinfo
  {eid} {052503} (\bibinfo {year} {2022})}\BibitemShut {NoStop}%
\bibitem [{\citenamefont {{Li}}\ \emph {et~al.}(2023)\citenamefont {{Li}},
  \citenamefont {{Xu}}, \citenamefont {{Qu}}, \citenamefont {{Lin}},
  \citenamefont {{Dong}}, \citenamefont {{Peng}},\ and\ \citenamefont
  {{Li}}}]{LiNF23}%
  \BibitemOpen
  \bibfield  {author} {\bibinfo {author} {\bibfnamefont {J.}~\bibnamefont
  {{Li}}}, \bibinfo {author} {\bibfnamefont {J.~Q.}\ \bibnamefont {{Xu}}},
  \bibinfo {author} {\bibfnamefont {Y.~R.}\ \bibnamefont {{Qu}}}, \bibinfo
  {author} {\bibfnamefont {Z.}~\bibnamefont {{Lin}}}, \bibinfo {author}
  {\bibfnamefont {J.~Q.}\ \bibnamefont {{Dong}}}, \bibinfo {author}
  {\bibfnamefont {X.~D.}\ \bibnamefont {{Peng}}},\ and\ \bibinfo {author}
  {\bibfnamefont {J.~Q.}\ \bibnamefont {{Li}}},\ }\bibfield  {title} {\enquote
  {\bibinfo {title} {{Global gyrokinetic simulations of the impact of magnetic
  island on ion temperature gradient driven turbulence}},}\ }\href
  {https://doi.org/10.1088/1741-4326/ace461} {\bibfield  {journal} {\bibinfo
  {journal} {Nuclear Fusion}\ }\textbf {\bibinfo {volume} {63}},\ \bibinfo
  {eid} {096005} (\bibinfo {year} {2023})},\ \Eprint
  {https://arxiv.org/abs/2306.05607} {arXiv:2306.05607 [physics.plasm-ph]}
  \BibitemShut {NoStop}%
\bibitem [{\citenamefont {{Hornsby}}\ \emph
  {et~al.}(2015{\natexlab{a}})\citenamefont {{Hornsby}}, \citenamefont
  {{Migliano}}, \citenamefont {{Buchholz}}, \citenamefont {{Zarzoso}},
  \citenamefont {{Casson}}, \citenamefont {{Poli}},\ and\ \citenamefont
  {{Peeters}}}]{HornsbyPPCF15}%
  \BibitemOpen
  \bibfield  {author} {\bibinfo {author} {\bibfnamefont {W.~A.}\ \bibnamefont
  {{Hornsby}}}, \bibinfo {author} {\bibfnamefont {P.}~\bibnamefont
  {{Migliano}}}, \bibinfo {author} {\bibfnamefont {R.}~\bibnamefont
  {{Buchholz}}}, \bibinfo {author} {\bibfnamefont {D.}~\bibnamefont
  {{Zarzoso}}}, \bibinfo {author} {\bibfnamefont {F.~J.}\ \bibnamefont
  {{Casson}}}, \bibinfo {author} {\bibfnamefont {E.}~\bibnamefont {{Poli}}},\
  and\ \bibinfo {author} {\bibfnamefont {A.~G.}\ \bibnamefont {{Peeters}}},\
  }\bibfield  {title} {\enquote {\bibinfo {title} {{On seed island generation
  and the non-linear self-consistent interaction of the tearing mode With
  electromagnetic gyro-kinetic turbulence}},}\ }\href
  {https://doi.org/10.1088/0741-3335/57/5/054018} {\bibfield  {journal}
  {\bibinfo  {journal} {Plasma Physics and Controlled Fusion}\ }\textbf
  {\bibinfo {volume} {57}},\ \bibinfo {eid} {054018} (\bibinfo {year}
  {2015}{\natexlab{a}})},\ \Eprint {https://arxiv.org/abs/1409.0648}
  {arXiv:1409.0648 [physics.plasm-ph]} \BibitemShut {NoStop}%
\bibitem [{\citenamefont {{Hornsby}}\ \emph {et~al.}(2016)\citenamefont
  {{Hornsby}}, \citenamefont {{Migliano}}, \citenamefont {{Buchholz}},
  \citenamefont {{Grosshauser}}, \citenamefont {{Weikl}}, \citenamefont
  {{Zarzoso}}, \citenamefont {{Casson}}, \citenamefont {{Poli}},\ and\
  \citenamefont {{Peeters}}}]{HornsbyPPCF16}%
  \BibitemOpen
  \bibfield  {author} {\bibinfo {author} {\bibfnamefont {W.~A.}\ \bibnamefont
  {{Hornsby}}}, \bibinfo {author} {\bibfnamefont {P.}~\bibnamefont
  {{Migliano}}}, \bibinfo {author} {\bibfnamefont {R.}~\bibnamefont
  {{Buchholz}}}, \bibinfo {author} {\bibfnamefont {S.}~\bibnamefont
  {{Grosshauser}}}, \bibinfo {author} {\bibfnamefont {A.}~\bibnamefont
  {{Weikl}}}, \bibinfo {author} {\bibfnamefont {D.}~\bibnamefont {{Zarzoso}}},
  \bibinfo {author} {\bibfnamefont {F.~J.}\ \bibnamefont {{Casson}}}, \bibinfo
  {author} {\bibfnamefont {E.}~\bibnamefont {{Poli}}},\ and\ \bibinfo {author}
  {\bibfnamefont {A.~G.}\ \bibnamefont {{Peeters}}},\ }\bibfield  {title}
  {\enquote {\bibinfo {title} {{The non-linear evolution of the tearing mode in
  electromagnetic turbulence using gyrokinetic simulations}},}\ }\href
  {https://doi.org/10.1088/0741-3335/58/1/014028} {\bibfield  {journal}
  {\bibinfo  {journal} {Plasma Physics and Controlled Fusion}\ }\textbf
  {\bibinfo {volume} {58}},\ \bibinfo {eid} {014028} (\bibinfo {year}
  {2016})},\ \Eprint {https://arxiv.org/abs/1507.02841} {arXiv:1507.02841
  [physics.plasm-ph]} \BibitemShut {NoStop}%
\bibitem [{\citenamefont {{Zarzoso}}\ \emph {et~al.}(2019)\citenamefont
  {{Zarzoso}}, \citenamefont {{Nasr}}, \citenamefont {{Garbet}}, \citenamefont
  {{Smolyakov}},\ and\ \citenamefont {{Benkadda}}}]{ZarzosoPoP19}%
  \BibitemOpen
  \bibfield  {author} {\bibinfo {author} {\bibfnamefont {D.}~\bibnamefont
  {{Zarzoso}}}, \bibinfo {author} {\bibfnamefont {S.}~\bibnamefont {{Nasr}}},
  \bibinfo {author} {\bibfnamefont {X.}~\bibnamefont {{Garbet}}}, \bibinfo
  {author} {\bibfnamefont {A.~I.}\ \bibnamefont {{Smolyakov}}},\ and\ \bibinfo
  {author} {\bibfnamefont {S.}~\bibnamefont {{Benkadda}}},\ }\bibfield  {title}
  {\enquote {\bibinfo {title} {{Gyro-kinetic theory and global simulations of
  the collisionless tearing instability: The impact of trapped particles
  through the magnetic field curvature}},}\ }\href
  {https://doi.org/10.1063/1.5109947} {\bibfield  {journal} {\bibinfo
  {journal} {Physics of Plasmas}\ }\textbf {\bibinfo {volume} {26}},\ \bibinfo
  {eid} {112112} (\bibinfo {year} {2019})}\BibitemShut {NoStop}%
\bibitem [{\citenamefont {{Jitsuk}}\ \emph {et~al.}(2024)\citenamefont
  {{Jitsuk}}, \citenamefont {{Di Siena}}, \citenamefont {{Pueschel}},
  \citenamefont {{Terry}}, \citenamefont {{Widmer}}, \citenamefont {{Poli}},\
  and\ \citenamefont {{Sarff}}}]{JitsukNF24}%
  \BibitemOpen
  \bibfield  {author} {\bibinfo {author} {\bibfnamefont {T.}~\bibnamefont
  {{Jitsuk}}}, \bibinfo {author} {\bibfnamefont {A.}~\bibnamefont {{Di
  Siena}}}, \bibinfo {author} {\bibfnamefont {M.~J.}\ \bibnamefont
  {{Pueschel}}}, \bibinfo {author} {\bibfnamefont {P.~W.}\ \bibnamefont
  {{Terry}}}, \bibinfo {author} {\bibfnamefont {F.}~\bibnamefont {{Widmer}}},
  \bibinfo {author} {\bibfnamefont {E.}~\bibnamefont {{Poli}}},\ and\ \bibinfo
  {author} {\bibfnamefont {J.~S.}\ \bibnamefont {{Sarff}}},\ }\bibfield
  {title} {\enquote {\bibinfo {title} {{Global linear and nonlinear gyrokinetic
  simulations of tearing modes}},}\ }\href
  {https://doi.org/10.1088/1741-4326/ad279b} {\bibfield  {journal} {\bibinfo
  {journal} {Nuclear Fusion}\ }\textbf {\bibinfo {volume} {64}},\ \bibinfo
  {eid} {046005} (\bibinfo {year} {2024})},\ \Eprint
  {https://arxiv.org/abs/2308.16345} {arXiv:2308.16345 [physics.plasm-ph]}
  \BibitemShut {NoStop}%
\bibitem [{\citenamefont {{Wan}}, \citenamefont {{Chen}},\ and\ \citenamefont
  {{Parker}}(2005)}]{WanPoP05}%
  \BibitemOpen
  \bibfield  {author} {\bibinfo {author} {\bibfnamefont {W.}~\bibnamefont
  {{Wan}}}, \bibinfo {author} {\bibfnamefont {Y.}~\bibnamefont {{Chen}}},\ and\
  \bibinfo {author} {\bibfnamefont {S.~E.}\ \bibnamefont {{Parker}}},\
  }\bibfield  {title} {\enquote {\bibinfo {title} {{Gyrokinetic
  {\ensuremath{\delta}}f simulation of the collisionless and semicollisional
  tearing mode instability}},}\ }\href {https://doi.org/10.1063/1.1827216}
  {\bibfield  {journal} {\bibinfo  {journal} {Physics of Plasmas}\ }\textbf
  {\bibinfo {volume} {12}},\ \bibinfo {eid} {012311} (\bibinfo {year}
  {2005})}\BibitemShut {NoStop}%
\bibitem [{\citenamefont {Rogers}\ \emph {et~al.}(2007)\citenamefont {Rogers},
  \citenamefont {Kobayashi}, \citenamefont {Ricci}, \citenamefont {Dorland},
  \citenamefont {Drake},\ and\ \citenamefont {Tatsuno}}]{RogersPoP07}%
  \BibitemOpen
  \bibfield  {author} {\bibinfo {author} {\bibfnamefont {B.~N.}\ \bibnamefont
  {Rogers}}, \bibinfo {author} {\bibfnamefont {S.}~\bibnamefont {Kobayashi}},
  \bibinfo {author} {\bibfnamefont {P.}~\bibnamefont {Ricci}}, \bibinfo
  {author} {\bibfnamefont {W.}~\bibnamefont {Dorland}}, \bibinfo {author}
  {\bibfnamefont {J.}~\bibnamefont {Drake}},\ and\ \bibinfo {author}
  {\bibfnamefont {T.}~\bibnamefont {Tatsuno}},\ }\bibfield  {title} {\enquote
  {\bibinfo {title} {{Gyrokinetic simulations of collisionless magnetic
  reconnection}},}\ }\href {https://doi.org/10.1063/1.2774003} {\bibfield
  {journal} {\bibinfo  {journal} {Physics of Plasmas}\ }\textbf {\bibinfo
  {volume} {14}} (\bibinfo {year} {2007}),\ 10.1063/1.2774003}\BibitemShut
  {NoStop}%
\bibitem [{\citenamefont {{Pueschel}}\ \emph {et~al.}(2011)\citenamefont
  {{Pueschel}}, \citenamefont {{Jenko}}, \citenamefont {{Told}},\ and\
  \citenamefont {{B{\"u}chner}}}]{PueschelPoP11}%
  \BibitemOpen
  \bibfield  {author} {\bibinfo {author} {\bibfnamefont {M.~J.}\ \bibnamefont
  {{Pueschel}}}, \bibinfo {author} {\bibfnamefont {F.}~\bibnamefont {{Jenko}}},
  \bibinfo {author} {\bibfnamefont {D.}~\bibnamefont {{Told}}},\ and\ \bibinfo
  {author} {\bibfnamefont {J.}~\bibnamefont {{B{\"u}chner}}},\ }\bibfield
  {title} {\enquote {\bibinfo {title} {{Gyrokinetic simulations of magnetic
  reconnection}},}\ }\href {https://doi.org/10.1063/1.3656965} {\bibfield
  {journal} {\bibinfo  {journal} {Physics of Plasmas}\ }\textbf {\bibinfo
  {volume} {18}},\ \bibinfo {eid} {112102} (\bibinfo {year}
  {2011})}\BibitemShut {NoStop}%
\bibitem [{\citenamefont {{Zacharias}}, \citenamefont {{Kleiber}},\ and\
  \citenamefont {{Hatzky}}(2012)}]{Zacharias12}%
  \BibitemOpen
  \bibfield  {author} {\bibinfo {author} {\bibfnamefont {O.}~\bibnamefont
  {{Zacharias}}}, \bibinfo {author} {\bibfnamefont {R.}~\bibnamefont
  {{Kleiber}}},\ and\ \bibinfo {author} {\bibfnamefont {R.}~\bibnamefont
  {{Hatzky}}},\ }\bibfield  {title} {\enquote {\bibinfo {title} {{Gyrokinetic
  simulations of collisionless tearing modes}},}\ }in\ \href
  {https://doi.org/10.1088/1742-6596/401/1/012026} {\emph {\bibinfo {booktitle}
  {Journal of Physics Conference Series}}},\ \bibinfo {series} {Journal of
  Physics Conference Series}, Vol.\ \bibinfo {volume} {401}\ (\bibinfo {year}
  {2012})\ p.\ \bibinfo {pages} {012026}\BibitemShut {NoStop}%
\bibitem [{\citenamefont {{Yao}}\ \emph {et~al.}(2021)\citenamefont {{Yao}},
  \citenamefont {{Lin}}, \citenamefont {{Dong}}, \citenamefont {{Shi}},
  \citenamefont {{Liu}},\ and\ \citenamefont {{Li}}}]{YaoPLA21}%
  \BibitemOpen
  \bibfield  {author} {\bibinfo {author} {\bibfnamefont {Y.}~\bibnamefont
  {{Yao}}}, \bibinfo {author} {\bibfnamefont {Z.}~\bibnamefont {{Lin}}},
  \bibinfo {author} {\bibfnamefont {J.~Q.}\ \bibnamefont {{Dong}}}, \bibinfo
  {author} {\bibfnamefont {P.}~\bibnamefont {{Shi}}}, \bibinfo {author}
  {\bibfnamefont {S.~F.}\ \bibnamefont {{Liu}}},\ and\ \bibinfo {author}
  {\bibfnamefont {J.}~\bibnamefont {{Li}}},\ }\bibfield  {title} {\enquote
  {\bibinfo {title} {{Gyrokinetic simulations of double tearing modes in
  toroidal plasma}},}\ }\href {https://doi.org/10.1016/j.physleta.2021.127681}
  {\bibfield  {journal} {\bibinfo  {journal} {Physics Letters A}\ }\textbf
  {\bibinfo {volume} {417}},\ \bibinfo {eid} {127681} (\bibinfo {year}
  {2021})}\BibitemShut {NoStop}%
\bibitem [{\citenamefont {{Mishchenko}}\ \emph {et~al.}(2022)\citenamefont
  {{Mishchenko}}, \citenamefont {{Bottino}}, \citenamefont
  {{Hayward-Schneider}}, \citenamefont {{Poli}}, \citenamefont {{Wang}},
  \citenamefont {{Kleiber}}, \citenamefont {{Borchardt}}, \citenamefont
  {{N{\"u}hrenberg}}, \citenamefont {{Biancalani}}, \citenamefont
  {{K{\"o}nies}}, \citenamefont {{Lanti}}, \citenamefont {{Lauber}},
  \citenamefont {{Hatzky}}, \citenamefont {{Vannini}}, \citenamefont
  {{Villard}},\ and\ \citenamefont {{Widmer}}}]{MishchenkoPPCF22}%
  \BibitemOpen
  \bibfield  {author} {\bibinfo {author} {\bibfnamefont {A.}~\bibnamefont
  {{Mishchenko}}}, \bibinfo {author} {\bibfnamefont {A.}~\bibnamefont
  {{Bottino}}}, \bibinfo {author} {\bibfnamefont {T.}~\bibnamefont
  {{Hayward-Schneider}}}, \bibinfo {author} {\bibfnamefont {E.}~\bibnamefont
  {{Poli}}}, \bibinfo {author} {\bibfnamefont {X.}~\bibnamefont {{Wang}}},
  \bibinfo {author} {\bibfnamefont {R.}~\bibnamefont {{Kleiber}}}, \bibinfo
  {author} {\bibfnamefont {M.}~\bibnamefont {{Borchardt}}}, \bibinfo {author}
  {\bibfnamefont {C.}~\bibnamefont {{N{\"u}hrenberg}}}, \bibinfo {author}
  {\bibfnamefont {A.}~\bibnamefont {{Biancalani}}}, \bibinfo {author}
  {\bibfnamefont {A.}~\bibnamefont {{K{\"o}nies}}}, \bibinfo {author}
  {\bibfnamefont {E.}~\bibnamefont {{Lanti}}}, \bibinfo {author} {\bibfnamefont
  {P.}~\bibnamefont {{Lauber}}}, \bibinfo {author} {\bibfnamefont
  {R.}~\bibnamefont {{Hatzky}}}, \bibinfo {author} {\bibfnamefont
  {F.}~\bibnamefont {{Vannini}}}, \bibinfo {author} {\bibfnamefont
  {L.}~\bibnamefont {{Villard}}},\ and\ \bibinfo {author} {\bibfnamefont
  {F.}~\bibnamefont {{Widmer}}},\ }\bibfield  {title} {\enquote {\bibinfo
  {title} {{Gyrokinetic particle-in-cell simulations of electromagnetic
  turbulence in the presence of fast particles and global modes}},}\ }\href
  {https://doi.org/10.1088/1361-6587/ac8dbc} {\bibfield  {journal} {\bibinfo
  {journal} {Plasma Physics and Controlled Fusion}\ }\textbf {\bibinfo {volume}
  {64}},\ \bibinfo {eid} {104009} (\bibinfo {year} {2022})},\ \Eprint
  {https://arxiv.org/abs/2203.11983} {arXiv:2203.11983 [physics.plasm-ph]}
  \BibitemShut {NoStop}%
\bibitem [{\citenamefont {Jolliet}\ \emph {et~al.}(2007)\citenamefont
  {Jolliet}, \citenamefont {Bottino}, \citenamefont {Angelino}, \citenamefont
  {Hatzky}, \citenamefont {Tran}, \citenamefont {Mcmillan}, \citenamefont
  {Sauter}, \citenamefont {Appert}, \citenamefont {Idomura},\ and\
  \citenamefont {Villard}}]{JollietCPC2007}%
  \BibitemOpen
  \bibfield  {author} {\bibinfo {author} {\bibfnamefont {S.}~\bibnamefont
  {Jolliet}}, \bibinfo {author} {\bibfnamefont {A.}~\bibnamefont {Bottino}},
  \bibinfo {author} {\bibfnamefont {P.}~\bibnamefont {Angelino}}, \bibinfo
  {author} {\bibfnamefont {R.}~\bibnamefont {Hatzky}}, \bibinfo {author}
  {\bibfnamefont {T.}~\bibnamefont {Tran}}, \bibinfo {author} {\bibfnamefont
  {B.}~\bibnamefont {Mcmillan}}, \bibinfo {author} {\bibfnamefont
  {O.}~\bibnamefont {Sauter}}, \bibinfo {author} {\bibfnamefont
  {K.}~\bibnamefont {Appert}}, \bibinfo {author} {\bibfnamefont
  {Y.}~\bibnamefont {Idomura}},\ and\ \bibinfo {author} {\bibfnamefont
  {L.}~\bibnamefont {Villard}},\ }\bibfield  {title} {\enquote {\bibinfo
  {title} {{A global collisionless PIC code in magnetic coordinates}},}\ }\href
  {https://doi.org/10.1016/j.cpc.2007.04.006} {\bibfield  {journal} {\bibinfo
  {journal} {Computer Physics Communications}\ }\textbf {\bibinfo {volume}
  {177}},\ \bibinfo {pages} {409--425} (\bibinfo {year} {2007})}\BibitemShut
  {NoStop}%
\bibitem [{\citenamefont {Lanti}\ \emph {et~al.}(2018)\citenamefont {Lanti},
  \citenamefont {McMillan}, \citenamefont {Brunner}, \citenamefont {Ohana},\
  and\ \citenamefont {Villard}}]{LantiJPCS2018}%
  \BibitemOpen
  \bibfield  {author} {\bibinfo {author} {\bibfnamefont {E.}~\bibnamefont
  {Lanti}}, \bibinfo {author} {\bibfnamefont {B.~F.}\ \bibnamefont {McMillan}},
  \bibinfo {author} {\bibfnamefont {S.}~\bibnamefont {Brunner}}, \bibinfo
  {author} {\bibfnamefont {N.}~\bibnamefont {Ohana}},\ and\ \bibinfo {author}
  {\bibfnamefont {L.}~\bibnamefont {Villard}},\ }\bibfield  {title} {\enquote
  {\bibinfo {title} {{Gradient- and flux-driven global gyrokinetic simulations
  of ITG and TEM turbulence with an improved hybrid kinetic electron model}},}\
  }\href {https://doi.org/10.1088/1742-6596/1125/1/012014} {\bibfield
  {journal} {\bibinfo  {journal} {Journal of Physics: Conference Series}\
  }\textbf {\bibinfo {volume} {1125}},\ \bibinfo {pages} {012014} (\bibinfo
  {year} {2018})}\BibitemShut {NoStop}%
\bibitem [{\citenamefont {Wilson}\ \emph {et~al.}(1996)\citenamefont {Wilson},
  \citenamefont {Connor}, \citenamefont {Hastie},\ and\ \citenamefont
  {Hegna}}]{Wilson1996a}%
  \BibitemOpen
  \bibfield  {author} {\bibinfo {author} {\bibfnamefont {H.~R.}\ \bibnamefont
  {Wilson}}, \bibinfo {author} {\bibfnamefont {J.~W.}\ \bibnamefont {Connor}},
  \bibinfo {author} {\bibfnamefont {R.~J.}\ \bibnamefont {Hastie}},\ and\
  \bibinfo {author} {\bibfnamefont {C.~C.}\ \bibnamefont {Hegna}},\ }\bibfield
  {title} {\enquote {\bibinfo {title} {{Threshold for neoclassical magnetic
  islands in a low collision frequency tokamak}},}\ }\href
  {https://doi.org/10.1063/1.871830} {\bibfield  {journal} {\bibinfo  {journal}
  {Physics of Plasmas}\ }\textbf {\bibinfo {volume} {3}},\ \bibinfo {pages}
  {248--265} (\bibinfo {year} {1996})}\BibitemShut {NoStop}%
\bibitem [{\citenamefont {{G{\"u}nter}}\ \emph {et~al.}(2003)\citenamefont
  {{G{\"u}nter}}, \citenamefont {{Gantenbein}}, \citenamefont {{Gude}},
  \citenamefont {{Igochine}}, \citenamefont {{Maraschek}}, \citenamefont
  {{M{\"u}ck}}, \citenamefont {{Saarelma}}, \citenamefont {{Sauter}},
  \citenamefont {{Sips}}, \citenamefont {{Zohm}},\ and\ \citenamefont {{ASDEX
  Upgrade Team}}}]{GuenterNF03}%
  \BibitemOpen
  \bibfield  {author} {\bibinfo {author} {\bibfnamefont {S.}~\bibnamefont
  {{G{\"u}nter}}}, \bibinfo {author} {\bibfnamefont {G.}~\bibnamefont
  {{Gantenbein}}}, \bibinfo {author} {\bibfnamefont {A.}~\bibnamefont
  {{Gude}}}, \bibinfo {author} {\bibfnamefont {V.}~\bibnamefont {{Igochine}}},
  \bibinfo {author} {\bibfnamefont {M.}~\bibnamefont {{Maraschek}}}, \bibinfo
  {author} {\bibfnamefont {A.}~\bibnamefont {{M{\"u}ck}}}, \bibinfo {author}
  {\bibfnamefont {S.}~\bibnamefont {{Saarelma}}}, \bibinfo {author}
  {\bibfnamefont {O.}~\bibnamefont {{Sauter}}}, \bibinfo {author}
  {\bibfnamefont {A.~C.~C.}\ \bibnamefont {{Sips}}}, \bibinfo {author}
  {\bibfnamefont {H.}~\bibnamefont {{Zohm}}},\ and\ \bibinfo {author}
  {\bibnamefont {{ASDEX Upgrade Team}}},\ }\bibfield  {title} {\enquote
  {\bibinfo {title} {{Neoclassical tearing modes on ASDEX Upgrade: improved
  scaling laws, high confinement at high {\ensuremath{\beta}}$_{N}$ and new
  stabilization experiments}},}\ }\href
  {https://doi.org/10.1088/0029-5515/43/3/301} {\bibfield  {journal} {\bibinfo
  {journal} {Nuclear Fusion}\ }\textbf {\bibinfo {volume} {43}},\ \bibinfo
  {pages} {161--167} (\bibinfo {year} {2003})}\BibitemShut {NoStop}%
\bibitem [{\citenamefont {Buttery}\ \emph {et~al.}(2003)\citenamefont
  {Buttery}, \citenamefont {Hender}, \citenamefont {Howell}, \citenamefont
  {Haye}, \citenamefont {Sauter}, \citenamefont {Testa},\ and\ \citenamefont
  {{contributors to the EFDA-JET Workprogramme}}}]{Buttery_NF2003}%
  \BibitemOpen
  \bibfield  {author} {\bibinfo {author} {\bibfnamefont {R.~J.}\ \bibnamefont
  {Buttery}}, \bibinfo {author} {\bibfnamefont {T.~C.}\ \bibnamefont {Hender}},
  \bibinfo {author} {\bibfnamefont {D.~F.}\ \bibnamefont {Howell}}, \bibinfo
  {author} {\bibfnamefont {R.~J.~L.}\ \bibnamefont {Haye}}, \bibinfo {author}
  {\bibfnamefont {O.}~\bibnamefont {Sauter}}, \bibinfo {author} {\bibfnamefont
  {D.}~\bibnamefont {Testa}},\ and\ \bibinfo {author} {\bibnamefont
  {{contributors to the EFDA-JET Workprogramme}}},\ }\bibfield  {title}
  {\enquote {\bibinfo {title} {{Onset of neoclassical tearing modes on
  {\{}JET{\}}}},}\ }\href@noop {} {\bibfield  {journal} {\bibinfo  {journal}
  {Nuclear Fusion}\ }\textbf {\bibinfo {volume} {43}},\ \bibinfo {pages} {69}
  (\bibinfo {year} {2003})}\BibitemShut {NoStop}%
\bibitem [{\citenamefont {{Poli}}\ \emph {et~al.}(2005)\citenamefont {{Poli}},
  \citenamefont {{Bergmann}}, \citenamefont {{Peeters}}, \citenamefont
  {{Appel}},\ and\ \citenamefont {{Pinches}}}]{PoliNF05}%
  \BibitemOpen
  \bibfield  {author} {\bibinfo {author} {\bibfnamefont {E.}~\bibnamefont
  {{Poli}}}, \bibinfo {author} {\bibfnamefont {A.}~\bibnamefont {{Bergmann}}},
  \bibinfo {author} {\bibfnamefont {A.~G.}\ \bibnamefont {{Peeters}}}, \bibinfo
  {author} {\bibfnamefont {L.~C.}\ \bibnamefont {{Appel}}},\ and\ \bibinfo
  {author} {\bibfnamefont {S.~D.}\ \bibnamefont {{Pinches}}},\ }\bibfield
  {title} {\enquote {\bibinfo {title} {{Kinetic calculation of the polarization
  current in the presence of a neoclassical tearing mode}},}\ }\href
  {https://doi.org/10.1088/0029-5515/45/5/009} {\bibfield  {journal} {\bibinfo
  {journal} {Nuclear Fusion}\ }\textbf {\bibinfo {volume} {45}},\ \bibinfo
  {pages} {384--390} (\bibinfo {year} {2005})}\BibitemShut {NoStop}%
\bibitem [{\citenamefont {Waelbroeck}\ \emph {et~al.}(2009)\citenamefont
  {Waelbroeck}, \citenamefont {Militello}, \citenamefont {Fitzpatrick},\ and\
  \citenamefont {Horton}}]{Waelbroeck_PPCF2008}%
  \BibitemOpen
  \bibfield  {author} {\bibinfo {author} {\bibfnamefont {F.~L.}\ \bibnamefont
  {Waelbroeck}}, \bibinfo {author} {\bibfnamefont {F.}~\bibnamefont
  {Militello}}, \bibinfo {author} {\bibfnamefont {R.}~\bibnamefont
  {Fitzpatrick}},\ and\ \bibinfo {author} {\bibfnamefont {W.}~\bibnamefont
  {Horton}},\ }\bibfield  {title} {\enquote {\bibinfo {title} {{Effect of
  electrostatic turbulence on magnetic islands}},}\ }\href@noop {} {\bibfield
  {journal} {\bibinfo  {journal} {Plasma Physics and Controlled Fusion}\
  }\textbf {\bibinfo {volume} {51}},\ \bibinfo {pages} {15015} (\bibinfo {year}
  {2009})}\BibitemShut {NoStop}%
\bibitem [{\citenamefont {{Ishizawa}}\ and\ \citenamefont
  {{Waelbroeck}}(2013)}]{IshizawaPoP13}%
  \BibitemOpen
  \bibfield  {author} {\bibinfo {author} {\bibfnamefont {A.}~\bibnamefont
  {{Ishizawa}}}\ and\ \bibinfo {author} {\bibfnamefont {F.~L.}\ \bibnamefont
  {{Waelbroeck}}},\ }\bibfield  {title} {\enquote {\bibinfo {title} {{Magnetic
  island evolution in the presence of ion-temperature gradient-driven
  turbulence}},}\ }\href {https://doi.org/10.1063/1.4838176} {\bibfield
  {journal} {\bibinfo  {journal} {Physics of Plasmas}\ }\textbf {\bibinfo
  {volume} {20}},\ \bibinfo {eid} {122301} (\bibinfo {year}
  {2013})}\BibitemShut {NoStop}%
\bibitem [{\citenamefont {{Bergmann}}, \citenamefont {{Poli}},\ and\
  \citenamefont {{Peeters}}(2009)}]{Bergmann_PoP09}%
  \BibitemOpen
  \bibfield  {author} {\bibinfo {author} {\bibfnamefont {A.}~\bibnamefont
  {{Bergmann}}}, \bibinfo {author} {\bibfnamefont {E.}~\bibnamefont {{Poli}}},\
  and\ \bibinfo {author} {\bibfnamefont {A.~G.}\ \bibnamefont {{Peeters}}},\
  }\bibfield  {title} {\enquote {\bibinfo {title} {{The bootstrap current in
  small rotating magnetic islands}},}\ }\href
  {https://doi.org/10.1063/1.3234252} {\bibfield  {journal} {\bibinfo
  {journal} {Physics of Plasmas}\ }\textbf {\bibinfo {volume} {16}},\ \bibinfo
  {eid} {092507} (\bibinfo {year} {2009})}\BibitemShut {NoStop}%
\bibitem [{\citenamefont {{Siccinio}}\ \emph {et~al.}(2011)\citenamefont
  {{Siccinio}}, \citenamefont {{Poli}}, \citenamefont {{Casson}}, \citenamefont
  {{Hornsby}},\ and\ \citenamefont {{Peeters}}}]{SiccinoPoP11}%
  \BibitemOpen
  \bibfield  {author} {\bibinfo {author} {\bibfnamefont {M.}~\bibnamefont
  {{Siccinio}}}, \bibinfo {author} {\bibfnamefont {E.}~\bibnamefont {{Poli}}},
  \bibinfo {author} {\bibfnamefont {F.~J.}\ \bibnamefont {{Casson}}}, \bibinfo
  {author} {\bibfnamefont {W.~A.}\ \bibnamefont {{Hornsby}}},\ and\ \bibinfo
  {author} {\bibfnamefont {A.~G.}\ \bibnamefont {{Peeters}}},\ }\bibfield
  {title} {\enquote {\bibinfo {title} {{Gyrokinetic determination of the
  electrostatic potential of rotating magnetic islands in tokamaks}},}\ }\href
  {https://doi.org/10.1063/1.3671964} {\bibfield  {journal} {\bibinfo
  {journal} {Physics of Plasmas}\ }\textbf {\bibinfo {volume} {18}},\ \bibinfo
  {eid} {122506} (\bibinfo {year} {2011})}\BibitemShut {NoStop}%
\bibitem [{\citenamefont {{Coppi}}(1964)}]{Coppi_PoF64}%
  \BibitemOpen
  \bibfield  {author} {\bibinfo {author} {\bibfnamefont {B.}~\bibnamefont
  {{Coppi}}},\ }\bibfield  {title} {\enquote {\bibinfo {title} {{Influence of
  Gyration Radius and Collisions on Hydromagnetic Stability}},}\ }\href
  {https://doi.org/10.1063/1.1711405} {\bibfield  {journal} {\bibinfo
  {journal} {Physics of Fluids}\ }\textbf {\bibinfo {volume} {7}},\ \bibinfo
  {pages} {1501--1516} (\bibinfo {year} {1964})}\BibitemShut {NoStop}%
\bibitem [{\citenamefont {{Biskamp}}(1978)}]{Biskamp_NF78}%
  \BibitemOpen
  \bibfield  {author} {\bibinfo {author} {\bibfnamefont {D.}~\bibnamefont
  {{Biskamp}}},\ }\bibfield  {title} {\enquote {\bibinfo {title}
  {{Drift-tearing modes in a tokamak plasma}},}\ }\href@noop {} {\bibfield
  {journal} {\bibinfo  {journal} {Nuclear Fusion}\ }\textbf {\bibinfo {volume}
  {18}},\ \bibinfo {pages} {1059--1068} (\bibinfo {year} {1978})}\BibitemShut
  {NoStop}%
\bibitem [{\citenamefont {Smolyakov}\ \emph {et~al.}(1993)\citenamefont
  {Smolyakov}, \citenamefont {Search}, \citenamefont {Journals}, \citenamefont
  {Contact}, \citenamefont {Iopscience}, \citenamefont {Phys},\ and\
  \citenamefont {Address}}]{Smolyakov_PPCF1993}%
  \BibitemOpen
  \bibfield  {author} {\bibinfo {author} {\bibfnamefont {A.~I.}\ \bibnamefont
  {Smolyakov}}, \bibinfo {author} {\bibfnamefont {H.}~\bibnamefont {Search}},
  \bibinfo {author} {\bibfnamefont {C.}~\bibnamefont {Journals}}, \bibinfo
  {author} {\bibfnamefont {A.}~\bibnamefont {Contact}}, \bibinfo {author}
  {\bibfnamefont {M.}~\bibnamefont {Iopscience}}, \bibinfo {author}
  {\bibfnamefont {P.}~\bibnamefont {Phys}},\ and\ \bibinfo {author}
  {\bibfnamefont {I.~P.}\ \bibnamefont {Address}},\ }\bibfield  {title}
  {\enquote {\bibinfo {title} {{Nonlinear evolution of tearing modes in
  inhomogeneous plasmas}},}\ }\href
  {http://stacks.iop.org/0741-3335/35/i=6/a=002} {\bibfield  {journal}
  {\bibinfo  {journal} {Plasma Physics and Controlled Fusion}\ }\textbf
  {\bibinfo {volume} {35}},\ \bibinfo {pages} {657} (\bibinfo {year}
  {1993})}\BibitemShut {NoStop}%
\bibitem [{\citenamefont {Waelbroeck}(2009)}]{Waelbroeck2009}%
  \BibitemOpen
  \bibfield  {author} {\bibinfo {author} {\bibfnamefont {F.~L.}\ \bibnamefont
  {Waelbroeck}},\ }\bibfield  {title} {\enquote {\bibinfo {title} {{Theory and
  observations of magnetic islands}},}\ }\href
  {https://doi.org/10.1088/0029-5515/49/10/104025} {\bibfield  {journal}
  {\bibinfo  {journal} {Nuclear Fusion}\ }\textbf {\bibinfo {volume} {49}},\
  \bibinfo {pages} {104025} (\bibinfo {year} {2009})}\BibitemShut {NoStop}%
\bibitem [{\citenamefont {Nishimura}\ \emph {et~al.}(2008)\citenamefont
  {Nishimura}, \citenamefont {Benkadda}, \citenamefont {Yagi}, \citenamefont
  {Itoh},\ and\ \citenamefont {Itoh}}]{Nishimura2008}%
  \BibitemOpen
  \bibfield  {author} {\bibinfo {author} {\bibfnamefont {S.}~\bibnamefont
  {Nishimura}}, \bibinfo {author} {\bibfnamefont {S.}~\bibnamefont {Benkadda}},
  \bibinfo {author} {\bibfnamefont {M.}~\bibnamefont {Yagi}}, \bibinfo {author}
  {\bibfnamefont {S.~I.}\ \bibnamefont {Itoh}},\ and\ \bibinfo {author}
  {\bibfnamefont {K.}~\bibnamefont {Itoh}},\ }\bibfield  {title} {\enquote
  {\bibinfo {title} {{Nonlinear dynamics of rotating drift-tearing modes in
  tokamak plasmas}},}\ }\href {https://doi.org/10.1063/1.2980286} {\bibfield
  {journal} {\bibinfo  {journal} {Physics of Plasmas}\ }\textbf {\bibinfo
  {volume} {15}},\ \bibinfo {pages} {1--10} (\bibinfo {year}
  {2008})}\BibitemShut {NoStop}%
\bibitem [{\citenamefont {Muraglia}\ \emph {et~al.}(2009)\citenamefont
  {Muraglia}, \citenamefont {Agullo}, \citenamefont {Yagi}, \citenamefont
  {Benkadda}, \citenamefont {Beyer}, \citenamefont {Garbet}, \citenamefont
  {Itoh}, \citenamefont {Itoh},\ and\ \citenamefont {Sen}}]{MuragliaNF09}%
  \BibitemOpen
  \bibfield  {author} {\bibinfo {author} {\bibfnamefont {M.}~\bibnamefont
  {Muraglia}}, \bibinfo {author} {\bibfnamefont {O.}~\bibnamefont {Agullo}},
  \bibinfo {author} {\bibfnamefont {M.}~\bibnamefont {Yagi}}, \bibinfo {author}
  {\bibfnamefont {S.}~\bibnamefont {Benkadda}}, \bibinfo {author}
  {\bibfnamefont {P.}~\bibnamefont {Beyer}}, \bibinfo {author} {\bibfnamefont
  {X.}~\bibnamefont {Garbet}}, \bibinfo {author} {\bibfnamefont {S.-I.}\
  \bibnamefont {Itoh}}, \bibinfo {author} {\bibfnamefont {K.}~\bibnamefont
  {Itoh}},\ and\ \bibinfo {author} {\bibfnamefont {A.}~\bibnamefont {Sen}},\
  }\bibfield  {title} {\enquote {\bibinfo {title} {{Effect of the curvature and
  the $\beta$ parameter on the nonlinear dynamics of a drift tearing magnetic
  island}},}\ }\href {https://doi.org/10.1088/0029-5515/49/5/055016} {\bibfield
   {journal} {\bibinfo  {journal} {Nuclear Fusion}\ }\textbf {\bibinfo {volume}
  {49}},\ \bibinfo {pages} {055016} (\bibinfo {year} {2009})}\BibitemShut
  {NoStop}%
\bibitem [{\citenamefont {Uzawa}, \citenamefont {Ishizawa},\ and\ \citenamefont
  {Nakajima}(2010)}]{Uzawa_PoP2010}%
  \BibitemOpen
  \bibfield  {author} {\bibinfo {author} {\bibfnamefont {K.}~\bibnamefont
  {Uzawa}}, \bibinfo {author} {\bibfnamefont {A.}~\bibnamefont {Ishizawa}},\
  and\ \bibinfo {author} {\bibfnamefont {N.}~\bibnamefont {Nakajima}},\
  }\bibfield  {title} {\enquote {\bibinfo {title} {{Propagation of magnetic
  island due to self-induced zonal flow}},}\ }\href
  {https://doi.org/10.1063/1.3368047} {\bibfield  {journal} {\bibinfo
  {journal} {Physics of Plasmas}\ }\textbf {\bibinfo {volume} {17}},\ \bibinfo
  {pages} {42508} (\bibinfo {year} {2010})}\BibitemShut {NoStop}%
\bibitem [{\citenamefont {Ishizawa}\ \emph {et~al.}(2012)\citenamefont
  {Ishizawa}, \citenamefont {Waelbroeck}, \citenamefont {Fitzpatrick},
  \citenamefont {Horton},\ and\ \citenamefont {Nakajima}}]{Ishizawa2012a}%
  \BibitemOpen
  \bibfield  {author} {\bibinfo {author} {\bibfnamefont {A.}~\bibnamefont
  {Ishizawa}}, \bibinfo {author} {\bibfnamefont {F.~L.}\ \bibnamefont
  {Waelbroeck}}, \bibinfo {author} {\bibfnamefont {R.}~\bibnamefont
  {Fitzpatrick}}, \bibinfo {author} {\bibfnamefont {W.}~\bibnamefont
  {Horton}},\ and\ \bibinfo {author} {\bibfnamefont {N.}~\bibnamefont
  {Nakajima}},\ }\bibfield  {title} {\enquote {\bibinfo {title} {{Magnetic
  island evolution in hot ion plasmas}},}\ }\href
  {https://doi.org/10.1063/1.4739291} {\bibfield  {journal} {\bibinfo
  {journal} {Physics of Plasmas}\ }\textbf {\bibinfo {volume} {19}} (\bibinfo
  {year} {2012}),\ 10.1063/1.4739291}\BibitemShut {NoStop}%
\bibitem [{\citenamefont {{Drake}}\ and\ \citenamefont
  {{Lee}}(1977)}]{Drake77}%
  \BibitemOpen
  \bibfield  {author} {\bibinfo {author} {\bibfnamefont {J.~F.}\ \bibnamefont
  {{Drake}}}\ and\ \bibinfo {author} {\bibfnamefont {Y.~C.}\ \bibnamefont
  {{Lee}}},\ }\bibfield  {title} {\enquote {\bibinfo {title} {{Kinetic theory
  of tearing instabilities}},}\ }\href {https://doi.org/10.1063/1.862017}
  {\bibfield  {journal} {\bibinfo  {journal} {Physics of Fluids}\ }\textbf
  {\bibinfo {volume} {20}},\ \bibinfo {pages} {1341--1353} (\bibinfo {year}
  {1977})}\BibitemShut {NoStop}%
\bibitem [{\citenamefont {Drake}(1983)}]{Drake1983}%
  \BibitemOpen
  \bibfield  {author} {\bibinfo {author} {\bibfnamefont {J.~F.}\ \bibnamefont
  {Drake}},\ }\bibfield  {title} {\enquote {\bibinfo {title} {{Stabilization of
  the tearing mode in high-temperature plasma}},}\ }\href
  {https://doi.org/10.1063/1.864441} {\bibfield  {journal} {\bibinfo  {journal}
  {Physics of Fluids}\ }\textbf {\bibinfo {volume} {26}},\ \bibinfo {pages}
  {2509} (\bibinfo {year} {1983})}\BibitemShut {NoStop}%
\bibitem [{\citenamefont {{Nishimura}}\ \emph {et~al.}(2007)\citenamefont
  {{Nishimura}}, \citenamefont {{Yagi}}, \citenamefont {{Itoh}}, \citenamefont
  {{Azumi}},\ and\ \citenamefont {{Itoh}}}]{NishimuraJPSJ07}%
  \BibitemOpen
  \bibfield  {author} {\bibinfo {author} {\bibfnamefont {S.}~\bibnamefont
  {{Nishimura}}}, \bibinfo {author} {\bibfnamefont {M.}~\bibnamefont {{Yagi}}},
  \bibinfo {author} {\bibfnamefont {S.-I.}\ \bibnamefont {{Itoh}}}, \bibinfo
  {author} {\bibfnamefont {M.}~\bibnamefont {{Azumi}}},\ and\ \bibinfo {author}
  {\bibfnamefont {K.}~\bibnamefont {{Itoh}}},\ }\bibfield  {title} {\enquote
  {\bibinfo {title} {{Thermal Transport Effects on Drift-Tearing Mode}},}\
  }\href {https://doi.org/10.1143/JPSJ.76.064501} {\bibfield  {journal}
  {\bibinfo  {journal} {Journal of the Physical Society of Japan}\ }\textbf
  {\bibinfo {volume} {76}},\ \bibinfo {pages} {064501} (\bibinfo {year}
  {2007})}\BibitemShut {NoStop}%
\bibitem [{\citenamefont {Smolyakov}\ \emph {et~al.}(1995)\citenamefont
  {Smolyakov}, \citenamefont {Hirose}, \citenamefont {Lazzaro}, \citenamefont
  {Re},\ and\ \citenamefont {Callen}}]{smolyakov_PoP1995}%
  \BibitemOpen
  \bibfield  {author} {\bibinfo {author} {\bibfnamefont {A.~I.}\ \bibnamefont
  {Smolyakov}}, \bibinfo {author} {\bibfnamefont {A.}~\bibnamefont {Hirose}},
  \bibinfo {author} {\bibfnamefont {E.}~\bibnamefont {Lazzaro}}, \bibinfo
  {author} {\bibfnamefont {G.~B.}\ \bibnamefont {Re}},\ and\ \bibinfo {author}
  {\bibfnamefont {J.~D.}\ \bibnamefont {Callen}},\ }\bibfield  {title}
  {\enquote {\bibinfo {title} {{Rotating nonlinear magnetic islands in a
  tokamak plasma}},}\ }\href {https://doi.org/10.1063/1.871308} {\bibfield
  {journal} {\bibinfo  {journal} {Physics of Plasmas}\ }\textbf {\bibinfo
  {volume} {2}},\ \bibinfo {pages} {1581--1598} (\bibinfo {year}
  {1995})}\BibitemShut {NoStop}%
\bibitem [{\citenamefont {{Sato}}\ and\ \citenamefont
  {{Ishizawa}}(2017)}]{Sato_PoP17}%
  \BibitemOpen
  \bibfield  {author} {\bibinfo {author} {\bibfnamefont {M.}~\bibnamefont
  {{Sato}}}\ and\ \bibinfo {author} {\bibfnamefont {A.}~\bibnamefont
  {{Ishizawa}}},\ }\bibfield  {title} {\enquote {\bibinfo {title} {{Nonlinear
  parity mixtures controlling the propagation of interchange modes}},}\ }\href
  {https://doi.org/10.1063/1.4993472]} {\bibfield  {journal} {\bibinfo
  {journal} {Physics of Plasmas}\ }\textbf {\bibinfo {volume} {24}},\ \bibinfo
  {pages} {082501} (\bibinfo {year} {2017})}\BibitemShut {NoStop}%
\bibitem [{\citenamefont {{Lanti}}\ \emph {et~al.}(2020)\citenamefont
  {{Lanti}}, \citenamefont {{Ohana}}, \citenamefont {{Tronko}}, \citenamefont
  {{Hayward-Schneider}}, \citenamefont {{Bottino}}, \citenamefont {{McMillan}},
  \citenamefont {{Mishchenko}}, \citenamefont {{Scheinberg}}, \citenamefont
  {{Biancalani}}, \citenamefont {{Angelino}}, \citenamefont {{Brunner}},
  \citenamefont {{Dominski}}, \citenamefont {{Donnel}}, \citenamefont
  {{Gheller}}, \citenamefont {{Hatzky}}, \citenamefont {{Jocksch}},
  \citenamefont {{Jolliet}}, \citenamefont {{Lu}}, \citenamefont {{Martin
  Collar}}, \citenamefont {{Novikau}}, \citenamefont {{Sonnendr{\"u}cker}},
  \citenamefont {{Vernay}},\ and\ \citenamefont {{Villard}}}]{Lanti2020}%
  \BibitemOpen
  \bibfield  {author} {\bibinfo {author} {\bibfnamefont {E.}~\bibnamefont
  {{Lanti}}}, \bibinfo {author} {\bibfnamefont {N.}~\bibnamefont {{Ohana}}},
  \bibinfo {author} {\bibfnamefont {N.}~\bibnamefont {{Tronko}}}, \bibinfo
  {author} {\bibfnamefont {T.}~\bibnamefont {{Hayward-Schneider}}}, \bibinfo
  {author} {\bibfnamefont {A.}~\bibnamefont {{Bottino}}}, \bibinfo {author}
  {\bibfnamefont {B.~F.}\ \bibnamefont {{McMillan}}}, \bibinfo {author}
  {\bibfnamefont {A.}~\bibnamefont {{Mishchenko}}}, \bibinfo {author}
  {\bibfnamefont {A.}~\bibnamefont {{Scheinberg}}}, \bibinfo {author}
  {\bibfnamefont {A.}~\bibnamefont {{Biancalani}}}, \bibinfo {author}
  {\bibfnamefont {P.}~\bibnamefont {{Angelino}}}, \bibinfo {author}
  {\bibfnamefont {S.}~\bibnamefont {{Brunner}}}, \bibinfo {author}
  {\bibfnamefont {J.}~\bibnamefont {{Dominski}}}, \bibinfo {author}
  {\bibfnamefont {P.}~\bibnamefont {{Donnel}}}, \bibinfo {author}
  {\bibfnamefont {C.}~\bibnamefont {{Gheller}}}, \bibinfo {author}
  {\bibfnamefont {R.}~\bibnamefont {{Hatzky}}}, \bibinfo {author}
  {\bibfnamefont {A.}~\bibnamefont {{Jocksch}}}, \bibinfo {author}
  {\bibfnamefont {S.}~\bibnamefont {{Jolliet}}}, \bibinfo {author}
  {\bibfnamefont {Z.~X.}\ \bibnamefont {{Lu}}}, \bibinfo {author}
  {\bibfnamefont {J.~P.}\ \bibnamefont {{Martin Collar}}}, \bibinfo {author}
  {\bibfnamefont {I.}~\bibnamefont {{Novikau}}}, \bibinfo {author}
  {\bibfnamefont {E.}~\bibnamefont {{Sonnendr{\"u}cker}}}, \bibinfo {author}
  {\bibfnamefont {T.}~\bibnamefont {{Vernay}}},\ and\ \bibinfo {author}
  {\bibfnamefont {L.}~\bibnamefont {{Villard}}},\ }\bibfield  {title} {\enquote
  {\bibinfo {title} {{ORB5: A global electromagnetic gyrokinetic code using the
  PIC approach in toroidal geometry}},}\ }\href
  {https://doi.org/10.1016/j.cpc.2019.107072} {\bibfield  {journal} {\bibinfo
  {journal} {Computer Physics Communications}\ }\textbf {\bibinfo {volume}
  {251}},\ \bibinfo {eid} {107072} (\bibinfo {year} {2020})},\ \Eprint
  {https://arxiv.org/abs/1905.01906} {arXiv:1905.01906 [physics.plasm-ph]}
  \BibitemShut {NoStop}%
\bibitem [{\citenamefont {Mishchenko}\ \emph {et~al.}(2019)\citenamefont
  {Mishchenko}, \citenamefont {Bottino}, \citenamefont {Biancalani},
  \citenamefont {Hatzky}, \citenamefont {Hayward-Schneider}, \citenamefont
  {Ohana}, \citenamefont {Lanti}, \citenamefont {Brunner}, \citenamefont
  {Villard}, \citenamefont {Borchardt}, \citenamefont {Kleiber},\ and\
  \citenamefont {K{\"{o}}nies}}]{MishchenkoCPC19}%
  \BibitemOpen
  \bibfield  {author} {\bibinfo {author} {\bibfnamefont {A.}~\bibnamefont
  {Mishchenko}}, \bibinfo {author} {\bibfnamefont {A.}~\bibnamefont {Bottino}},
  \bibinfo {author} {\bibfnamefont {A.}~\bibnamefont {Biancalani}}, \bibinfo
  {author} {\bibfnamefont {R.}~\bibnamefont {Hatzky}}, \bibinfo {author}
  {\bibfnamefont {T.}~\bibnamefont {Hayward-Schneider}}, \bibinfo {author}
  {\bibfnamefont {N.}~\bibnamefont {Ohana}}, \bibinfo {author} {\bibfnamefont
  {E.}~\bibnamefont {Lanti}}, \bibinfo {author} {\bibfnamefont
  {S.}~\bibnamefont {Brunner}}, \bibinfo {author} {\bibfnamefont
  {L.}~\bibnamefont {Villard}}, \bibinfo {author} {\bibfnamefont
  {M.}~\bibnamefont {Borchardt}}, \bibinfo {author} {\bibfnamefont
  {R.}~\bibnamefont {Kleiber}},\ and\ \bibinfo {author} {\bibfnamefont
  {A.}~\bibnamefont {K{\"{o}}nies}},\ }\bibfield  {title} {\enquote {\bibinfo
  {title} {{Pullback scheme implementation in ORB5}},}\ }\href
  {https://doi.org/10.1016/j.cpc.2018.12.002} {\bibfield  {journal} {\bibinfo
  {journal} {Computer Physics Communications}\ }\textbf {\bibinfo {volume}
  {238}},\ \bibinfo {pages} {194--202} (\bibinfo {year} {2019})},\ \Eprint
  {https://arxiv.org/abs/1811.05346} {arXiv:1811.05346} \BibitemShut {NoStop}%
\bibitem [{\citenamefont {{Chen}}\ and\ \citenamefont
  {{Parker}}(2001)}]{ChenPoP01}%
  \BibitemOpen
  \bibfield  {author} {\bibinfo {author} {\bibfnamefont {Y.}~\bibnamefont
  {{Chen}}}\ and\ \bibinfo {author} {\bibfnamefont {S.}~\bibnamefont
  {{Parker}}},\ }\bibfield  {title} {\enquote {\bibinfo {title} {{Gyrokinetic
  turbulence simulations with kinetic electrons}},}\ }\href
  {https://doi.org/10.1063/1.1351828} {\bibfield  {journal} {\bibinfo
  {journal} {Physics of Plasmas}\ }\textbf {\bibinfo {volume} {8}},\ \bibinfo
  {pages} {2095--2100} (\bibinfo {year} {2001})}\BibitemShut {NoStop}%
\bibitem [{\citenamefont {{Mishchenko}}, \citenamefont {{Hatzky}},\ and\
  \citenamefont {{K{\"o}nies}}(2004)}]{MishchenkoPoP04}%
  \BibitemOpen
  \bibfield  {author} {\bibinfo {author} {\bibfnamefont {A.}~\bibnamefont
  {{Mishchenko}}}, \bibinfo {author} {\bibfnamefont {R.}~\bibnamefont
  {{Hatzky}}},\ and\ \bibinfo {author} {\bibfnamefont {A.}~\bibnamefont
  {{K{\"o}nies}}},\ }\bibfield  {title} {\enquote {\bibinfo {title}
  {{Conventional {\ensuremath{\delta}}f-particle simulations of electromagnetic
  perturbations with finite elements}},}\ }\href
  {https://doi.org/10.1063/1.1812275} {\bibfield  {journal} {\bibinfo
  {journal} {Physics of Plasmas}\ }\textbf {\bibinfo {volume} {11}},\ \bibinfo
  {pages} {5480--5486} (\bibinfo {year} {2004})}\BibitemShut {NoStop}%
\bibitem [{\citenamefont {{Bottino}}\ \emph {et~al.}(2011)\citenamefont
  {{Bottino}}, \citenamefont {{Vernay}}, \citenamefont {{Scott}}, \citenamefont
  {{Brunner}}, \citenamefont {{Hatzky}}, \citenamefont {{Jolliet}},
  \citenamefont {{McMillan}}, \citenamefont {{Tran}},\ and\ \citenamefont
  {{Villard}}}]{BottinoPPCF11}%
  \BibitemOpen
  \bibfield  {author} {\bibinfo {author} {\bibfnamefont {A.}~\bibnamefont
  {{Bottino}}}, \bibinfo {author} {\bibfnamefont {T.}~\bibnamefont {{Vernay}}},
  \bibinfo {author} {\bibfnamefont {B.}~\bibnamefont {{Scott}}}, \bibinfo
  {author} {\bibfnamefont {S.}~\bibnamefont {{Brunner}}}, \bibinfo {author}
  {\bibfnamefont {R.}~\bibnamefont {{Hatzky}}}, \bibinfo {author}
  {\bibfnamefont {S.}~\bibnamefont {{Jolliet}}}, \bibinfo {author}
  {\bibfnamefont {B.~F.}\ \bibnamefont {{McMillan}}}, \bibinfo {author}
  {\bibfnamefont {T.~M.}\ \bibnamefont {{Tran}}},\ and\ \bibinfo {author}
  {\bibfnamefont {L.}~\bibnamefont {{Villard}}},\ }\bibfield  {title} {\enquote
  {\bibinfo {title} {{Global simulations of tokamak microturbulence:
  finite-{\ensuremath{\beta}} effects and collisions}},}\ }\href
  {https://doi.org/10.1088/0741-3335/53/12/124027} {\bibfield  {journal}
  {\bibinfo  {journal} {Plasma Physics and Controlled Fusion}\ }\textbf
  {\bibinfo {volume} {53}},\ \bibinfo {eid} {124027} (\bibinfo {year}
  {2011})}\BibitemShut {NoStop}%
\bibitem [{\citenamefont {{Mishchenko}}\ \emph
  {et~al.}(2014{\natexlab{a}})\citenamefont {{Mishchenko}}, \citenamefont
  {{Cole}}, \citenamefont {{Kleiber}},\ and\ \citenamefont
  {{K{\"o}nies}}}]{MishchenkoPoP14a}%
  \BibitemOpen
  \bibfield  {author} {\bibinfo {author} {\bibfnamefont {A.}~\bibnamefont
  {{Mishchenko}}}, \bibinfo {author} {\bibfnamefont {M.}~\bibnamefont
  {{Cole}}}, \bibinfo {author} {\bibfnamefont {R.}~\bibnamefont {{Kleiber}}},\
  and\ \bibinfo {author} {\bibfnamefont {A.}~\bibnamefont {{K{\"o}nies}}},\
  }\bibfield  {title} {\enquote {\bibinfo {title} {{New variables for
  gyrokinetic electromagnetic simulations}},}\ }\href
  {https://doi.org/10.1063/1.4880560} {\bibfield  {journal} {\bibinfo
  {journal} {Physics of Plasmas}\ }\textbf {\bibinfo {volume} {21}},\ \bibinfo
  {eid} {052113} (\bibinfo {year} {2014}{\natexlab{a}})}\BibitemShut {NoStop}%
\bibitem [{\citenamefont {{Mishchenko}}\ \emph
  {et~al.}(2014{\natexlab{b}})\citenamefont {{Mishchenko}}, \citenamefont
  {{K{\"o}nies}}, \citenamefont {{Kleiber}},\ and\ \citenamefont
  {{Cole}}}]{MishchenkoPoP14b}%
  \BibitemOpen
  \bibfield  {author} {\bibinfo {author} {\bibfnamefont {A.}~\bibnamefont
  {{Mishchenko}}}, \bibinfo {author} {\bibfnamefont {A.}~\bibnamefont
  {{K{\"o}nies}}}, \bibinfo {author} {\bibfnamefont {R.}~\bibnamefont
  {{Kleiber}}},\ and\ \bibinfo {author} {\bibfnamefont {M.}~\bibnamefont
  {{Cole}}},\ }\bibfield  {title} {\enquote {\bibinfo {title} {{Pullback
  transformation in gyrokinetic electromagnetic simulations}},}\ }\href
  {https://doi.org/10.1063/1.4895501} {\bibfield  {journal} {\bibinfo
  {journal} {Physics of Plasmas}\ }\textbf {\bibinfo {volume} {21}},\ \bibinfo
  {eid} {092110} (\bibinfo {year} {2014}{\natexlab{b}})},\ \Eprint
  {https://arxiv.org/abs/1407.3992} {arXiv:1407.3992 [physics.plasm-ph]}
  \BibitemShut {NoStop}%
\bibitem [{\citenamefont {Wesson}\ and\ \citenamefont
  {Campbell}(2011)}]{WessonBook}%
  \BibitemOpen
  \bibfield  {author} {\bibinfo {author} {\bibfnamefont {J.}~\bibnamefont
  {Wesson}}\ and\ \bibinfo {author} {\bibfnamefont {D.}~\bibnamefont
  {Campbell}},\ }\href@noop {} {\emph {\bibinfo {title} {Tokamaks 4th
  Edition}}},\ International Series of Monographs on Physics\ (\bibinfo
  {publisher} {OUP Oxford},\ \bibinfo {year} {2011})\BibitemShut {NoStop}%
\bibitem [{NoteI()}]{NoteI}%
  \BibitemOpen
  \bibinfo {note} {The number is found to be sufficient by a convergence
  test.}\BibitemShut {Stop}%
\bibitem [{\citenamefont {{Kleva}}, \citenamefont {{Drake}},\ and\
  \citenamefont {{Waelbroeck}}(1995)}]{KlevaPoP95}%
  \BibitemOpen
  \bibfield  {author} {\bibinfo {author} {\bibfnamefont {R.~G.}\ \bibnamefont
  {{Kleva}}}, \bibinfo {author} {\bibfnamefont {J.~F.}\ \bibnamefont
  {{Drake}}},\ and\ \bibinfo {author} {\bibfnamefont {F.~L.}\ \bibnamefont
  {{Waelbroeck}}},\ }\bibfield  {title} {\enquote {\bibinfo {title} {{Fast
  reconnection in high temperature plasmas}},}\ }\href
  {https://doi.org/10.1063/1.871095} {\bibfield  {journal} {\bibinfo  {journal}
  {Physics of Plasmas}\ }\textbf {\bibinfo {volume} {2}},\ \bibinfo {pages}
  {23--34} (\bibinfo {year} {1995})}\BibitemShut {NoStop}%
\bibitem [{\citenamefont {{Jain}}, \citenamefont {{B{\"u}chner}},\ and\
  \citenamefont {{Mu{\~n}oz}}(2017)}]{JainPoP17}%
  \BibitemOpen
  \bibfield  {author} {\bibinfo {author} {\bibfnamefont {N.}~\bibnamefont
  {{Jain}}}, \bibinfo {author} {\bibfnamefont {J.}~\bibnamefont
  {{B{\"u}chner}}},\ and\ \bibinfo {author} {\bibfnamefont {P.~A.}\
  \bibnamefont {{Mu{\~n}oz}}},\ }\bibfield  {title} {\enquote {\bibinfo {title}
  {{Nonlinear evolution of electron shear flow instabilities in the presence of
  an external guide magnetic field}},}\ }\href
  {https://doi.org/10.1063/1.4977528} {\bibfield  {journal} {\bibinfo
  {journal} {Physics of Plasmas}\ }\textbf {\bibinfo {volume} {24}},\ \bibinfo
  {eid} {032303} (\bibinfo {year} {2017})},\ \Eprint
  {https://arxiv.org/abs/1608.02403} {arXiv:1608.02403 [physics.plasm-ph]}
  \BibitemShut {NoStop}%
\bibitem [{\citenamefont {Pueschel}\ \emph {et~al.}(2015)\citenamefont
  {Pueschel}, \citenamefont {Terry}, \citenamefont {Told},\ and\ \citenamefont
  {Jenko}}]{Pueschel_PoP2015}%
  \BibitemOpen
  \bibfield  {author} {\bibinfo {author} {\bibfnamefont {M.~J.}\ \bibnamefont
  {Pueschel}}, \bibinfo {author} {\bibfnamefont {P.~W.}\ \bibnamefont {Terry}},
  \bibinfo {author} {\bibfnamefont {D.}~\bibnamefont {Told}},\ and\ \bibinfo
  {author} {\bibfnamefont {F.}~\bibnamefont {Jenko}},\ }\bibfield  {title}
  {\enquote {\bibinfo {title} {Enhanced magnetic reconnection in the presence
  of pressure gradients},}\ }\href {https://doi.org/10.1063/1.4922064}
  {\bibfield  {journal} {\bibinfo  {journal} {Physics of Plasmas}\ }\textbf
  {\bibinfo {volume} {22}} (\bibinfo {year} {2015}),\
  10.1063/1.4922064}\BibitemShut {NoStop}%
\bibitem [{\citenamefont {Ishizawa}\ and\ \citenamefont
  {Nakajima}(2009)}]{Ishizawa2009}%
  \BibitemOpen
  \bibfield  {author} {\bibinfo {author} {\bibfnamefont {A.}~\bibnamefont
  {Ishizawa}}\ and\ \bibinfo {author} {\bibfnamefont {N.}~\bibnamefont
  {Nakajima}},\ }\bibfield  {title} {\enquote {\bibinfo {title} {{Thermal
  transport due to turbulence including magnetic fluctuation in externally
  heated plasma}},}\ }\href {https://doi.org/10.1088/0029-5515/49/5/055015}
  {\bibfield  {journal} {\bibinfo  {journal} {Nuclear Fusion}\ }\textbf
  {\bibinfo {volume} {49}} (\bibinfo {year} {2009}),\
  10.1088/0029-5515/49/5/055015}\BibitemShut {NoStop}%
\bibitem [{\citenamefont {{Arcis}}, \citenamefont {{Escande}},\ and\
  \citenamefont {{Ottaviani}}(2005)}]{ArcisPLA05}%
  \BibitemOpen
  \bibfield  {author} {\bibinfo {author} {\bibfnamefont {N.}~\bibnamefont
  {{Arcis}}}, \bibinfo {author} {\bibfnamefont {D.~F.}\ \bibnamefont
  {{Escande}}},\ and\ \bibinfo {author} {\bibfnamefont {M.}~\bibnamefont
  {{Ottaviani}}},\ }\bibfield  {title} {\enquote {\bibinfo {title} {{Nonlinear
  dynamics of the tearing mode for any current gradient [rapid
  communication]}},}\ }\href {https://doi.org/10.1016/j.physleta.2005.08.027}
  {\bibfield  {journal} {\bibinfo  {journal} {Physics Letters A}\ }\textbf
  {\bibinfo {volume} {347}},\ \bibinfo {pages} {241--247} (\bibinfo {year}
  {2005})}\BibitemShut {NoStop}%
\bibitem [{\citenamefont {{Arcis}}, \citenamefont {{Escande}},\ and\
  \citenamefont {{Ottaviani}}(2006)}]{ArcisPoP06}%
  \BibitemOpen
  \bibfield  {author} {\bibinfo {author} {\bibfnamefont {N.}~\bibnamefont
  {{Arcis}}}, \bibinfo {author} {\bibfnamefont {D.~F.}\ \bibnamefont
  {{Escande}}},\ and\ \bibinfo {author} {\bibfnamefont {M.}~\bibnamefont
  {{Ottaviani}}},\ }\bibfield  {title} {\enquote {\bibinfo {title} {{Rigorous
  approach to the nonlinear saturation of the tearing mode in cylindrical and
  slab geometry}},}\ }\href {https://doi.org/10.1063/1.2199208} {\bibfield
  {journal} {\bibinfo  {journal} {Physics of Plasmas}\ }\textbf {\bibinfo
  {volume} {13}},\ \bibinfo {eid} {052305} (\bibinfo {year} {2006})},\ \Eprint
  {https://arxiv.org/abs/physics/0603204} {arXiv:physics/0603204
  [physics.plasm-ph]} \BibitemShut {NoStop}%
\bibitem [{\citenamefont {{Hornsby}}\ \emph
  {et~al.}(2015{\natexlab{b}})\citenamefont {{Hornsby}}, \citenamefont
  {{Migliano}}, \citenamefont {{Buchholz}}, \citenamefont {{Kroenert}},
  \citenamefont {{Weikl}}, \citenamefont {{Peeters}}, \citenamefont
  {{Zarzoso}}, \citenamefont {{Poli}},\ and\ \citenamefont
  {{Casson}}}]{HornsbyPoP15}%
  \BibitemOpen
  \bibfield  {author} {\bibinfo {author} {\bibfnamefont {W.~A.}\ \bibnamefont
  {{Hornsby}}}, \bibinfo {author} {\bibfnamefont {P.}~\bibnamefont
  {{Migliano}}}, \bibinfo {author} {\bibfnamefont {R.}~\bibnamefont
  {{Buchholz}}}, \bibinfo {author} {\bibfnamefont {L.}~\bibnamefont
  {{Kroenert}}}, \bibinfo {author} {\bibfnamefont {A.}~\bibnamefont {{Weikl}}},
  \bibinfo {author} {\bibfnamefont {A.~G.}\ \bibnamefont {{Peeters}}}, \bibinfo
  {author} {\bibfnamefont {D.}~\bibnamefont {{Zarzoso}}}, \bibinfo {author}
  {\bibfnamefont {E.}~\bibnamefont {{Poli}}},\ and\ \bibinfo {author}
  {\bibfnamefont {F.~J.}\ \bibnamefont {{Casson}}},\ }\bibfield  {title}
  {\enquote {\bibinfo {title} {{The linear tearing instability in three
  dimensional, toroidal gyro-kinetic simulations}},}\ }\href
  {https://doi.org/10.1063/1.4907900} {\bibfield  {journal} {\bibinfo
  {journal} {Physics of Plasmas}\ }\textbf {\bibinfo {volume} {22}},\ \bibinfo
  {eid} {022118} (\bibinfo {year} {2015}{\natexlab{b}})},\ \Eprint
  {https://arxiv.org/abs/1408.0112} {arXiv:1408.0112 [physics.plasm-ph]}
  \BibitemShut {NoStop}%
\bibitem [{\citenamefont {{Jenko}}\ \emph {et~al.}(2000)\citenamefont
  {{Jenko}}, \citenamefont {{Dorland}}, \citenamefont {{Kotschenreuther}},\
  and\ \citenamefont {{Rogers}}}]{JenkoPoP00}%
  \BibitemOpen
  \bibfield  {author} {\bibinfo {author} {\bibfnamefont {F.}~\bibnamefont
  {{Jenko}}}, \bibinfo {author} {\bibfnamefont {W.}~\bibnamefont {{Dorland}}},
  \bibinfo {author} {\bibfnamefont {M.}~\bibnamefont {{Kotschenreuther}}},\
  and\ \bibinfo {author} {\bibfnamefont {B.~N.}\ \bibnamefont {{Rogers}}},\
  }\bibfield  {title} {\enquote {\bibinfo {title} {{Electron temperature
  gradient driven turbulence}},}\ }\href {https://doi.org/10.1063/1.874014}
  {\bibfield  {journal} {\bibinfo  {journal} {Physics of Plasmas}\ }\textbf
  {\bibinfo {volume} {7}},\ \bibinfo {pages} {1904--1910} (\bibinfo {year}
  {2000})}\BibitemShut {NoStop}%
\bibitem [{NoteII()}]{NoteII}%
  \BibitemOpen
  \bibinfo {note} {Although the equilibrium pressure is flat in these
  simulations, we refer to the diamagnetic direction as the direction in which
  a given particle species would rotate in the presence of a finite
  gradient.}\BibitemShut {Stop}%
\bibitem [{\citenamefont {Hornsby}\ \emph {et~al.}(2016)\citenamefont
  {Hornsby}, \citenamefont {Migliano}, \citenamefont {Buchholz}, \citenamefont
  {Grosshauser}, \citenamefont {Weikl}, \citenamefont {Zarzoso}, \citenamefont
  {Casson}, \citenamefont {Poli},\ and\ \citenamefont {Peeters}}]{HornsbyNF16}%
  \BibitemOpen
  \bibfield  {author} {\bibinfo {author} {\bibfnamefont {W.~A.}\ \bibnamefont
  {Hornsby}}, \bibinfo {author} {\bibfnamefont {P.}~\bibnamefont {Migliano}},
  \bibinfo {author} {\bibfnamefont {R.}~\bibnamefont {Buchholz}}, \bibinfo
  {author} {\bibfnamefont {S.}~\bibnamefont {Grosshauser}}, \bibinfo {author}
  {\bibfnamefont {A.}~\bibnamefont {Weikl}}, \bibinfo {author} {\bibfnamefont
  {D.}~\bibnamefont {Zarzoso}}, \bibinfo {author} {\bibfnamefont {F.~J.}\
  \bibnamefont {Casson}}, \bibinfo {author} {\bibfnamefont {E.}~\bibnamefont
  {Poli}},\ and\ \bibinfo {author} {\bibfnamefont {A.~G.}\ \bibnamefont
  {Peeters}},\ }\bibfield  {title} {\enquote {\bibinfo {title} {{The non-linear
  evolution of the tearing mode in electromagnetic turbulence using gyrokinetic
  simulations}},}\ }\href {https://doi.org/10.1088/0741-3335/58/1/014028}
  {\bibfield  {journal} {\bibinfo  {journal} {Plasma Physics and Controlled
  Fusion}\ }\textbf {\bibinfo {volume} {58}},\ \bibinfo {pages} {014028}
  (\bibinfo {year} {2016})}\BibitemShut {NoStop}%
\bibitem [{\citenamefont {{Grasso}}\ \emph {et~al.}(2020)\citenamefont
  {{Grasso}}, \citenamefont {{Borgogno}}, \citenamefont {{Tassi}},\ and\
  \citenamefont {{Perona}}}]{GrassoPoP20}%
  \BibitemOpen
  \bibfield  {author} {\bibinfo {author} {\bibfnamefont {D.}~\bibnamefont
  {{Grasso}}}, \bibinfo {author} {\bibfnamefont {D.}~\bibnamefont
  {{Borgogno}}}, \bibinfo {author} {\bibfnamefont {E.}~\bibnamefont
  {{Tassi}}},\ and\ \bibinfo {author} {\bibfnamefont {A.}~\bibnamefont
  {{Perona}}},\ }\bibfield  {title} {\enquote {\bibinfo {title} {{Asymmetry
  effects driving secondary instabilities in two-dimensional collisionless
  magnetic reconnection}},}\ }\href {https://doi.org/10.1063/1.5125122}
  {\bibfield  {journal} {\bibinfo  {journal} {Physics of Plasmas}\ }\textbf
  {\bibinfo {volume} {27}},\ \bibinfo {eid} {012302} (\bibinfo {year}
  {2020})}\BibitemShut {NoStop}%
\bibitem [{\citenamefont {{Borgogno}}\ \emph {et~al.}(2022)\citenamefont
  {{Borgogno}}, \citenamefont {{Grasso}}, \citenamefont {{Achilli}},
  \citenamefont {{Rom{\'e}}},\ and\ \citenamefont {{Comisso}}}]{BorgognoAPJ22}%
  \BibitemOpen
  \bibfield  {author} {\bibinfo {author} {\bibfnamefont {D.}~\bibnamefont
  {{Borgogno}}}, \bibinfo {author} {\bibfnamefont {D.}~\bibnamefont
  {{Grasso}}}, \bibinfo {author} {\bibfnamefont {B.}~\bibnamefont {{Achilli}}},
  \bibinfo {author} {\bibfnamefont {M.}~\bibnamefont {{Rom{\'e}}}},\ and\
  \bibinfo {author} {\bibfnamefont {L.}~\bibnamefont {{Comisso}}},\ }\bibfield
  {title} {\enquote {\bibinfo {title} {{Coexistence of Plasmoid and
  Kelvin-Helmholtz Instabilities in Collisionless Plasma Turbulence}},}\ }\href
  {https://doi.org/10.3847/1538-4357/ac582f} {\bibfield  {journal} {\bibinfo
  {journal} {apj}\ }\textbf {\bibinfo {volume} {929}},\ \bibinfo {eid} {62}
  (\bibinfo {year} {2022})}\BibitemShut {NoStop}%
\bibitem [{\citenamefont {{Granier}}\ \emph {et~al.}(2024)\citenamefont
  {{Granier}}, \citenamefont {{Tassi}}, \citenamefont {{Laveder}},
  \citenamefont {{Passot}},\ and\ \citenamefont {{Sulem}}}]{GranierPoP24}%
  \BibitemOpen
  \bibfield  {author} {\bibinfo {author} {\bibfnamefont {C.}~\bibnamefont
  {{Granier}}}, \bibinfo {author} {\bibfnamefont {E.}~\bibnamefont {{Tassi}}},
  \bibinfo {author} {\bibfnamefont {D.}~\bibnamefont {{Laveder}}}, \bibinfo
  {author} {\bibfnamefont {T.}~\bibnamefont {{Passot}}},\ and\ \bibinfo
  {author} {\bibfnamefont {P.~L.}\ \bibnamefont {{Sulem}}},\ }\bibfield
  {title} {\enquote {\bibinfo {title} {{Influence of ion-to-electron
  temperature ratio on tearing instability and resulting subion-scale
  turbulence in a low-{\ensuremath{\beta}}$_{e}$ collisionless plasma}},}\
  }\href {https://doi.org/10.1063/5.0185897} {\bibfield  {journal} {\bibinfo
  {journal} {Physics of Plasmas}\ }\textbf {\bibinfo {volume} {31}},\ \bibinfo
  {eid} {032115} (\bibinfo {year} {2024})},\ \Eprint
  {https://arxiv.org/abs/2311.01539} {arXiv:2311.01539 [physics.plasm-ph]}
  \BibitemShut {NoStop}%
\bibitem [{\citenamefont {{Fermo}}, \citenamefont {{Drake}},\ and\
  \citenamefont {{Swisdak}}(2012)}]{FermoPRL12}%
  \BibitemOpen
  \bibfield  {author} {\bibinfo {author} {\bibfnamefont {R.~L.}\ \bibnamefont
  {{Fermo}}}, \bibinfo {author} {\bibfnamefont {J.~F.}\ \bibnamefont
  {{Drake}}},\ and\ \bibinfo {author} {\bibfnamefont {M.}~\bibnamefont
  {{Swisdak}}},\ }\bibfield  {title} {\enquote {\bibinfo {title} {{Secondary
  Magnetic Islands Generated by the Kelvin-Helmholtz Instability in a
  Reconnecting Current Sheet}},}\ }\href
  {https://doi.org/10.1103/PhysRevLett.108.255005} {\bibfield  {journal}
  {\bibinfo  {journal} {prl}\ }\textbf {\bibinfo {volume} {108}},\ \bibinfo
  {eid} {255005} (\bibinfo {year} {2012})}\BibitemShut {NoStop}%
\bibitem [{\citenamefont {{Porcelli}}(1991)}]{Porcelli_PRL91}%
  \BibitemOpen
  \bibfield  {author} {\bibinfo {author} {\bibfnamefont {F.}~\bibnamefont
  {{Porcelli}}},\ }\bibfield  {title} {\enquote {\bibinfo {title}
  {{Collisionless m=1 tearing mode}},}\ }\href
  {https://doi.org/10.1103/PhysRevLett.66.425} {\bibfield  {journal} {\bibinfo
  {journal} {prl}\ }\textbf {\bibinfo {volume} {66}},\ \bibinfo {pages}
  {425--428} (\bibinfo {year} {1991})}\BibitemShut {NoStop}%
\bibitem [{\citenamefont {{Tassi}}, \citenamefont {{Waelbroeck}},\ and\
  \citenamefont {{Grasso}}(2010)}]{Tassi_JPhCS10}%
  \BibitemOpen
  \bibfield  {author} {\bibinfo {author} {\bibfnamefont {E.}~\bibnamefont
  {{Tassi}}}, \bibinfo {author} {\bibfnamefont {F.~L.}\ \bibnamefont
  {{Waelbroeck}}},\ and\ \bibinfo {author} {\bibfnamefont {D.}~\bibnamefont
  {{Grasso}}},\ }\bibfield  {title} {\enquote {\bibinfo {title} {{Gyrofluid
  simulations of collisionless reconnection in the presence of diamagnetic
  effects}},}\ }in\ \href {https://doi.org/10.1088/1742-6596/260/1/012020}
  {\emph {\bibinfo {booktitle} {Journal of Physics Conference Series}}},\
  \bibinfo {series} {Journal of Physics Conference Series}, Vol.\ \bibinfo
  {volume} {260}\ (\bibinfo {year} {2010})\ p.\ \bibinfo {pages} {012020},\
  \Eprint {https://arxiv.org/abs/1109.0092} {arXiv:1109.0092
  [physics.plasm-ph]} \BibitemShut {NoStop}%
\bibitem [{\citenamefont {Maeyama}\ \emph {et~al.}(2014)\citenamefont
  {Maeyama}, \citenamefont {Ishizawa}, \citenamefont {Watanabe}, \citenamefont
  {Nakata}, \citenamefont {Miyato}, \citenamefont {Yagi},\ and\ \citenamefont
  {Idomura}}]{Maeyama_PoP2014}%
  \BibitemOpen
  \bibfield  {author} {\bibinfo {author} {\bibfnamefont {S.}~\bibnamefont
  {Maeyama}}, \bibinfo {author} {\bibfnamefont {A.}~\bibnamefont {Ishizawa}},
  \bibinfo {author} {\bibfnamefont {T.~H.}\ \bibnamefont {Watanabe}}, \bibinfo
  {author} {\bibfnamefont {M.}~\bibnamefont {Nakata}}, \bibinfo {author}
  {\bibfnamefont {N.}~\bibnamefont {Miyato}}, \bibinfo {author} {\bibfnamefont
  {M.}~\bibnamefont {Yagi}},\ and\ \bibinfo {author} {\bibfnamefont
  {Y.}~\bibnamefont {Idomura}},\ }\bibfield  {title} {\enquote {\bibinfo
  {title} {{Comparison between kinetic-ballooning-mode-driven turbulence and
  ion-temperature-gradient-driven turbulence}},}\ }\href
  {https://doi.org/10.1063/1.4873379} {\bibfield  {journal} {\bibinfo
  {journal} {Physics of Plasmas}\ }\textbf {\bibinfo {volume} {21}} (\bibinfo
  {year} {2014}),\ 10.1063/1.4873379}\BibitemShut {NoStop}%
\bibitem [{\citenamefont {Ishizawa}\ \emph {et~al.}(2015)\citenamefont
  {Ishizawa}, \citenamefont {Maeyama}, \citenamefont {Watanabe}, \citenamefont
  {Sugama},\ and\ \citenamefont {Nakajima}}]{IshizawaJPP15}%
  \BibitemOpen
  \bibfield  {author} {\bibinfo {author} {\bibfnamefont {A.}~\bibnamefont
  {Ishizawa}}, \bibinfo {author} {\bibfnamefont {S.}~\bibnamefont {Maeyama}},
  \bibinfo {author} {\bibfnamefont {T.~H.}\ \bibnamefont {Watanabe}}, \bibinfo
  {author} {\bibfnamefont {H.}~\bibnamefont {Sugama}},\ and\ \bibinfo {author}
  {\bibfnamefont {N.}~\bibnamefont {Nakajima}},\ }\bibfield  {title} {\enquote
  {\bibinfo {title} {{Electromagnetic gyrokinetic simulation of turbulence in
  torus plasmas}},}\ }\href {https://doi.org/10.1017/S0022377815000100}
  {\bibfield  {journal} {\bibinfo  {journal} {Journal of Plasma Physics}\
  }\textbf {\bibinfo {volume} {81}} (\bibinfo {year} {2015}),\
  10.1017/S0022377815000100}\BibitemShut {NoStop}%
\bibitem [{\citenamefont {{Zonca}}\ \emph {et~al.}(1999)\citenamefont
  {{Zonca}}, \citenamefont {{Chen}}, \citenamefont {{Dong}},\ and\
  \citenamefont {{Santoro}}}]{ZoncaPoP99}%
  \BibitemOpen
  \bibfield  {author} {\bibinfo {author} {\bibfnamefont {F.}~\bibnamefont
  {{Zonca}}}, \bibinfo {author} {\bibfnamefont {L.}~\bibnamefont {{Chen}}},
  \bibinfo {author} {\bibfnamefont {J.~Q.}\ \bibnamefont {{Dong}}},\ and\
  \bibinfo {author} {\bibfnamefont {R.~A.}\ \bibnamefont {{Santoro}}},\
  }\bibfield  {title} {\enquote {\bibinfo {title} {{Existence of ion
  temperature gradient driven shear Alfv{\'e}n instabilities in tokamaks}},}\
  }\href {https://doi.org/10.1063/1.873449} {\bibfield  {journal} {\bibinfo
  {journal} {Physics of Plasmas}\ }\textbf {\bibinfo {volume} {6}},\ \bibinfo
  {pages} {1917--1924} (\bibinfo {year} {1999})}\BibitemShut {NoStop}%
\bibitem [{\citenamefont {{Falchetto}}, \citenamefont {{Vaclavik}},\ and\
  \citenamefont {{Villard}}(2003)}]{FalchettoPoP03}%
  \BibitemOpen
  \bibfield  {author} {\bibinfo {author} {\bibfnamefont {G.~L.}\ \bibnamefont
  {{Falchetto}}}, \bibinfo {author} {\bibfnamefont {J.}~\bibnamefont
  {{Vaclavik}}},\ and\ \bibinfo {author} {\bibfnamefont {L.}~\bibnamefont
  {{Villard}}},\ }\bibfield  {title} {\enquote {\bibinfo {title}
  {{Global-gyrokinetic study of finite {\ensuremath{\beta}} effects on linear
  microinstabilities}},}\ }\href {https://doi.org/10.1063/1.1566028} {\bibfield
   {journal} {\bibinfo  {journal} {Physics of Plasmas}\ }\textbf {\bibinfo
  {volume} {10}},\ \bibinfo {pages} {1424--1436} (\bibinfo {year}
  {2003})}\BibitemShut {NoStop}%
\bibitem [{\citenamefont {Waelbroeck}(2005)}]{waelbroeck_PRL2005}%
  \BibitemOpen
  \bibfield  {author} {\bibinfo {author} {\bibfnamefont {F.~L.}\ \bibnamefont
  {Waelbroeck}},\ }\bibfield  {title} {\enquote {\bibinfo {title} {{Natural
  Velocity of Magnetic Islands}},}\ }\href
  {https://doi.org/10.1103/PhysRevLett.95.035002} {\bibfield  {journal}
  {\bibinfo  {journal} {Physical Review Letters}\ }\textbf {\bibinfo {volume}
  {95}},\ \bibinfo {pages} {35002} (\bibinfo {year} {2005})}\BibitemShut
  {NoStop}%
\bibitem [{\citenamefont {Haye}\ \emph {et~al.}(2003)\citenamefont {Haye},
  \citenamefont {Petty}, \citenamefont {Strait}, \citenamefont {Waelbroeck},\
  and\ \citenamefont {Wilson}}]{lahaye_PoP2003}%
  \BibitemOpen
  \bibfield  {author} {\bibinfo {author} {\bibfnamefont {R.~J.~L.}\
  \bibnamefont {Haye}}, \bibinfo {author} {\bibfnamefont {C.~C.}\ \bibnamefont
  {Petty}}, \bibinfo {author} {\bibfnamefont {E.~J.}\ \bibnamefont {Strait}},
  \bibinfo {author} {\bibfnamefont {F.~L.}\ \bibnamefont {Waelbroeck}},\ and\
  \bibinfo {author} {\bibfnamefont {H.~R.}\ \bibnamefont {Wilson}},\ }\bibfield
   {title} {\enquote {\bibinfo {title} {{Propagation of magnetic islands in the
  Er = 0 frame of co-injected neutral beam driven discharges in the DIII-D
  tokamak}},}\ }\href {https://doi.org/10.1063/1.1602452} {\bibfield  {journal}
  {\bibinfo  {journal} {Physics of Plasmas}\ }\textbf {\bibinfo {volume}
  {10}},\ \bibinfo {pages} {3644--3648} (\bibinfo {year} {2003})}\BibitemShut
  {NoStop}%
\bibitem [{\citenamefont {Horton}(1999)}]{Horton_RMP99}%
  \BibitemOpen
  \bibfield  {author} {\bibinfo {author} {\bibfnamefont {W.}~\bibnamefont
  {Horton}},\ }\bibfield  {title} {\enquote {\bibinfo {title} {{Drift waves and
  transport}},}\ }\href {https://doi.org/10.1103/RevModPhys.71.735} {\bibfield
  {journal} {\bibinfo  {journal} {Reviews of Modern Physics}\ }\textbf
  {\bibinfo {volume} {71}},\ \bibinfo {pages} {735--778} (\bibinfo {year}
  {1999})}\BibitemShut {NoStop}%
\bibitem [{\citenamefont {Ryter}\ \emph {et~al.}(2005)\citenamefont {Ryter},
  \citenamefont {Angioni}, \citenamefont {Peeters}, \citenamefont {Leuterer},
  \citenamefont {Fahrbach},\ and\ \citenamefont {Suttrop}}]{RyterPRL05}%
  \BibitemOpen
  \bibfield  {author} {\bibinfo {author} {\bibfnamefont {F.}~\bibnamefont
  {Ryter}}, \bibinfo {author} {\bibfnamefont {C.}~\bibnamefont {Angioni}},
  \bibinfo {author} {\bibfnamefont {A.~G.}\ \bibnamefont {Peeters}}, \bibinfo
  {author} {\bibfnamefont {F.}~\bibnamefont {Leuterer}}, \bibinfo {author}
  {\bibfnamefont {H.-U.}\ \bibnamefont {Fahrbach}},\ and\ \bibinfo {author}
  {\bibfnamefont {W.}~\bibnamefont {Suttrop}} (\bibinfo {collaboration} {ASDEX
  Upgrade Team}),\ }\bibfield  {title} {\enquote {\bibinfo {title}
  {Experimental study of trapped-electron-mode properties in tokamaks:
  Threshold and stabilization by collisions},}\ }\href
  {https://doi.org/10.1103/PhysRevLett.95.085001} {\bibfield  {journal}
  {\bibinfo  {journal} {Phys. Rev. Lett.}\ }\textbf {\bibinfo {volume} {95}},\
  \bibinfo {pages} {085001} (\bibinfo {year} {2005})}\BibitemShut {NoStop}%
\bibitem [{\citenamefont {{Dannert}}\ and\ \citenamefont
  {{Jenko}}(2005)}]{DannertPoP2005}%
  \BibitemOpen
  \bibfield  {author} {\bibinfo {author} {\bibfnamefont {T.}~\bibnamefont
  {{Dannert}}}\ and\ \bibinfo {author} {\bibfnamefont {F.}~\bibnamefont
  {{Jenko}}},\ }\bibfield  {title} {\enquote {\bibinfo {title} {{Gyrokinetic
  simulation of collisionless trapped-electron mode turbulence}},}\ }\href
  {https://doi.org/10.1063/1.1947447} {\bibfield  {journal} {\bibinfo
  {journal} {Physics of Plasmas}\ }\textbf {\bibinfo {volume} {12}},\ \bibinfo
  {eid} {072309} (\bibinfo {year} {2005})}\BibitemShut {NoStop}%
\bibitem [{\citenamefont {{Zarzoso}}\ \emph
  {et~al.}(2015{\natexlab{b}})\citenamefont {{Zarzoso}}, \citenamefont
  {{Hornsby}}, \citenamefont {{Poli}}, \citenamefont {{Casson}}, \citenamefont
  {{Peeters}},\ and\ \citenamefont {{Nasr}}}]{ZarzosoNF15}%
  \BibitemOpen
  \bibfield  {author} {\bibinfo {author} {\bibfnamefont {D.}~\bibnamefont
  {{Zarzoso}}}, \bibinfo {author} {\bibfnamefont {W.~A.}\ \bibnamefont
  {{Hornsby}}}, \bibinfo {author} {\bibfnamefont {E.}~\bibnamefont {{Poli}}},
  \bibinfo {author} {\bibfnamefont {F.~J.}\ \bibnamefont {{Casson}}}, \bibinfo
  {author} {\bibfnamefont {A.~G.}\ \bibnamefont {{Peeters}}},\ and\ \bibinfo
  {author} {\bibfnamefont {S.}~\bibnamefont {{Nasr}}},\ }\bibfield  {title}
  {\enquote {\bibinfo {title} {{Impact of rotating magnetic islands on density
  profile flattening and turbulent transport}},}\ }\href
  {https://doi.org/10.1088/0029-5515/55/11/113018} {\bibfield  {journal}
  {\bibinfo  {journal} {Nuclear Fusion}\ }\textbf {\bibinfo {volume} {55}},\
  \bibinfo {eid} {113018} (\bibinfo {year} {2015}{\natexlab{b}})}\BibitemShut
  {NoStop}%
\bibitem [{\citenamefont {{Smolyakov}}\ \emph {et~al.}(1995)\citenamefont
  {{Smolyakov}}, \citenamefont {{Hirose}}, \citenamefont {{Lazzaro}},
  \citenamefont {{Re}},\ and\ \citenamefont {{Callen}}}]{SmolyakovPoP95}%
  \BibitemOpen
  \bibfield  {author} {\bibinfo {author} {\bibfnamefont {A.~I.}\ \bibnamefont
  {{Smolyakov}}}, \bibinfo {author} {\bibfnamefont {A.}~\bibnamefont
  {{Hirose}}}, \bibinfo {author} {\bibfnamefont {E.}~\bibnamefont {{Lazzaro}}},
  \bibinfo {author} {\bibfnamefont {G.~B.}\ \bibnamefont {{Re}}},\ and\
  \bibinfo {author} {\bibfnamefont {J.~D.}\ \bibnamefont {{Callen}}},\
  }\bibfield  {title} {\enquote {\bibinfo {title} {{Rotating nonlinear magnetic
  islands in a tokamak plasma}},}\ }\href {https://doi.org/10.1063/1.871308}
  {\bibfield  {journal} {\bibinfo  {journal} {Physics of Plasmas}\ }\textbf
  {\bibinfo {volume} {2}},\ \bibinfo {pages} {1581--1598} (\bibinfo {year}
  {1995})}\BibitemShut {NoStop}%
\bibitem [{\citenamefont {{Biskamp}}\ and\ \citenamefont
  {{Sato}}(1997)}]{BiskampPoP97}%
  \BibitemOpen
  \bibfield  {author} {\bibinfo {author} {\bibfnamefont {D.}~\bibnamefont
  {{Biskamp}}}\ and\ \bibinfo {author} {\bibfnamefont {T.}~\bibnamefont
  {{Sato}}},\ }\bibfield  {title} {\enquote {\bibinfo {title} {{Partial
  reconnection in the nonlinear internal kink mode}},}\ }\href
  {https://doi.org/10.1063/1.872308} {\bibfield  {journal} {\bibinfo  {journal}
  {Physics of Plasmas}\ }\textbf {\bibinfo {volume} {4}},\ \bibinfo {pages}
  {1326--1329} (\bibinfo {year} {1997})}\BibitemShut {NoStop}%
\bibitem [{\citenamefont {{Dudkovskaia}}\ \emph {et~al.}(2021)\citenamefont
  {{Dudkovskaia}}, \citenamefont {{Connor}}, \citenamefont {{Dickinson}},
  \citenamefont {{Hill}}, \citenamefont {K.}, \citenamefont {{Leigh}},\ and\
  \citenamefont {{Wilson}}}]{DudkovskaiaPPCF21}%
  \BibitemOpen
  \bibfield  {author} {\bibinfo {author} {\bibfnamefont {A.~V.}\ \bibnamefont
  {{Dudkovskaia}}}, \bibinfo {author} {\bibfnamefont {J.~W.}\ \bibnamefont
  {{Connor}}}, \bibinfo {author} {\bibfnamefont {D.}~\bibnamefont
  {{Dickinson}}}, \bibinfo {author} {\bibfnamefont {P.}~\bibnamefont {{Hill}}},
  \bibinfo {author} {\bibfnamefont {I.}~\bibnamefont {K.}}, \bibinfo {author}
  {\bibfnamefont {S.}~\bibnamefont {{Leigh}}},\ and\ \bibinfo {author}
  {\bibfnamefont {H.~R.}\ \bibnamefont {{Wilson}}},\ }\bibfield  {title}
  {\enquote {\bibinfo {title} {{Drift kinetic theory of neoclassical tearing
  modes in a low collisionality tokamak plasma: magnetic island threshold
  physics}},}\ }\href {https://doi.org/10.1088/1361-6587/abea2e} {\bibfield
  {journal} {\bibinfo  {journal} {Plasma Physics and Controlled Fusion}\
  }\textbf {\bibinfo {volume} {63}},\ \bibinfo {pages} {054001} (\bibinfo
  {year} {2021})}\BibitemShut {NoStop}%
\end{thebibliography}%

\end{document}